\documentclass[12pt,english]{article}
\usepackage{subfigure}
\usepackage{graphicx}
\usepackage{rotating}

\usepackage{hyperref}
\usepackage{amssymb}

\providecommand{\tabularnewline}{\\}

\makeatletter

\usepackage{babel}
\makeatother
\usepackage{geometry}
\usepackage{babel}
\usepackage{graphics}
\def\beq{\begin{equation}}
\def\eeq{\end{equation}}

\def\beqn{\begin{eqnarray}}
\def\eeqn{\end{eqnarray}}
\def\ba{\begin{eqnarray}}
\def\ea{\end{eqnarray}}

\def\atp{\frac{\alpha_s(Q^2)}{2\pi}}

\newcommand{\beqa}{\begin{eqnarray}}
\newcommand{\eeqa}{\end{eqnarray}}

\makeatletter

\providecommand{\LyX}{L\kern-.1667em\lower.25em\hbox{Y}\kern-.125emX\@}

\makeatother

\begin{document}

\begin{center}
\vspace{2.cm}
{\bf \large NNLO Logarithmic Expansions and Precise Determinations of the
Neutral Currents near the Z Resonance at the LHC:\\ The Drell-Yan case}

\vspace{.5cm}

 {\large \bf $^{1,3}$Alessandro Cafarella, $^{2,3}$Claudio Corian\`{o},}
\large {\bf  $^{2}$Marco Guzzi\\}
\vspace{0.5cm}
{\it $^1$Institute of Nuclear Physics, NCSR ``Demokritos'', 15310 Athens, Greece\\}
\vspace{.5cm}
{\it $^2$Dipartimento di Fisica, Universit\`{a} del Salento\\
 and
INFN Sezione di Lecce,
Via per Arnesano, 73100 Lecce, Italy\\}

\vspace{0.5cm}
{\it $^3$Department of Physics and Institute of Plasma Physics, University of
Crete, GR-710 03 Heraklion, Crete, Greece\\}
\vspace{0.5cm}

\end{center}
\vspace{0cm}
\begin{abstract}

We present a comparative study of the invariant mass and
rapidity distributions in Drell-Yan lepton pair production, with particular
emphasis on the role played by
the QCD evolution. We focus our study around the Z resonance
($50 <Q < 200$ GeV) and perform a general analysis of the
factorization/renormalization scale dependence of the cross sections,
with the two scales included both in the evolution and in the hard scatterings.
We also present the variations of the
cross sections due to the errors on the parton distributions (pdf's) and
an analysis of the corresponding $K$-factors.
Predictions from several sets of pdf's, evolved by MRST and Alekhin are
compared with those generated using \textsc{Candia}, a NNLO evolution program that implements
the theory of the logarithmic expansions, developed in a previous work.
These expansions allow to select truncated
solutions of varying accuracy using the method of the $x$-space iterates.
The evolved parton distributions are in good agreement 
with other approaches. The study can be generalized 
for high precision searches of extra neutral gauge interactions at the LHC.

\end{abstract}
\newpage

\section{Introduction}
Accurate determinations of the QCD background at the LHC, especially for some selected hadronic cross sections,
 are going to be very important in order to increase our potential for
new discoveries. For this reason it is necessary to know the size of the radiative corrections to some selected processes at higher orders. At the same time, the quantification of the impact of the errors in the determination of these observables is going to be critical in order to enhance our confidence on the reliability of the perturbative
expansion. It is particularly so in the search for extra Z', which are ubiquitous in extensions of the Standard Model \cite{Langacker} \cite{Daleo} \cite{Leike},
for instance in models derived from the
string construction \cite{Alon} or with extra neutral interactions
modified by an anomaly inflow \cite{CorianoIrgesMorelli}, where mass-dependent and anomaly-related corrections require
very high accuracy to be properly identified and separated from the large QCD background. 

With these motivations in mind we have proceeded with an independent analysis of the next-to-next-to-leading order (NNLO) QCD corrections, starting 
from an accurate investigation of the impact of the evolution on the physical observables at the LHC. 
In this context, the role played by the Drell Yan cross section is particularly important. In fact, the possibility of discovering extra neutral currents at the new collider may be related to the determination 
of this cross section with very high accuracy far beyond the Z peak, at values of the 
invariant mass of the lepton pair up to 5 TeV \cite{Langacker}, which is commonly thought to be the upper 
limit in searches of this type. For this reason, a precise determination of the pdf's at any value of the Bjorken variable $x$ \cite{Amanda} is needed (see \cite{Sterman} and refs. therein). Issues of resummation of 
the perturbative expansions become also critical for the correct determinations of several 
distributions at the edge of phase space (see for example \cite{Nason}). 

At this time, while our knowledge of the role played by the coefficients of
the QCD hard scatterings in some key partonic processes is quite satisfactory, 
that of the behavior of the pdf's
is not of a comparable level,
and the model-dependence of the various parameterizations is still large.
Recent parameterizations of the pdf's come with the quantification of their errors,
presented to next-to-leading order (NLO) and NNLO, whenever possible,
obtained by the fits used by various groups (we limit our analysis to \cite{Alekhin} and \cite{MRST1}) to match several sets of experimental 
data in pp collisions, such as Drell-Yan and Deep Inelastic Scattering (DIS). These errors, which estimate the
goodness of a fit, are naturally thought of being of experimental origin. But there are also other sources of
indetermination, mostly of theoretical origin, which need to be taken into consideration.
One of them, apparently of more trivial nature, is related to the way the evolution is implemented through NNLO.

At a first look this last point might be misinterpreted and the corresponding ``error'' coming to be attributed to the 
``model dependence'' of a given parameterization set, while it 
amounts to a theoretical indetermination, intrinsic to perturbation theory, since it is going to be there for any 
chosen model of pdf's. The reason is simple and also quite immediate: there is not a unique approach to solve the DGLAP equation, and, again, not for a numerical/algorithmic reason, related 
to the limited numerical precision of a given algorithm. In fact, a given solution, of a typical accuracy, 
organizes the logarithmic corrections in a specific and unique way. These are summed or, eventually, 
resummed if exact or, instead, accurate (truncated) solutions are selected. Therefore, the issue of determining the best possible way to solve the evolution does 
not seem to have a unique answer, 
being directly related to the possibility of choosing among different theoretical approaches, all equally acceptable. 
The goal of this work is threefold: to test the accuracy of the logarithmic expansions proposed by us in a previous 
work by comparing with other methods of solution; to quantify the experimental errors on the DY 
cross section coming from the pdf's and to  analyze the differences between accurate and ``exact'' solutions of the evolution equations. The 
method used by us to solve the DGLAP equations is based on a simple perturbative 
organization of the logarithmic corrections that we call ``the NNLO {\em logarithmic expansion} in $x$-space''. These expansions work for kernels given to all orders.

Thanks to the availability of the two-loop evolution kernels \cite{vogt1} and of the
corresponding hard scattering coefficients, which in the Drell-Yan case 
have been known for some time \cite{Van_Neerven1}, complete NNLO analysis are now possible and allow to perform sophisticated tests that 
give to us the opportunity to quantify the effects that we have just mentioned.
In the case of Drell-Yan both the total cross section and, with a modification,
the rapidity distributions of the lepton pair on the final state are at reach.
The latter have been presented recently \cite{Anastasiou}, together with a dependence of the predictions
on the factorization scale, which is important in order
to monitor the overall stability of the perturbative series in $\alpha_s $, the strong coupling constant.
This requires the determination of the cross section for a varying factorization scale and of
the relevant $K$-factors at various energies. However, in order to 
determine in a robust way the size of the NNLO corrections and the role that they 
play in some important predictions, we believe that it is mandatory to perform 
an independent analysis of the evolution, defining benchmarks for the evolution of the 
pdf's from different perspectives respect to the common ones. 
These are based either on numerical discretizations 
({so called {\em brute force} methods), which are affected by contributions of all-orders in the strong coupling, or on the methods of the Mellin moments, using special 
expansions. Accuracy means that we define the perturbative solution so to include only parts of the corrections, in a certain expansion in the strong coupling \cite{CCG1}, given our limited knowledge of the perturbative expansion of the kernels.

It is well known that the 
issue of accuracy in the choice of the solution of the equation 
has never been fully addressed in the previous literature. While this issue is less important 
at NLO, given the size of the $K$-factors which are about $20 \% $ in the region that we 
explore and at the energy that we select ($\sqrt{S}=14$ TeV) for a p-p collider, 
things become more subtle at NNLO, where the relative $K$-factors relating the NLO to the NNLO 
cross section are determined to be much smaller. We will show that it is the QCD evolution 
to drive the NNLO cross section to an overall reduction in the region that we have analyzed.
As a result of our analysis we are able to quantify the theoretical error implicit
in the various choices of the evolution scheme, which are smaller respect
to the errors in the pdf's, but not
insignificant for the rest. However, if one intends to take into considerations the impact of
possible resummations on the pdf's in a quantitative way, then the issues that we raise become crucial for obtaining a
correct quantitative answer. In fact, it is also quite likely that, in the long run, with the large data flow from the LHC, these sources of errors that we investigate will become more 
significant in order to obtain more precise parameterizations of the pdf's in the future. Our approach is part of an ongoing effort to develop a complete numerical program, \textsc{Candia}, 
that we hope will include not only an analysis of the evolution, with applications to DY and the Higgs sector 
through NNLO, with the inclusion of resummation effects, and so on. At this time \textsc{Candia} contains the DY and Higgs total 
cross sections beside, of course, the evolution.

\subsection{Comparison with the previous literature}

Regarding the constraints on our analysis, some comments are in order.
In comparing the various results one possible source of disagreement
lays in the treatment of the heavy flavours.
\textsc{Candia} and \textsc{Pegasus} treat the heavy flavours following
the varying flavor number scheme (VFN) as in \cite{Smith}
\footnote{We will denote with the acronym VFN the varying flavour number scheme that follows
the treatment of \cite{Smith}.}.
MRST and Alekhin instead, follow different prescriptions, respectively described in \cite{Thorne} and
\cite{Alekhin2}.

We have also found that other finer issues, in general not discussed in the previous literature,
introduce systematic differences. For instance, MRST give a parameterized form for the input
distributions at $\mu_0^2=1$ GeV$^2$, while the lowest value of $\mu_f^2$ in their grid is $1.25$ GeV$^2$.
In Alekhin's case, this author gives an analytic
parameterization at 9 GeV$^2$ without including the charm quark, even if the initial 
scale is above the charm threshold. The charm contribution
is instead present in his evolved pdf's, available on a grid.
The differences induced on the cross sections by the two methods are not negligible.
A source of disagreement between \textsc{Candia} and \textsc{Pegasus} can be attributed to the
fact that a given initial condition has to be fitted to a certain
functional form in Mellin space, if one solves the equations using Mellin moments.
These limitations are absent if one works directly in $x$-space \cite{CCG1}.

\subsection{Our approach}
As we have already mentioned, we base our analysis entirely on
the implementation of an algorithm that solves the DGLAP equations
directly in $x$-space and uses an ansatz based on various logarithmic expansions. These expansions have
been shown to be related either to exact or to truncated solutions of the renormalization group equations
(RGE's), characterized by coefficients which are determined recursively. Notice that these
expansions are also typical of Mellin space \cite{EllisKunstLevin}, \cite{Pegasus}.

The structure of the recursion
relations solved by this method is fixed by the choice of the original DGLAP equation and by the
approximations performed on its right-hand-side (rhs), justified in a perturbative fashion.
Beside the logarithmic expansions, also exact solutions are available for
the nonsinglet sector up to NNLO, as we have shown, \cite{CCG1}, which are useful in order to establish the
convergence of the expansion toward the exact solution, whenever, of course, this is available. The method allows to bypass the appearance of commutators in the definition of
the iterated solution from Mellin space, which is an unavoidable step in the singlet sector,
as done in all the previous literature \cite{EllisKunstLevin},\cite{Pegasus},\cite{Petronzio}.
The logarithmic expansions, in both sectors, instead, take exactly the same simple
form, being either scalar or matrix-valued. The relations for the unknown coefficient
functions, introduced by the ansatz, and which are determined recursively, are also the same in both cases.

In this paper we elaborate on the main features of these expansions by performing a
thorough numerical analysis of the various truncated solutions introduced in our previous paper \cite{CCG1}, discussing their behavior. In the singlet sector we also
show the fast numerical convergence of the expansion and compare our results with
those of the Les-Houches benchmarks which are based on a toy model of initial conditions.
Comparisons are done both for
a fixed and for a varying flavor number. The anomalous dimensions involving the
heavy flavors have been implemented as in \cite{Smith, Pegasus}.

We are going to see that variations induced by the choice of the solution 
induces variations on the cross section of the order of $1 \%$ or so at NNLO, and clearly affect also the 
NNLO $K$-factor for the total cross section. In our determination, the change
in the value of the cross section from NLO to NNLO is around $4 \%$ on the Z 
peak, while the MRST and the Alekhin determinations are $2.6 \%$ 
and about $1.5 \%$ respectively. While these variations appear to be more modest compared to 
the analogous ones at a lower order (which are of the order of $20 \%$ or so), they are nevertheless important 
for the discovery of extra neutral currents at large invariant mass of the lepton pair in DY, given the fast falling cross section at those large values. 
However, as we are going to show, the errors on the pdf's induce percentile variations of the cross section 
as we move from NLO to NNLO of the order of $4 \%$ around
the best-fit result, reducing the NNLO cross section compared to the
NLO prediction and rendering these results compatible.

\section{Initial sets}

Our comparisons are performed using an implementation of the theory in a code 
written by us and called {\textsc{Candia}} that will be documented elsewhere. 
The other implementation that we can directly compare to is the code written by A. Vogt, \textsc{Pegasus}
\cite{Pegasus}, which implements the Mellin-transform method. 
This is the only NNLO code which is of public domain at this time. 
\textsc{Pegasus} can be run in different modes and allows to select numerical 
solutions of a given accuracy. Our evolution is also compared to the MRST evolution. 
We clarify in our results if we have used in our implementations of the hard scatterings
the MRST input evolved by MRST or our evolution of the same input.
For the evolution of the MRST parametric input with \textsc{Candia}
we have worked in the same VFN scheme, but we have used a slightly different prescription,
and the comparisons are performed either starting from their values on the grids
or from the parametric input provided by the same authors.

As we will show below,
the grid and the parametric inputs (the first at $\mu_0^2=1.25$ GeV$^2$,
the second at $\mu_0^2=1$ GeV$^2$) give results which differ at the percent level.

The closeness between
our predictions, obtained using the asymptotic solution, whose nature we clarify below,
and their distributions is also quite evident, but the variations are such to generate differences 
at the percent level 
in the cross sections. \textsc{Candia} generates numerical outputs of the exact solutions up to NNLO
for the nonsinglet sector and truncated solutions with arbitrary
order of accuracy in the singlet sector. In our case the term ``arbitrary order of accuracy'', referred to a solution, means that we use logarithmic expansions in $\log(\alpha_s)$ multiplied by coefficients of a certain power of $\alpha_s$ rather than $\log(\chi(\alpha_s))$, where $\chi(\alpha_s)$ is a 
typical NLO function of the running coupling or some other non-trivial (composed) 
function generated at higher orders. We remark that the simple $\log(\alpha_s)$ expansion converges 
after very few iterates (7-8), with a precision of 4 to 5 significant digits. Clearly, if one 
is searching for exact solutions, the iterates converge rather slowly 
to give the ``exact'' numerical solution since the leading order solution is not 
factored out in the ansatz, as in the case of the $U$-ansatz of \cite{EllisKunstLevin,Pegasus}, as we will clarify below. The differences in the cross sections are very small (0.1 \%),
if an asymptotic truncated solution (with $\kappa^{\prime}=7$ or $8$) replaces the ``exact'' solution, 
or {\em brute force} solution. But the theoretical indetermination remains: 
at NNLO even the second truncated solution is a solution and the differences on the observables, as we are going to show, become more substantial than the fraction 
of a percent obtained using the asymptotic solution. 

As we are going to show next,
the results produced by our implementation are in excellent
agreement up to NNLO with those obtained with
\textsc{Pegasus} and the Les Houches benchmarks (see refs.
\cite{LesHouches02}, \cite{LesHouches05}). Regarding the computation of the
errors on the pdf's, these are not available for all the
most popular sets and through all orders. For instance, the Alekhin and MRST fits are presented up to NNLO,
but only one of them, the set \cite{Alekhin} presents errors through NNLO.

Coming to describe the region that we have studied, our numerical investigation 
covers both the resonance region around the peak of the Z - this being useful
in order to assess the impact of the corrections at the various orders -
and the remaining regions of faster fall-off.
We remark that these studies can also be extended to the search for
extra $Z^{\prime}$ in extensions of the Standard Model, and as such provide a clear indication of the role played
by these corrections in a precise determination of the QCD/electroweak background, useful
for potential discoveries of additional neutral currents at the LHC using this process.

\subsection{A classification of the possible solutions}

We start our study by briefly reviewing the nature of the
NNLO solutions and the level of accuracy which is intrinsic 
to any solution. 

As we have already mentioned, there are essentially three types of solutions that one can extract from the evolution equations. We have decided to classify 
them as follows:  1) the accurate solutions with few iterates, 
2) the iterated solutions with a large number of iterates, eventually combined 
with exact analytical solutions of the nonsinglet sector and
3) the brute force solutions. 

Solutions of type-1 are built using a very simple ansatz which can be showed 
to be accurate through NNLO. 
The ansatz contains all the terms of the form $\log^n(\alpha)$, $\alpha_s \log^n(\alpha)$ and 
$\alpha_s^2 \log^n(\alpha)$. Solutions of type-2 include terms of arbitrary 
higher order in the logarithmic expansion. They are characterized by one or two indices, the first 
$(\kappa^{\prime})$ which defines the accuracy and the index $\kappa$ which defines the order of the truncation of the right-hand-side (rhs) of the evolution equation. The two indices $\kappa,\kappa^{\prime}$ can be taken as label of a given truncated solution. They are built starting from one of the two forms of 
the evolution equations (form-1 and form-2), as we will discuss below. The index $\kappa$ appears 
in form-2, when the DGLAP equations are written directly in terms of the logarithmic derivative 
of the running coupling $\log(\alpha_s)$.

 Solutions of type-3 are not accurate since they are affected by contributions of all orders 
and can be obtained using {\em brute force} methods, by discretization of the evolution equations.  
They exceed the level of accuracy typical of a perturbative expansions where 
the kernels are only known up to NNLO ($\alpha_s^2$).

 Truncated solutions, instead, are obtained, as we have just mentioned, by truncating the rhs of
the evolutions equations at a given order and then searching for solutions of these equations
with a certain accuracy. 
Also in this case the truncated equations can be solved exactly, 
giving solutions which exceed the level of accuracy of the expansion. This happens in the 
nonsinglet sector. If we use the DGLAP written in terms of a logarithmic derivative of $Q$ rather 
than of the coupling (form-1), then the NNLO exact nonsinglet solution is available. Similar 
solutions are also available for the form-2 of the equations in the same sector for $\kappa \leq 2$. 
In this second case these solutions do not have, however, a well defined meaning, since they are not accurate nor converge, in the limit $\kappa^{\prime}\to \infty$, to the exact solution of the exact equation 
(as for the ansatz\"e written for form-1, for instance).  Therefore, also in this case an accurate solution is obtained by an additional
expansion in the strong coupling and retaining only terms up to a certain order, which is the 
order of the selected accuracy.

\subsection{The two forms of the evolution}
We proceed by illustrating the two forms that the equations can take. 

We recall that the perturbative expansion of the DGLAP splitting
functions and of the $\beta$-function take the generic form (to the m-th order)
\ba
\label{kernel_NmLO}
&&{P^{N^m LO}(x,Q^2)}=\sum_{i=0}^{m}\left(\atp\right)^{i+1} P^{(i)}(x,Q^2),
\nonumber\\
&&\frac{\partial \beta^{N^m LO}(\alpha_s(Q^2))}{\partial \log{Q^2}}=
\sum_{k=0}^{m}\left(\frac{\alpha_s(Q^2)}{4\pi}\right)^{k+2}\beta_{k},
\ea
with $\beta_k$ being the corresponding coefficients of the $\beta$ function which have been summarized in
\cite{CCG1}. Leading, next to leading and NNLO correspond to the cases $m=0,1,2$ respectively.

The equation can be written either as
\ba
\label{exact2NNLO}
\frac{\partial f(x,\alpha_s(Q^2))}{\partial \log (Q^2)}&=& P^{N^m LO}(x,Q^2)\otimes f(x,\alpha_s(Q^2))
\ea
(form-1)
or, equivalently, as

\ba
\label{exactNNLO}
\frac{\partial f(x,\alpha_s)}{\partial\alpha_s}&=& \frac{P^{N^m LO}(x,Q^2)}{\beta^{N^m LO}}\otimes f(x,\alpha_s)
\ea
(form-2).
While the two forms are equivalent, form-2 needs an expansion of the $1/\beta$ factor. 
This generates on the rhs of the expanded equation an infinite set of truncated equations characterized 
by a parameter of accuracy ($\kappa$). This parameter has to be sent to infinity in order for form-2 
to be equivalent to form-1. When we search for solutions of the DGLAP in the form-1 and we need
to compare with the form-2, the recursion relations for the solution ansatz start to differ after the 
order $\kappa$. Notice that once that we have introduced an expansion of the rhs, such as in form-2, 
we may search either for truncated solutions of this truncated equation or for exact solutions. These 
are options that increase the type of possible solutions, all of them of different theoretical accuracy. To illustrate this point we start from the form-2 of the equations and choose $m=2$ (NNLO), obtaining the truncated equation
\ba
\frac{\partial f(\alpha_s,x)}{\partial\alpha_s}=\frac{1}{\alpha_s}\left[R_0(x)+\alpha_s R_1(x)+
\alpha_s^2 R_2(x)+\cdots+\alpha_s^{\kappa} R_{\kappa}(x)\right]\otimes f(\alpha_s,x).\,
\label{kappa_trunc}
\ea
 The explicit form of the operators $R_\kappa$ can be easily worked out, 
at any perturbative order. For instance at NNLO the first few terms are given by
\ba
&&R_0=-\frac{2}{\beta_0} P^{(0)} \nonumber\\
&&R_1=-\frac{1}{\pi\beta_0} P^{(1)} -\frac{b_1}{(4\pi)}P^{(0)}\nonumber\\
&&R_2=-\frac{1}{\pi\beta_0} P^{(2)} -\frac{b_2}{(4\pi)^2}R_0-\frac{b_1}{(4\pi)}R_1\nonumber\\
&&R_3=-\frac{b_2}{(4\pi)^2}R_1 -\frac{b_1}{(4\pi)}R_2\nonumber\\
&&R_4=-\frac{b_1 b_2}{(4\pi)^3}R_1 +\frac{b_1^2}{(4\pi)^2}R_2-\frac{b_2}{(4\pi)^2}R_2\nonumber\\
&&R_5=-\frac{b_1^2}{(4\pi)^2}R_3 +\frac{b_2^2}{(4\pi)^4}R_1+\frac{b_1 b_2}{(4\pi)^2 (2\pi)}R_2\nonumber\\
&&\vdots\nonumber\\
\ea
where $b_i=\beta_i/\beta_0$, valid in the nonsinglet case.
A similar expansion holds also for the singlet sector, although
in this case the recursion relations involve some commutators of 
the matrix-valued kernels. The unknown operators that define the ansatz 
need to be identified by solving the related recursion relations.

In Mellin space, the ansatz that solves the Eq.~(\ref{kappa_trunc}), is chosen to be of the form \cite{EllisKunstLevin,Pegasus}
\ba
f(N,\alpha_s)&=&U(N,\alpha_s)f_{LO}(N,\alpha_s,\alpha_0)U^{-1}(N,\alpha_0)
\nonumber\\
&=&\left[1+\sum_{\kappa^{\prime}=1}^{+\infty}U_{\kappa^{\prime}}(N)\alpha_s^{\kappa^{\prime}}\right]
f_{LO}(N,\alpha_s,\alpha_0)\left[1+\sum_{\kappa^{\prime}=1}^{+\infty}U_{\kappa^{\prime}}(N)\alpha_0^{\kappa^{\prime}}\right]^{-1},
\label{Uexact}
\ea
that we call, for convenience, {\em the $U$-ansatz},
and inserting this expression in Eq.~(\ref{kappa_trunc}) it generates a chain of recursion relations which allow us to determine the matrices
$U_{\kappa^{\prime}}(N)$ in terms of the operators $R_{\kappa}$.

We remark that even if we take a fixed value of $\kappa$ in the
truncated equation, the running
index $\kappa^{\prime}$ in Eq.~(\ref{Uexact}) is still free.
The case $\kappa^{\prime}\geq \kappa$ with $\kappa^{\prime}\rightarrow \infty$ \footnote{ This corresponds
to the option IMODEV $=2$ in \textsc{Pegasus} while the case $\kappa^{\prime}=\kappa$ with $\kappa,\kappa^{\prime}\rightarrow \infty$ corresponds to the
option IMODEV $=1$.} allows to find 
the exact solution of the $\kappa$-truncated equation ~(\ref{kappa_trunc}). 

A third option corresponds to the choice $\kappa^{\prime}<\kappa$. This gives an approximate solution
of the $\kappa$-truncated NNLO equation accurate at the order $\alpha_s^{\kappa^{\prime}}$. Notice that 
if we start from the first form of the evolution (form-1) and use a recursive ansatz 
to solve this equation (either in moment space or in $x$-space) this solution has to agree with the 
solutions of the truncated equation considered above, once we perform an expansion of that solution 
in $\alpha_s$ and $\alpha_0$, as we have showed in \cite{CCG1}.

As an example, the most accurate NNLO solution is generated by the choice 
$\kappa^{\prime}=\kappa=m=2$. In this case we can write the $\alpha_s^2$-truncated solution
of the truncated equation with $\kappa=2$ as follows
\ba
\label{solNNLO}
&&f(N,\alpha_s)=\left(\frac{\alpha_s}{\alpha_0}\right)^{-\frac{2}{\beta_0}P^{(0)}}
\left[1+\left(\alpha_s-\alpha_0\right)U_1(N) +\alpha_s^2 U_2(N)\right.\nonumber\\
&&\hspace{4.5cm}\left.-\alpha_s\alpha_0 U_1^2(N)
+\alpha_0^2\left(U_1^2(N)-U_2(N)\right)\right]f(N,\alpha_0).\,\nonumber\\
\ea

At this retained accuracy $(m=2)$ of the evolution integral,
the truncated solution of the corresponding (truncated) DGLAP equation can be easily found,
in moment space, as
\ba
&&f(N,\alpha_s)=f(N,\alpha_0)\left(\frac{\alpha_s}{\alpha_0}\right)^{-2 \frac{P^{(0)}}{\beta_0}}
\left\{1 + \left(\alpha_s-\alpha_0\right)\left[-\frac{P^{(1)}}{\pi\beta_0}
+\frac{P^{(0)}\beta_1}{2\pi\beta_0^2}\right] \right.\nonumber\\
&&\left.\hspace{1.6cm}+\alpha_s^2\left[\frac{{P^{(1)}}^2}{2\pi^2\beta_0^2}
-\frac{P^{(2)}}{4\pi^2\beta_0}-\frac{P^{(0)}P^{(1)}\beta_1}{2\pi^2\beta_0^2}
+\frac{P^{(1)}\beta_1}{8\pi^2\beta_0^2}+\frac{{P^{(0)}}^2\beta_1^2}{8\pi^2\beta_0^4}
-\frac{P^{(0)}\beta_1^2}{16\pi^2\beta_0^3}+\frac{P^{(0)}\beta_2}{16\pi^2\beta_0^2}
\right]\right.\nonumber\\
&&\left.\hspace{1.6cm}+\alpha_0^2\left[\frac{{P^{(1)}}^2}{2\pi^2\beta_0^2}
+\frac{P^{(2)}}{4\pi^2\beta_0}-\frac{P^{(0)}P^{(1)}\beta_1}{2\pi^2\beta_0^2}
-\frac{P^{(1)}\beta_1}{8\pi^2\beta_0^2}+\frac{{P^{(0)}}^2\beta_1^2}{8\pi^2\beta_0^4}
+\frac{P^{(0)}\beta_1^2}{16\pi^2\beta_0^3}-\frac{P^{(0)}\beta_2}{16\pi^2\beta_0^2}
\right]\right.\nonumber\\
&&\left.\hspace{1.6cm}+\alpha_0\alpha_s\left[-\frac{{P^{(1)}}^2}{\pi^2\beta_0^2}
+\frac{P^{(0)}P^{(1)}\beta_1}{\pi^2\beta_0^3}-
\frac{{P^{(0)}}^2\beta_1^2}{4\pi^2\beta_0^4}\right]\right\}\,.
\label{para}
\ea
These are solutions of type-1. They coincide with the first few terms of the 
exact NNLO solution of the DGLAP
equation, obtained by a
double expansion in the couplings and retaining only the $O(\alpha_s^2)$ terms, 
as can be explicitly 
checked in the nonsinglet sector for the equation given in form-1. Therefore
the solution is organized effectively as a double expansion
in $\alpha_s$ and $\alpha_0$. This approach remains valid also in the singlet case,
when the equations assume a matrix form, though an exact solution, in the form of an ansatz, similar 
to that of the nonsinglet sector (Eq.~\ref{chain1} below), is not available in this case.

\subsection{The logarithmic expansions for the form-1 of the evolution}
Our previous analysis has involved form-2 of the equations and we have presented an ansatz 
that solves this equation. We intend now to show how to construct an ansatz directly starting 
from form-1. 

The advantage of solving the equations directly in form-1 is that 
one has a single ansatz for the entire equation and the accuracy is just determined by 
the order of the chosen ansatz, differently from form-2. 
If we are interested in an accurate solution of order $\kappa^{\prime}$,
for instance, we use the ansatz
\beq
f_{N^{\kappa' LO}}(x,\alpha_s)\big|_{O(\alpha_s^{\kappa'})}=
\sum_{n=0}^\infty\left(A^0_n(x) + \alpha_s A^1_n(x)+ \alpha_s^2 A^2_n(x) + \dots +
\alpha_s^{\kappa'} A^{\kappa'}_n(x)\right)
\left[\ln{\left(\frac{\alpha_s(Q^2)}{\alpha_s(Q^2_0)}\right)}\right]^{n},
\label{truncatedseries}
\eeq
which can be correctly defined to be a {\em truncated solution of order $\kappa'$}
of the DGLAP in form-1. As we are going to show next, we will monitor the numerical
behavior of this expansion and its convergence. 
Sending the index $\kappa^{\prime}$ in the logarithmic expansion of (\ref{truncatedseries}) to infinity, then the ansatz that accompanies this choice becomes
\beq
f_{N^mLO}(N,\alpha_s)=\sum_{n=0}^\infty\left(\sum_{l=0}^{\infty}
\alpha_s^{l} A_n^{l}(x)\right)
\left[\ln{\left(\frac{\alpha_s(Q^2)}{\alpha_s(Q^2_0)}\right)}\right]^{n},
\label{truncatedseries1}
\eeq
and converges, in principle, to the exact solution of the equation given in 
form-1. In practice, however, this convergence is hampered by the factorial 
suppression. For this reason is it convenient to use the term 
``asymptotic solutions'' rather than ``exact solution'' for these iterates of 
larger index.   

\subsubsection{Exact Solutions in the nonsinglet case at NNLO}

The search for exact NNLO solutions in the nonsinglet sector proceeds similarly. This has been analyzed in \cite{CCG1}. We define the following functions
\begin{eqnarray}
\mathcal{L} & = & \log\frac{\alpha_s}{\alpha_{0}},\\
\mathcal{M} & = & \log\frac{16\pi^{2}\beta_{0}+4\pi\alpha_s\beta_{1}
+\alpha_s^{2}\beta_{2}}{16\pi^{2}\beta_{0}+4\pi\alpha_{0}\beta_{1}+\alpha_{0}^{2}\beta_{2}},\\
\mathcal{Q} & = & \frac{1}{\sqrt{4\beta_{0}\beta_{2}-\beta_{1}^{2}}}\arctan \chi,\\
a(N) & = & -\frac{2P^{(0)}(N)}{\beta_{0}},\\
b(N) & = & \frac{P^{(0)}(N)}{\beta_{0}}-\frac{4P^{(2)}(N)}{\beta_{2}},\\
c(N) & = & \frac{2\beta_{1}}{\beta_{0}}P^{(0)}(N)-8P^{(1)}(N)
+\frac{8\beta_{1}}{\beta_{2}}P^{(2)}(N),
\end{eqnarray}
where for $n_{f}=6$ the solution has a branch point since
$4\beta_{0}\beta_{2}-\beta_{1}^{2}<0$. If we increase $n_f$ as we
step up in the factorization scale, for $n_f=6$
$\mathcal{Q}$ is replaced by its analytic continuation
\begin{equation}
\mathcal{Q}=\frac{1}{\sqrt{\beta_{1}^{2}-4\beta_{0}\beta_{2}}}
\textrm{arctanh} \chi
\end{equation}
where
\ba
\chi &=&\frac{2\pi(\alpha_s-\alpha_{0})\sqrt{4\beta_{0}\beta_{2}-
\beta_{1}^{2}}}{2\pi(8\pi\beta_{0}+(\alpha_s+\alpha_{0})\beta_{1})+
\alpha_s\alpha_{0}\beta_{2}} \nonumber \\
\textrm{arctanh} \chi &=& \frac{1}{2} \log \left(\frac{1 + \chi}{1 - \chi}\right).
\ea
Clearly all the (nontrivial) dependence on the coupling constants
$\alpha_s$ is contained
in the 3 functions $\mathcal{L}, \mathcal{M}$ and $\mathcal{Q}$.
The general solution can be written in terms of $A^{\prime}_n, B^{\prime}_n, C^{\prime}_n $,
coefficients that will be calculated by a chain of recursion
relations \cite{CCG1} giving
\begin{eqnarray}
f(x,Q^{2}) & = & \left(\sum_{n=0}^{\infty}\frac{A'_{n}(x)}{n!}\mathcal{L}^{n}\right)_\otimes
\left(\sum_{m=0}^{\infty}\frac{B'_{m}(x)}{m!}\mathcal{M}^{m}\right)_\otimes
\left(\sum_{p=0}^{\infty}\frac{C'_{p}(x)}{p!}\mathcal{Q}^{p}\right)_\otimes f(x,Q_0^2)\nonumber \\
& = & \sum_{s=0}^{\infty}\sum_{t=0}^{s}\sum_{n=0}^{t}\frac{A'_{n}(x)\otimes
B'_{t-n}(x)\otimes {C'_{s-t}(x)}}{n!(t-n)!(s-t)!}
\otimes f(x,Q_0^2)\,\mathcal{L}^{n}\mathcal{M}^{t-n}\mathcal{Q}^{s-t}\nonumber \\
& = & \sum_{s=0}^{\infty}\sum_{t=0}^{s}\sum_{n=0}^{t}
\frac{ D_{t,n}^{s}(x)}{n!(t-n)!(s-t)!}\mathcal{L}^{n}\mathcal{M}^{t-n}\mathcal{Q}^{s-t},
\label{eq:NNLOansatz}
\end{eqnarray}
and where
\beq
D_{t,n}^{s}(x)= A'_{n}(x)\otimes
B'_{t-n}(x)\otimes C'_{s-t}(x)\otimes f(x,Q_0^2).
\eeq
Solving the chain of recursion relations,
the above solution in $x$-space can be simply written as
\ba
&&f(x,\alpha_s(Q^2))= \exp\left\{\left[-\frac{2}{\beta_0}P^{(0)}(x)
\log{\left(\frac{\alpha_s}{\alpha_0}\right)}\right]\right\}\otimes\nonumber\\
&&\hspace{2cm}\exp\left\{\log\left(\frac{16\pi^{2}\beta_{0}+4\pi\alpha_s\beta_{1}
+\alpha_s^{2}\beta_{2}}{16\pi^{2}\beta_{0}+4\pi\alpha_{0}\beta_{1}+\alpha_{0}^{2}\beta_{2}}
\right)\left[\frac{P^{(0)}(x)}{\beta_{0}}-\frac{4P^{(2)}(x)}{\beta_{2}}\right]\otimes\right\}
\nonumber\\
&&\hspace{2cm}\exp\left\{
\Bigg(\frac{1}{\sqrt{4\beta_{0}\beta_{2}-\beta_{1}^{2}}}
\arctan\frac{2\pi(\alpha_s-\alpha_{0})\sqrt{4\beta_{0}\beta_{2}-
\beta_{1}^{2}}}{2\pi(8\pi\beta_{0}+(\alpha_s+\alpha_{0})\beta_{1})+
\alpha_s\alpha_{0}\beta_{2}}\Bigg)\right.\nonumber\\
&&\left.\hspace{2cm}
\left[\frac{2\beta_{1}}{\beta_{0}}P^{(0)}(x)-8P^{(1)}(x)
+\frac{8\beta_{1}}{\beta_{2}}P^{(2)}(x)\right]\otimes\right\}D_{0,0}^{0}(x).
\label{NNLO_exact}
\ea
where $D_{0,0}^{0}(x)=f(x,Q_0^2)$. The possibility of finding an exact solution has, of
course, phenomenological implications, since the analytic solution {\em performs
a resummation} of the $\log(\alpha_s)$ which are generated to all orders by the various
truncations and by the corresponding logarithmic expansions. These are incorporated into the
functions $\mathcal{M}$, $\mathcal{Q}$ $(\chi)$.

We will get back to this point later.

\subsection{The Singlet Case}
Before we address the topic of the resummation/re-organization
of the logarithmic structure of the solution due to the choice of the different expansions, we move to analyze the extension of our previous reasonings to the singlet case.
One can start from form-1 or from form-2, obtaining solutions of overall
different accuracies.
In the singlet case, if we start from form-2, then one can consider a truncation
of this equation, for instance to second order, that can be written as

\ba
\label{NNLOsinglet2}
\frac{\partial{\vec{f}(N,\alpha_s)}}{\partial\alpha_s}=
\frac{1}{\alpha_s}\left[\hat{R_0}+\alpha_s \hat{R}_1 +\alpha_s^2
\hat{R}_2\right]\vec{f}(N,\alpha_s),\,\nonumber\\
\ea
where
\ba
&&\hat{R}_0=-\frac{2}{\beta_0}\hat{P}^{(0)}\nonumber\\
&&\hat{R}_1=-\frac{1}{2\pi\beta_0^2}\left[2 \beta_0 \hat{P}^{(1)}-\hat{P}^{(0)}\beta_1\right]\nonumber\\
&&\hat{R}_2=-\frac{1}{\pi}\left(\frac{\hat{P^{(2)}}}{2\pi\beta_0}
+\frac{\hat{R}_1 \beta_1}{4\beta_0}+\frac{\hat{R}_0\beta_2}{16\pi\beta_0}\right),
\ea
whose (exact) solution in Mellin  space is expected to be of the form 
(the $U$-ansatz)
\cite{EllisKunstLevin}
\ba
\label{NNLOans_sing1}
\vec{f}(N,\alpha_s)=\left[1+\alpha_s\hat{U}_{1}(N)+\alpha_s^2
\hat{U}_{2}(N)\right]\hat{L}(\alpha_s,\alpha_0,N)
\left[1+\alpha_0\hat{U}_{1}(N)+\alpha_0^2\hat{U}_{2}(N)\right]^{-1}
\vec{f}(N,\alpha_0),\,\nonumber\\
\label{Uansatz}
\ea
where
\ba
&&\left[\hat{R}_0,\hat{U}_1\right]=\hat{U}_1-\hat{R}_1,\nonumber\\
&&\left[\hat{R}_0,\hat{U}_2\right]=-\hat{R}_2 -\hat{R}_1\hat{U}_1 +2\hat{U}_2.\,
\ea
Using two projectors on the subspaces of the corresponding leading order (singlet) eigenvalues, denoted by ($e_\pm$) (see \cite{CCG1}), one can remove the commutators, obtaining
\ba
&&\hat{U}_2=\hat{U}_2^{++}+\hat{U}_2^{+-}+\hat{U}_2^{-+}+\hat{U}_2^{--}\,,
\ea
where
\ba
&&\hat{U}_2^{++}=\frac{1}{2}\left[\hat{R}_1^{++}\hat{R}_1^{++}
+\hat{R}_2^{++}-\frac{\hat{R}_1^{+-}
\hat{R}_1^{-+}}{r_- -r_+ -1}\right],\nonumber\\
&&\hat{U}_2^{--}=\frac{1}{2}\left[\hat{R}_1^{--}\hat{R}_1^{--}
+\hat{R}_2^{--}-\frac{\hat{R}_1^{-+}
\hat{R}_1^{+-}}{r_+ -r_- -1}\right],\nonumber\\
&&\hat{U}_2^{+-}=\frac{1}{r_+ -r_- -2}\left[-\hat{R}_1^{+-}\hat{R}_1^{--}
-\hat{R}_2^{+-}+\frac{\hat{R}_1^{++}\hat{R}_1^{+-}}{r_+ -r_- -1}
\right],\nonumber\\
&&\hat{U}_2^{-+}=\frac{1}{r_- -r_+ -2}\left[-\hat{R}_1^{-+}\hat{R}_1^{++}
-\hat{R}_2^{-+}+\frac{\hat{R}_1^{--}\hat{R}_1^{-+}}{r_- -r_+ -1}
\right],\,
\ea
and the formal solution from Mellin space can be simplified to 
\ba
\label{NNLOtrsolsing}
&&\vec{f}(N,\alpha_s)=\left[\hat{L}+\alpha_s\hat{U}_1\hat{L}
-\alpha_0\hat{L}\hat{U}_1\right.\nonumber\\
&&\hspace{2cm}\left.+\alpha_s^2 \hat{U}_2\hat{L}
-\alpha_s\alpha_0\hat{U}_1\hat{L}\hat{U}_1
+\alpha_0^2\hat{L}\left(\hat{U}_1^2-\hat{U}_2\right)
\right]\vec{f}(N,\alpha_0)\,
\ea
where the accuracy is kept through $O(\alpha_s^2)$. 

If we don't want to truncate the equation, then we work with form-1. 
We start constructing solutions of this equation using the logarithmic expansions introduced in 
\cite{CCG1} using few iterates. As we have mentioned, in this case there will be just one parameter 
appearing in the expansion, related to the desired accuracy, i.e. the terms retained in the 
ansatz, and by increasing the accuracy one expects the result to converge toward the exact solution.

The first truncated logarithmic ansatz that is expected to reproduce
(\ref{NNLOtrsolsing}) includes also an infinite set of new coefficients $\vec{C_n}$,
similar to the nonsinglet NNLO case
\ba
\vec{f}(N,\alpha_s)=\sum_{n=0}^{\infty}\frac{L^n}{n!}\left[{\bf A}_n +\alpha_s{\bf B}_n+
\alpha_s^2{\bf C}_n\right].
\label{altroans}
\ea
The ansatz can be generated to an arbitrarily high order.
If this order is $\kappa$,  we introduce the $\kappa$-truncated logarithmic ansatz
\ba
\vec{f}_{N^{\kappa LO}}(x,\alpha_s)\big|_{O(\alpha_s^{\kappa})}=
\sum_{n=0}^\infty\left({\bf O}^0_n(x) + \alpha_s {\bf O}^1_n(x)+ \alpha_s^2 {\bf O}^2_n(x) + \dots +
\alpha_s^{\kappa} {\bf O}^{\kappa}_n(x)\right)
\left[\ln{\left(\frac{\alpha_s(Q^2)}{\alpha_s(Q^2_0)}\right)}\right]^{n}
\label{ktruncatedseries}
\ea
in the NNLO DGLAP matrix equation and neglect the $O(\alpha_s^{\kappa+1})$ terms.
We obtain the following recursion relations which, in the NLO DGLAP case are
\ba
\label{genrecnlo}
&&{\bf O}_{n+1}^{0}(x)=-\frac{2}{\beta_{0}}\left[{\bf P}^{(0)}(x)\otimes {\bf O}_{n}^{0}(x)\right],
\nonumber\\
&&\vdots
\nonumber\\
&&{\bf O}_{n+1}^{\kappa}(x)= -\frac{2}{\beta_{0}}\left[{\bf P}^{(0)}\otimes {\bf O}_{n}^{\kappa}\right](x)
-\frac{1}{\pi\beta_{0}}\left[{\bf P}^{(1)}(x)\otimes {\bf O}_{n}^{\kappa-1}(x)\right]
\nonumber \\
&&\hspace{2cm}-\frac{\beta_{1}}{4\pi\beta_{0}}
{\bf O}_{n+1}^{\kappa-1}(x)-\kappa {\bf O}_{n}^{\kappa}(x)-(\kappa-1)
\frac{\beta_{1}}{4\pi\beta_{0}} {\bf O}_{n}^{\kappa-1}(x)\,,
\label{chain1}
\ea
while in the NNLO case become
\ba
\label{genrecnnlo}
&&{\bf O}_{n+1}^{0}(x)=-\frac{2}{\beta_{0}}\left[{\bf P}^{(0)}(x)\otimes {\bf O}_{n}^{0}(x)\right],
\nonumber\\
&& {\bf O}_{n+1}^{1}(x)= -\frac{2}{\beta_{0}}\left[{\bf P}^{(0)}(x)\otimes {\bf O}_{n}^{1}(x)\right]
-\frac{1}{\pi\beta_{0}}\left[{\bf P}^{(1)}(x)\otimes {\bf O}_{n}^{0}(x)\right]\nonumber \\
&&\hspace{2cm} -\frac{\beta_{1}}{4\pi\beta_{0}}{\bf O}_{n+1}^{0}(x)-{\bf O}_{n}^{1}(x),
\nonumber\\
&&\vdots
\nonumber\\
&& {\bf O}_{n+1}^{\kappa}(x) =  -\frac{2}{\beta_{0}}\left[{\bf P}^{(0)}
(x)\otimes {\bf O}_{n}^{\kappa}(x)\right]
-\frac{1}{\pi\beta_{0}}\left[{\bf P}^{(1)}(x)\otimes {\bf O}_{n}^{\kappa-1}(x)\right]\nonumber \\
&&\hspace{2cm} -\frac{1}{2\pi^{2}\beta_{0}}
\left[{\bf P}^{(2)}(x)\otimes {\bf O}_{n}^{\kappa-2}(x)\right]\nonumber \\
&&\hspace{2cm} -\frac{\beta_{1}}{4\pi\beta_{0}}{\bf O}_{n+1}^{\kappa-1}(x)
-\frac{\beta_{2}}{16\pi^{2}\beta_{0}} {\bf O}_{n+1}^{\kappa-2}(x)\nonumber \\
&&\hspace{2cm} -\kappa {\bf O}_{n}^{\kappa}(x)-(\kappa-1)
\frac{\beta_{1}}{4\pi\beta_{0}} {\bf O}_{n}^{\kappa-1}(x)-(\kappa-2)
\frac{\beta_{2}}{16\pi^{2}\beta_{0}} {\bf O}_{n}^{\kappa-2}(x)\,.
\label{chain2}
\ea

These relations hold both in the nonsinglet and singlet cases and they can be solved in $x$-space
and N-space in terms of the initial conditions $f(N,\alpha_0)={\bf O}_{0}^{0}(N)$. Since in the singlet sector
the recursion relations are in matrix form, we can solve them by the use of the projectors
$e_+$ and $e_-$.
A straightforward way to solve the matrix relations is first to
solve the relation for ${\bf O}_{n}^{0}(N)$ in terms of $e_+\,, e_-$ and $r_+^n\,,r_-^n$ as follows
\ba
{\bf O}_{n}^{0}(N)= e_+ r_+^n {\bf O}_{0}^{0}(N) + e_- r_-^n {\bf O}_{0}^{0}(N)
\ea
and use this result to solve the other relations. The ${\bf O}_{n}^{m}(N)$ operators 
can be decomposed in an $\mathbf{R}^{2}$ orthonormal basis $\left\{{\bf e}_{1},{\bf e}_{2}\right\}$ as
\ba
{\bf O}_{n}^{m}(N)={\bf e}_{1} O_{n}^{(1),m}(N)+{\bf e}_{2} O_{n}^{(2),m}(N)=
{\bf O}_{n}^{+,m}(N)+{\bf O}_{n}^{-,m}(N)\,.
\ea
Then, using the properties of the projectors we can write
\ba
&&\left(e_+ + e_- \right){\bf O}_{n}^{m}(N)=
e_+\left({\bf e}_{1} {\bf O}_{n}^{(1),m}(N)\right)
+e_+\left({\bf e}_{2} {\bf O}_{n}^{(2),m}(N)\right)
\nonumber\\
&&\hspace{3cm}+e_-\left({\bf e}_{1} {\bf O}_{n}^{(1),m}(N)\right)
+e_-\left({\bf e}_{2} {\bf O}_{n}^{(2),m}(N)\right)\,,
\ea
for the relations with $m>0$, and setting
\ba
&&{\bf O}_{n}^{++,m}(N)=e_+\left({\bf e}_{1} {\bf O}_{n}^{(1),m}(N)\right)
\nonumber\\
&&{\bf O}_{n}^{+-,m}(N)=e_+\left({\bf e}_{2} {\bf O}_{n}^{(2),m}(N)\right)
\nonumber\\
&&{\bf O}_{n}^{-+,m}(N)=e_-\left({\bf e}_{1} {\bf O}_{n}^{(1),m}(N)\right)
\nonumber\\
&&{\bf O}_{n}^{--,m}(N)=e_-\left({\bf e}_{2} {\bf O}_{n}^{(2),m}(N)\right)
\ea
we can derive some recursion relations.
For example, in the NLO case, which corresponds to the case $m=0,1$,
we have two recursion relations and having solved the ${\bf O}_{n}^{0}(N)$
as illustrated above, the $m=1$ relation can be decomposed
into four recursion relations as follows
\ba
&&{\bf O}_{n+1}^{++,1}(N)={\bf R}_{1}^{++}r_{+}^{n}{\bf O}_{0}^{0}(N)+(r_{+}-1){\bf O}_{n}^{++,1}(N)
\nonumber\\
&&{\bf O}_{n+1}^{+-,1}(N)={\bf R}_{1}^{+-}r_{+}^{n}{\bf O}_{0}^{0}(N)+(r_{-}-1){\bf O}_{n}^{+-,1}(N)
\nonumber\\
&&{\bf O}_{n+1}^{-+,1}(N)={\bf R}_{1}^{-+}r_{-}^{n}{\bf O}_{0}^{0}(N)+(r_{+}-1){\bf O}_{n}^{-+,1}(N)
\nonumber\\
&&{\bf O}_{n+1}^{++,1}(N)={\bf R}_{1}^{--}r_{-}^{n}{\bf O}_{0}^{0}(N)+(r_{-}-1){\bf O}_{n}^{--,1}(N).
\ea

This pattern can be extended to NNLO, in fact we have three sets of recursion
relations corresponding to the cases $m=0,1,2$.
Once we have solved all the relations corresponding to the cases $m=0,1$ we can proceed
to solve the following relations
\begin{eqnarray}
&&{\bf O}_{n+1}^{++,2}(N)=\left[{\bf R}_{2}^{++}+\frac{\beta_1}{\beta_0(4\pi)}{\bf R}_{1}^{++} \right]
r_{+}^n {\bf O}_{0}^{0}(N)+ \left[{\bf R}_{1}^{++} {\bf O}_{n}^{++,1}(N)
+{\bf R}_{1}^{+-}{\bf O}_{n}^{-+,1}(N)\right]
\nonumber\\
&&\hspace{1.5cm}
-\frac{\beta_1}{\beta_0(4\pi)}\left[{\bf O}_{n}^{++,1}(N)+{\bf O}_{n+1}^{++,1}(N)\right]
+\frac{\beta_1}{\beta_0(4\pi)}r_+ {\bf O}_{n}^{++,1}(N)+\left(r_+ -2 \right){\bf O}_{n}^{++,2}\,,
\nonumber\\ \nonumber\\
&&{\bf O}_{n+1}^{+-,2}(N)=\left[{\bf R}_{2}^{+-}+\frac{\beta_1}{\beta_0(4\pi)}{\bf R}_{1}^{+-} \right]
r_{-}^n {\bf O}_{0}^{0}(N)+ \left[{\bf R}_{1}^{+-} {\bf O}_{n}^{--,1}(N)
+{\bf R}_{1}^{++}{\bf O}_{n}^{+-,1}(N)\right]
\nonumber\\
&&\hspace{1.5cm}
-\frac{\beta_1}{\beta_0(4\pi)}\left[{\bf O}_{n}^{+-,1}(N)+{\bf O}_{n+1}^{+-,1}(N)\right]
+\frac{\beta_1}{\beta_0(4\pi)}r_+ {\bf O}_{n}^{+-,1}(N)+\left(r_+ -2 \right){\bf O}_{n}^{+-,2}\,,
\nonumber\\ \nonumber\\
&&{\bf O}_{n+1}^{-+,2}(N)=\left[{\bf R}_{2}^{-+}+\frac{\beta_1}{\beta_0(4\pi)}{\bf R}_{1}^{-+} \right]
r_{+}^n {\bf O}_{0}^{0}(N)+ \left[{\bf R}_{1}^{-+} {\bf O}_{n}^{++,1}(N)
+{\bf R}_{1}^{--}{\bf O}_{n}^{-+,1}(N)\right]
\nonumber\\
&&\hspace{1.5cm}
-\frac{\beta_1}{\beta_0(4\pi)}\left[{\bf O}_{n}^{-+,1}(N)+{\bf O}_{n+1}^{-+,1}(N)\right]
+\frac{\beta_1}{\beta_0(4\pi)}r_- {\bf O}_{n}^{-+,1}(N)+\left(r_- -2 \right){\bf O}_{n}^{-+,2}\,,
\nonumber\\ \nonumber\\
&&{\bf O}_{n+1}^{--,2}(N)=\left[{\bf R}_{2}^{--}+\frac{\beta_1}{\beta_0(4\pi)}{\bf R}_{1}^{--} \right]
r_{-}^n {\bf O}_{0}^{0}(N)+ \left[{\bf R}_{1}^{--} {\bf O}_{n}^{--,1}(N)
+{\bf R}_{1}^{-+}{\bf O}_{n}^{+-,1}(N)\right]
\nonumber\\
&&\hspace{1.5cm}
-\frac{\beta_1}{\beta_0(4\pi)}\left[{\bf O}_{n}^{--,1}(N)+{\bf O}_{n+1}^{--,1}(N)\right]
+\frac{\beta_1}{\beta_0(4\pi)}r_+ {\bf O}_{n}^{--,1}(N)+\left(r_- -2 \right){\bf O}_{n}^{--,2}\,
\nonumber\\
\label{recform1}
\end{eqnarray}
which can be implemented in a computer program, with a 
standard numerical inversion of the Mellin transform, being equivalent to 
(\ref{chain1}) and (\ref{chain2}). The $x$-space approach, as we are going 
to show, matches the numerical Mellin method with very high accuracy, since the asymptotic truncated solutions give the same answer. 
In the nonsinglet sector the exact solutions built by iterations as logarithms of composite functions of $\alpha_s$ are new and not present in the previous literature. These have been used in 
this sector to generate the corresponding exact solutions. 

\subsection{Relating the $U$-ansatz to the logarithmic expansion} 
It is important to compare the two expansions which 
are identical globally (that is to all orders) but that organize, at a certain fixed perturbative order, the logarithmic corrections in different ways. This 
can be easily shown in the nonsinglet sector, where the two expansions can be more easily mapped into one another. Let's see how this happens. 

The double Taylor-expansion of the solution of the Eq.~(\ref{Uansatz}) for
$(\alpha_s,\alpha_0)$ around $(0,0)$ up to order 4, for example,
has the following structure
\begin{eqnarray}
&&f(x,\alpha_s,\alpha_0)=\left(\frac{\alpha_s}{\alpha_0}\right)^{-\frac{2}{\beta_0} P^{(0)}}
\left[1 + \alpha_s a_1^{(1)}+ \alpha_0 a_2^{(1)} +\right.\nonumber\\
&&\hspace{2cm}\left.\alpha_s^2 a_1^{(2)} + \alpha_s \alpha_0 a_2^{(2)} + \alpha_0^2 a_3^{(2)} +
\right.\nonumber\\
&&\hspace{2cm}\left.\alpha_s^3 a_1^{(3)} + \alpha_s^2 \alpha_0 a_2^{(3)} + \alpha_s\alpha_0^2 a_3^{(3)}
+ \alpha_0^3 a_4^{(3)} + \alpha_s^2\alpha_0^2 a_1^{(4)}
\right.\nonumber\\
&&\hspace{2cm}\left.\alpha_s^4 a_2^{(4)} + \alpha_s \alpha_0^3 a_3^{(4)} + \alpha_s^2\alpha_0^2 a_4^{(4)}
+ \alpha_s\alpha_0^3 a_5^{(4)} +\alpha_0^4 a_6^{(4)}
+ \dots +\alpha_s^4\alpha_0^4a_1^{(8)} \right]f_0\,,
\nonumber\\
\label{comp_sol}
\end{eqnarray}
as we can see, the double expansion gives terms of higher order
of the type $\alpha_s^4 \alpha_0^4$. On the other end, for instance, the 
logarithmic expansion accurate to $\alpha_s^{4}$ is given by
\begin{eqnarray}
\tilde{f}(x,\alpha_s,\alpha_0)=\sum_{n=0}^{\infty}\left[
A_n(x)+\alpha_s B_n(x)+\alpha_s^2 C_n(x)+\alpha_s^3 D_n(x)+\alpha_s^4 E_n(x)
\right]
\frac{1}{n!}\log^n\left(\frac{\alpha_s}{\alpha_0}\right),\,
\nonumber\\
\label{ansatz}
\end{eqnarray}
that gives recursion relations for the coefficients $A_n,\dots,E_n$ which 
are solved and exponentiated, as we have shown in \cite{CCG1}. 
Once those coefficients have been determined, we substitute them into Eq.~(\ref{ansatz})
and rewrite $\tilde{f}$ as
\begin{eqnarray}
&&\tilde{f}(x,\alpha_s,\alpha_0)=\left(\frac{\alpha_s}{\alpha_0}\right)^{-\frac{2}{\beta_0} P^{(0)}}
\left[1 + \alpha_s c_1^{(1)}+ \alpha_0 c_2^{(1)} +\right.\nonumber\\
&&\hspace{2cm}\left.\alpha_s^2 c_1^{(2)} + \alpha_s \alpha_0 c_2^{(2)} + \alpha_0^2 c_3^{(2)} +
\right.\nonumber\\
&&\hspace{2cm}\left.\alpha_s^3 c_1^{(3)} + \alpha_s^2 \alpha_0 c_2^{(3)} + \alpha_s\alpha_0^2 c_3^{(3)}
+ \alpha_0^3 c_4^{(3)} +
\right.\nonumber\\
&&\hspace{2cm}\left.\alpha_s^4 c_1^{(4)} + \alpha_s \alpha_0^3 c_2^{(4)} + \alpha_s^2\alpha_0^2 c_3^{(4)}
+ \alpha_s\alpha_0^3 c_4^{(4)} +\alpha_0^4 c_5^{(4)}
\right]A_0\,.
\label{trunc_sol}
\end{eqnarray}
From the direct calculation of the coefficients $a_i^{(j)}$ and $c_i^{(j)}$
in the two Eqs.~(\ref{trunc_sol}) and (\ref{comp_sol}), 
we observe that they coincide only for those
terms which are of the same order in $\alpha_s\alpha_0$, but in general, 
the two expansions organize the corrections in different ways. 
For instance, in order to generate the terms of the type $\alpha_s^4\alpha_0^4$, we should take
the index $\kappa=8$ in (\ref{ktruncatedseries}). In this case we will 
reproduce all the coefficients up to 
$\alpha_s^4\alpha_0^4$, but we will also introduce terms of order 
$\alpha_s^8$ and $\alpha_0^8$ which were not present in the double expansion of (\ref{Uansatz}) arrested at order $\alpha_s^4$.
This is due to the fact that the Taylor expansion of (\ref{Uansatz}) in $(\alpha_s,\alpha_0)$
is a double expansion while the result of the logarithmic expansion corresponds
to a single expansion in $\alpha_s$ and the remaining power of $\alpha_0$
are introduced during the exponentiation procedure \cite{CCG1}. As we have 
mentioned, to establish the equivalence between the two approaches Eqs.~(\ref{Uansatz}) and 
(\ref{ktruncatedseries}) one needs to expand the leading order 
solution which appears as first factor in (\ref{Uansatz}), extracting all the logarithms of 
$\alpha_s$. The structure of the $U$-ansatz is such that in it the leading order 
solution is automatically factored out, while in the logarithmic expansions 
of type (\ref{altroans}) and, in general, (\ref{ktruncatedseries}), one 
needs to exponentiate the solution of the recursion relations to achieve 
the same result. Numerically this can't be done, but the two ansatz\"e, interpreted perturbatively both as ways to collect the logarithms of the solution 
of the evolution equations, become the same expansion as the order of the truncation grows.  
 
\section{Resummation and the exact solution}
It is interesting to compare the logarithmic corrections generated by the truncated solutions 
with the exact nonsinglet solutions obtained at the various perturbative orders. 
As we have already mentioned, the analytic solution resums the partial contributions 
coming from the truncates of various 
order introduced by the various ansatz\"e in $x$-space or in moment space. 
To illustrate this point, let's start the analysis from the NLO nonsinglet case and then we will generalize
the results to the NNLO case.

Solving NLO DGLAP nonsinglet equation in Mellin space
\ba
\frac{\partial f(N,\alpha_s)}{\partial \alpha_s}=\frac{P^{NLO}(N,\alpha_s)}{\beta_{NLO}(\alpha_s)}f(N,\alpha_s)
\ea
we obtain an exact solution which can be written as follows
\ba
f(\alpha_s,N)&=&\exp\left\{-\frac{2}{\beta_0}P^{(0)}(N)\log{\left(\frac{\alpha_s}
{\alpha_0}\right)}\right\}\nonumber \\
&& \times\exp\left\{\left[\frac{2}{\beta_0}P^{(0)}(N) -\frac{4}{\beta_1}P^{(1)}(N)\right]
\log\left(\frac{4\pi\beta_0+\alpha_s \beta_1}{4\pi\beta_0+\alpha_0 \beta_1}\right)\right\}f(\alpha_0,N)
\ea
in Mellin space, and as
\ba
f(\alpha_s,x)&=&\exp\left\{-\frac{2}{\beta_0}P^{(0)}(x)\log{\left(\frac{\alpha_s}{\alpha_0}\right)}
\right\}_\otimes \nonumber \\
&& \times\exp\left\{\left[\frac{2}{\beta_0}P^{(0)}(x) -\frac{4}{\beta_1}P^{(1)}(x)\right]
\log\left(\frac{4\pi\beta_0+\alpha_s \beta_1}{4\pi\beta_0+\alpha_0 \beta_1}\right)\right\}_\otimes f(\alpha_0,x)
\label{nlox1}
\ea
in $x$-space.

Expanding in terms of $\log{\left(\frac{\alpha_s}{\alpha_0}\right)}$
this solution we obtain 
\ba
&&f(\alpha_s,x)=\exp\left\{-\frac{2}{\beta_0}P^{(0)}(x)\log{\left(\frac{\alpha_s}{\alpha_0}\right)}\right\}_\otimes
\times \nonumber\\
&&\hspace{1cm}\exp\left\{\left[\frac{2}{\beta_0}P^{(0)}(x)-\frac{4}{\beta_1}P^{(1)}(x)\right]
\left[\frac{\alpha_0\beta_1}{4\pi \beta_0+\alpha_0\beta_1}
\log{\left(\frac{\alpha_s}{\alpha_0}\right)}+\dots\right]
\right\}_\otimes f(\alpha_0),\nonumber\\
\ea
with an analogous expression in moment space. The notations can be simplified by defining
\ba
&&\tilde{P}^{(0)}=-\frac{2}{\beta_0}P^{(0)}
\nonumber\\
&&\tilde{P}^{(1)}=\frac{2}{\beta_0}P^{(0)} -\frac{4}{\beta_1}P^{(1)}
\nonumber\\
&&g_1(\alpha_0)=\frac{\alpha_0\beta_1}{4\pi \beta_0+\alpha_0\beta_1}\,
\nonumber\\
&&t=\log{\left(\frac{\alpha_s}{\alpha_0}\right)}
\ea
and in $x$-space we can rewrite the solution in terms of t-iterates in the form
\ba
&&f(\alpha_s,x)=\exp\left\{\tilde{P}^{(0)}\,t\right\}_{\otimes}
\exp\left\{\tilde{P}^{(1)}\,t\,g_1(\alpha_0)+\tilde{P}^{(1)}\,t^2\,g_2(\alpha_0)
+\cdots\right\}_{\otimes}f(\alpha_0,x)
\nonumber\\
&&\hspace{1.8cm}\exp\left\{\tilde{P}^{(0)}\,t\right\}_{\otimes}
\exp\left\{\tilde{P}^{(1)}\,t\,g_1(\alpha_0)\right\}_{\otimes}
\exp\left\{\tilde{P}^{(1)}\,t^2\,g_2(\alpha_0)\right\}_{\otimes}\cdots\nonumber\\
&&\hspace{1.8cm}\exp\left\{\tilde{P}^{(1)}\,t^n\,g_n(\alpha_0)\right\}_{\otimes}f(\alpha_0,x)
\nonumber\\
\ea
where
\ba
&&g_2(\alpha_0)=\frac{1}{2}\left(g_1(\alpha_0)-g_1^2(\alpha_0)\right)\nonumber\\
&&g_3(\alpha_0)=\left(\frac{1}{6} g_1(\alpha_0)- \frac{1}{2} g_1^2(\alpha_0)
+\frac{1}{3} g_1^3(\alpha_0)\right)\nonumber\\
&&g_4(\alpha_0)=\left(\frac{1}{24} g_1(\alpha_0)- \frac{7}{24} g_1^2(\alpha_0)
+\frac{1}{2} g_1^3(\alpha_0)-\frac{1}{4}g_1^4(\alpha_0)\right)\nonumber\\
&&g_5(\alpha_0)=\left(\frac{1}{120} g_1(\alpha_0)- \frac{1}{8} g_1^2(\alpha_0)
+\frac{5}{12} g_1^3(\alpha_0)-\frac{1}{2}g_1^4(\alpha_0)+\frac{1}{5}g_1^5(\alpha_0)\right)\nonumber\\
&&\vdots
\ea
Finally, in the nonsinglet case we can re-arrange our solution in the form

\ba
&&f_{LO}(\alpha_s,x)=\exp\left\{\tilde{P}^{(0)}\,t\right\}_{\otimes}f(\alpha_0,x)
\nonumber\\
&&f(\alpha_s,x)=\exp\left\{\tilde{P}^{(1)}\,t\,g_1(\alpha_0)\right\}_{\otimes}
\exp\left\{\tilde{P}^{(1)}\,t^2\,g_2(\alpha_0)\right\}_{\otimes}\cdots
\exp\left\{\tilde{P}^{(1)}\,t^n\,g_n(\alpha_0)\right\}_{\otimes}f_{LO}(\alpha_s,x).
\nonumber\\
\label{nlox2}
\ea
It is interesting to note that the function $g_1(\alpha_0)$ is, in a sense, universal since 
it contains all the information about the initial conditions. A quick comparison between 
(\ref{nlox1}) and its expanded version (\ref{nlox2}) shows the features of the implicit 
resummation involved in moving from the second equation to the first. We will point out, 
in the numerical analysis presented below, that only a resummation can bring a logarithmic 
ansatz expressed in terms of $\log(\alpha_s)$ (either in Mellin space or in $x$-space) to reproduce numerically the exact solution. This is easy to 
show in the nonsinglet case, where both equations can be implemented as numerical iterations.

In a similar way we can proceed to re-arrange the exact solution in the nonsinglet
sector at NNLO. This can be rewritten as
\ba
&&f(x,\alpha_s)=
\exp\left\{\log\left(\frac{16\pi^{2}\beta_{0}+4\pi\alpha_s\beta_{1}
+\alpha_s^{2}\beta_{2}}{16\pi^{2}\beta_{0}+4\pi\alpha_{0}\beta_{1}+\alpha_{0}^{2}\beta_{2}}\right)
\left[\frac{P^{(0)}(x)}{\beta_{0}}-\frac{4P^{(2)}(x)}{\beta_{2}}\right] \right\}_{\otimes}
\nonumber\\
&&\hspace{2cm}\exp\left\{
\Bigg(\frac{1}{\sqrt{4\beta_{0}\beta_{2}-\beta_{1}^{2}}}
\arctan\frac{2\pi(\alpha_s-\alpha_{0})\sqrt{4\beta_{0}\beta_{2}-
\beta_{1}^{2}}}{2\pi(8\pi\beta_{0}+(\alpha_s+\alpha_{0})\beta_{1})+
\alpha_s\alpha_{0}\beta_{2}}\Bigg)\right.\nonumber\\
&&\left.\hspace{2cm}
\left[\frac{2\beta_{1}}{\beta_{0}}P^{(0)}(x)-8P^{(1)}(x)
+\frac{8\beta_{1}}{\beta_{2}}P^{(2)}(x)\right]\right\}_{\otimes}f_{LO}(x,\alpha_0).
\ea
Expanding in terms of the $\log$s, it is useful to define the following expressions
\ba
&&\tilde{P}^{(2)}_A=\left(4 P^{(2)}\beta_0-P^{(0)}\beta_2\right)\nonumber\\
&&\tilde{P}^{(2)}_B=\left(4 P^{(2)}\beta_0\beta_1-4 P^{(1)}\beta_0\beta_2+P^{(0)}\beta_1\beta_2
\right)\nonumber\\
&&G(\alpha_0)=\frac{1}{\beta_0\beta_2\left(16\pi^2\beta_0
+4\pi\alpha_0\beta_1+\alpha_0^2\beta_2 \right)}\,.
\ea
Then we get
\ba
&&f(x,\alpha_s)\simeq\exp\left\{\frac{t}{G(\alpha_0)}a_1(\alpha_0)\tilde{P}^{(2)}_A+
\frac{t^2}{G^2(\alpha_0)}a_2(\alpha_0)\tilde{P}^{(2)}_A+\cdots +
\frac{t^n}{G^n(\alpha_0)}a_n(\alpha_0)\tilde{P}^{(2)}_A\right\}_{\otimes}
\nonumber\\
&&\hspace{1cm}\exp\left\{\frac{t}{G(\alpha_0)}b_1(\alpha_0)\tilde{P}^{(2)}_B+
\frac{t^2}{G^2(\alpha_0)}b_2(\alpha_0)\tilde{P}^{(2)}_B+\cdots +
\frac{t^n}{G^n(\alpha_0)}b_n(\alpha_0)\tilde{P}^{(2)}_B\right\}_{\otimes}
f_{LO}(x,\alpha_0)\nonumber\\
\ea
where $G(\alpha_0)$ and the functions $a_1(\alpha_0),\dots,b_1(\alpha_0)\dots$,
are polynomial functions dependent on $\alpha_0$. We omit to give their
explicit expressions since they are not relevant for our discussion. With these definitions, the solution written in terms of 
simple logarithms of the coupling is summarized in $x$-space by the formal expression 

\ba
&&f(x,\alpha_s)\simeq\exp\left\{\frac{t}{G(\alpha_0)}a_1(\alpha_0)\tilde{P}^{(2)}_A
\right\}_{\otimes}
\exp\left\{\frac{t^2}{G^2(\alpha_0)}a_2(\alpha_0)\tilde{P}^{(2)}_A
\right\}_{\otimes}\cdots
\exp\left\{\frac{t^n}{G^n(\alpha_0)}a_n(\alpha_0)\tilde{P}^{(2)}_A\right\}_{\otimes}
\nonumber\\
&&\hspace{1.5cm}\exp\left\{\frac{t}{G(\alpha_0)}b_1(\alpha_0)\tilde{P}^{(2)}_B
\right\}_{\otimes}
\exp\left\{\frac{t^2}{G^2(\alpha_0)}b_2(\alpha_0)\tilde{P}^{(2)}_B
\right\}_{\otimes}\cdots\nonumber\\
&&\hspace{1.5cm}\otimes\exp\left\{\frac{t^n}{G^n(\alpha_0)}b_n(\alpha_0)\tilde{P}^{(2)}_B\right\}_{\otimes}
f_{LO}(x,\alpha_0).\nonumber\\
\ea

The relations between exact solutions and logarithmic expansions simplify considerably 
when one starts from the form-2 of the evolution equations. In fact, 
proceeding with the 1st truncated equation $(\kappa=1)$ this takes the form

\ba
\frac{\partial f(\alpha_s,x)}{\partial \alpha_s}=
\frac{1}{\alpha_s}\left[R_0+\alpha_s R_1\right]\otimes f(\alpha_s,x),
\ea
where we have set
\ba
R_0=-\frac{2}{\beta_0}P^{(0)} && R_1=-P^{(1)}\frac{1}{\pi\beta_0}
+P^{(0)}\frac{\beta_1}{2\pi\beta_0^2}.\,
\ea
In this specific case the exact solution is given by
\ba
\label{exaNLO}
f(\alpha_s,x)=\exp\left\{\left(\alpha_s-\alpha_0\right)R_1\right\}_{\otimes}
\exp\left\{t R_0\right\}_{\otimes}f(\alpha_0,x)\,
\label{exsol}
\ea
and using the relation 
\ba
f(\alpha_s,x)=\exp\left\{t R_0\right\}_{\otimes}\exp\left\{\alpha_0 t R_1\right\}_{\otimes}
\exp\left\{\alpha_0 \frac{t^2}{2!} R_1\right\}_{\otimes}\cdots
f(\alpha_0,x)
\ea
followed by a further expansion of the exponentials, the expression 
above can be re-organized in the form
\ba
f(\alpha_s,x)=\exp\left\{t R_0\right\} \otimes\left\{1+ R_1\alpha_0 t
+t^2 \left(R_1 \frac{\alpha_0}{2}+R_1\otimes R_1\frac{\alpha_0^2}{2}\right)+\cdots\right\}
\otimes f(\alpha_0,x).
\ea
If we want to preserve a certain accuracy in our solutions,
it is sufficient to do a Taylor expansion of (\ref{exsol}). 
For example, at NLO, the truncated solutions of order $\alpha_s$
of the truncated equation is 
\ba
f(\alpha_s,x)=\left[1+\left( \alpha_s-\alpha_0\right)R_1\right]\otimes f_{LO}(\alpha_s,x),\,
\ea
which takes the form originally given in \cite{Petronzio}.
Expanding this expression in terms of $\log(\alpha_s/\alpha_0)=t$ we obtain the traditional form of the solution
\ba
&&f(\alpha_s,x)=f_{LO}(\alpha_s,x)+R_1 \left[\alpha_0 t
+\frac{1}{2}\alpha_0 t^2 + \cdots\right]\otimes f_{LO}(\alpha_s,x).
\ea
Using this simple approach we can proceed to the determination of finite
accuracy $O(\alpha_s^{\kappa})$ solutions in the nonsinglet sector.

Increasing the value of $\kappa$, we can write the $\kappa$-th truncated NLO or 
NNLO equation as
\ba
\frac{\partial f(\alpha_s,x)}{\partial \alpha_s}=
\frac{1}{\alpha_s}\left[R_0+\alpha_s R_1+\alpha_s^2 R_2+\dots
+\alpha_s^{\kappa} R_{\kappa}\right]\otimes f(\alpha_s,x),
\ea
where all the coefficients $R_0(x),R_1(x),\dots,R_{\kappa}(x)$ are expressed
in terms of the $P^{(0)}$ and $P^{(1)}$ kernels in the NLO case, and
in terms of $P^{(0)},P^{(1)},P^{(2)}$ in the NNLO case.
In both cases the solution can be expanded in terms of $t$-logs as
\ba
&&f(\alpha_s,x)=\exp\left\{t R_0\right\} \otimes \exp\left\{
t\left( \alpha_0 R_1 c^{1}_1+\alpha_0^2 R_2 c^{1}_2 +\dots
+ \alpha_0^{\kappa} R_{\kappa} c^{1}_{\kappa}\right)\right\}
\otimes \nonumber\\
&&\hspace{2cm}\exp\left\{
t^2\left(\alpha_0 R_1 c^{2}_1+\alpha_0^2 R_2 c^{2}_2 +\dots
+ \alpha_0^{\kappa} R_{\kappa} c^{2}_{\kappa}\right)\right\}
\otimes\cdots \nonumber\\
&&\hspace{2cm}\otimes \exp\left\{t^n\left( \alpha_0 R_1 c^{n}_1+\alpha_0^2 R_2 c^{n}_2 +\dots
+ \alpha_0^{\kappa} R_{\kappa} c^{n}_{\kappa}\right)\right\}
\otimes f(\alpha_0,x),
\ea
being the coefficients $c^{n}_{\kappa}$ real numbers. 
After a further expansion one can cast the result in the form
\ba
&&f(\alpha_s,x)=
\left\{1+t \left(\alpha_0 R_1 c^{2}_1+\alpha_0^2 R_2 c^{2}_2 +\dots
+ \alpha_0^{\kappa} R_{\kappa} c^{2}_{\kappa}\right)
\right.\nonumber\\
&&\left.\hspace{4cm}+t^2 \left(\alpha_0 R_1 c^{2}_1+\alpha_0^2 R_2 c^{2}_2 +\dots
+ \alpha_0^{\kappa} R_{\kappa} c^{2}_{\kappa}\right)\otimes
\right.\nonumber\\
&&\left.\hspace{4cm}
\left(\alpha_0 R_1 c^{2}_1+\alpha_0^2 R_2 c^{2}_2 +\dots
+\alpha_0^{\kappa} R_{\kappa} c^{2}_{\kappa}\right)\otimes\cdots
\right\}\otimes f_{LO}(\alpha_0,x),\nonumber\\
\label{exa_exp}
\ea
having factored out the leading order solution.

One of the points that should be briefly taken into considerations concern the definition of 
the asymptotic solution. An asymptotic solution, in our terminology, identifies a solution which is the closest possible to the exact (brute force) solution. This means that while in the nonsinglet, for this solution, we will be using our exact ansatz, for the singlet we will let the number of iterates grow until the logarithmic series stabilizes. However, 
the absence of exact solutions in the singlet case shows 
that we will be surely differing from the brute force solution by some finite amount. Being \textsc{Candia}, or \textsc{Pegasus} based on analytical approaches rather than on discretizations, we are not able to compare with 
the exact solution and estimate the difference between our asymptotic solution and the exact one. 
We will quantify these difference rather accurately taking the Drell-Yan cross section as an example, but before coming to a numerical analysis we discuss the implementation of the renormalization scale 
dependence in our formalism.

\subsection{The treatment of the renormalization scale dependence and the implementation}
The scale dependence of the pdf's can be obtained by solving the modified equations
\ba
\label{evolution}
\frac{\partial}{\partial \ln \mu_F^2}\, f_i(x,\mu_F^2,\mu_R^2)=
P_{ij}(x,\mu_F^2,\mu_R^2) \otimes f_j(x,\mu_F^2,\mu_R^2)\,,
\ea
where $\mu_F$ is now a generic factorization scale. The explicit expression of these modified 
kernels are given below \cite{Pegasus}. This can be obtained by 
re-expressing the coupling 
constant, function of the factorization scale $\mu_F$, in terms of $\mu_R$ using the RGE for the running coupling at the corresponding order. 
Concerning the actual relation between the couplings at the two scales, this can be obtained by solving numerically the corresponding RGE for the running coupling at NLO and NNLO. We have also 
monitored the approximate solutions obtained  by the usual well-known asymptotic expansions in terms of $L=\ln(\mu_F^2/\mu_R^2)$. In the 
NLO case an implicit solution which allows to connect $\mu_F^2$ and $\mu_R^2$ is available
\ba
\label{implicit}
\frac{1}{a_s(\mu_F^2)}=\frac{1}{a_s(\mu_R^2)}
+\beta_0 \ln \left(\frac{\mu_F^2}{\mu_R^2} \right)
-b_1\ln\left\{\frac{a_s(\mu_F^2) \, [ 1 + b_1 a_s(\mu_R^2) ]}
{a_s(\mu_R^2) \, [ 1 + b_1 a_s(\mu_F^2) ]} \right\}
\ea
where $a_s(\mu^2)=\alpha_s(\mu^2)/(4\pi)$,
which can be solved as 
\ba
\label{alphas}
\alpha_s(\mu_F^2)=\alpha_s(\mu_R^2)-\left[\alpha_s^2(\mu_R^2)\frac{\beta_0 L}{4\pi}
+\frac{\alpha_s^3(\mu_R^2)}{(4\pi)^2}(-\beta_0^2 L^2+\beta_1 L)\right],
\ea
where the $\mu_F^2$ dependence is contained in the factor $L$, and we have used a 
$\beta$-function expanded up to NLO, involving $\beta_0$ and $\beta_1$. 
At NNLO implicit solutions such as (\ref{implicit}) are not available but one can derive the analogous 
of (\ref{alphas}). Both options, the exact and the asymptotic are present in \textsc{Candia}. The differences between the two determinations are quite small (see Tab.\ref{running}).

We have imposed logarithmic expansions on the equations with the kernels written in the form given below and derived recursion relations for these expressions. These reduce to the recursion relations 
discussed in the previous sections with the actual redefinitions 

\ba
\label{kern3}
&& \alpha_s(\mu_F^2)\to
\alpha_s(\mu_R^2)=\alpha_s(\mu_F^2)-\left[-\alpha_s^2(\mu_F^2)\frac{\beta_0
L}{4\pi}
+\frac{\alpha_s^3(\mu_F^2)}{(4\pi)^2}(-\beta_0^2 L^2-\beta_1
L)\right],\nonumber \\
&& P_{ij}^{(0)}(x) \to P_{ij}^{(0)}(x) \nonumber \\
&& P_{ij}^{(1)}(x) \to P_{ij}^{(1)}(x)
-\frac{\beta_0}{2}P_{ij}^{(0)}(x) L \nonumber \\
&& P_{ij}^{(2)}(x) \to P_{ij}^{(2)}(x)-
\beta_0 L P_{ij}^{(1)}(x)
-\left(\frac{\beta_1}{4} L - \frac{\beta_0^2}{4} L^2 \right) P_{ij}^{(0)}(x)
\ea
introduced into the equation expressed in form-1.

Concerning the implementation of the algorithm in \textsc{Candia}, we 
briefly illustrate the implementation of the flavor reconstruction.
We define
\ba
q_{i}^{(\pm)}=q_{i}\pm\overline{q}_{i},\qquad q^{(\pm)}=\sum_{i=1}^{n_{f}}q_{i}^{(\pm)}\,,
\label{notations}
\ea
then the general structure of the nonsinglet splitting functions is given by
\begin{equation}
P_{q_{i}q_{k}}=P_{\overline{q}_{i}\overline{q}_{k}}=\delta_{ik}P_{qq}^{V}+P_{qq}^{S},
\end{equation}

\begin{equation}
P_{q_{i}\overline{q}_{k}}=P_{\overline{q}_{i}q_{k}}=\delta_{ik}P_{q\bar{q}}^{V}+P_{q\bar{q}}^{S}.
\end{equation}

This leads to three independently evolving types of nonsinglet distributions:
the evolution of the flavor asymmetries
\begin{equation}
q_{NS,ik}^{(\pm)}=q_{i}^{(\pm)}-q_{k}^{(\pm)}\,,
\end{equation}
whose evolution is governed by
\begin{equation}
P_{NS}^{\pm}=P_{qq}^{V}\pm P_{q\bar{q}}^{V}\,,
\end{equation}
and the sum of the valence distributions of all flavors $q^{(-)}$ which 
evolves with
\begin{equation}
P_{NS}^{V}=P_{qq}^{V}-P_{q\bar{q}}^{V}+n_{f}\left(P_{qq}^{S}
-P_{q\bar{q}}^{S}\right)\equiv P_{NS}^{-}+P_{NS}^{S}.
\label{eq:PNSv}
\end{equation}
Notice that the quark-quark splitting function $P_{qq}$ can be expressed as
\begin{equation}
P_{qq}=P_{NS}^{+}+n_{f}\left(P_{qq}^{S}+P_{q\bar{q}}^{S}\right)\equiv P_{NS}^{+}+P_{ps}.
\label{eq:Pqq}
\end{equation}
It is important to observe that
the nonsinglet contribution is the most relevant one in 
Eq.~(\ref{eq:Pqq}) at large
$x$, where the \emph{pure singlet} term $P_{ps}=P_{qq}^{S}+P_{q\bar{q}}^{S}$
is very small. At small $x$, on the other hand, the latter contribution
takes over, as $xP_{ps}$ does not vanish for $x\rightarrow0$, unlike
$xP_{NS}^{+}$. 
The gluon-quark and quark-gluon entries are given by
\begin{equation}
P_{qg}=n_{f}P_{q_{i}g},
\end{equation}
\begin{equation}
P_{gq}=P_{gq_{i}}
\end{equation}
in terms of the flavor-independent splitting functions $P_{q_{i}g}=P_{\bar{q}_{i}g}$
and $P_{gq_{i}}=P_{g\bar{q}_{i}}$. With the exception of the first
order part of $P_{qg}$, neither of the quantities $xP_{qg}$, $xP_{gq}$
and $xP_{gg}$ vanish for $x\rightarrow0$.

In the expansion in powers of $\alpha_{s}$ of the evolution equations,
the flavor-diagonal (valence) quantity $P_{qq}^{V}$ is of order $\alpha_{s}$,
while $P_{q\bar{q}}^{V}$ and the flavor-independent (sea) contributions
$P_{qq}^{S}$ and $P_{q\bar{q}}^{S}$ are of order $\alpha_{s}^{2}$.
A non-vanishing difference $P_{qq}^{S}-P_{q\bar{q}}^{S}$ is present at 
order $\alpha_s^3$.

The next step is to choose a proper basis of nonsinglet distributions
that allows us to reconstruct, through linear combinations, the distribution
of each parton. The singlet evolution gives us 2 distributions,
$g$ and $q^{(+)}$, so we need to evolve $2n_{f}-1$ independent
nonsinglet distributions. We choose

\begin{enumerate}
\item $q^{(-)}$, evolving with $P_{NS}^{V}$;
\item $q_{NS,1i}^{(-)}=q_{1}^{(-)}-q_{i}^{(-)}$ (for 
$2\leq i\leq n_{f}$), evolving with $P_{NS}^{-}$;
\item $q_{NS,1i}^{(+)}=q_{1}^{(+)}-q_{i}^{(+)}$ (for 
$2\leq i\leq n_{f}$), evolving with $P_{NS}^{+}$,
\end{enumerate}
and use simple relations such as
\begin{equation}
q_{i}^{(\pm)}=\frac{1}{n_{f}}\left(q^{(\pm)}+\sum_{k=1,k\neq i}^{n_{f}}q_{NS,ik}^{(\pm)}\right)
\label{eq:comb_linNS}
\end{equation}
to perform the reconstructions of the various flavors.
Choosing $i=1$ in (\ref{eq:comb_linNS}), we compute $q_{1}^{(-)}$
from the evolved nonsinglets of type 1 and 2 and $q_{1}^{(+)}$ from
the evolved singlet $q^{(+)}$ and nonsinglet of type 3. Then from
the nonsinglets 2 and 3 we compute respectively $q_{i}^{(-)}$ and
$q_{i}^{(+)}$ for each $i$ such that $2\leq i\leq n_{f}$, and finally
$q_{i}$ and $\bar{q}_{i}$.

Moving from NNLO to NLO things simplify, as we have $P_{qq}^{S,(1)}=P_{q\bar{q}}^{S,(1)}$.
This implies (see Eq.~(\ref{eq:PNSv})) that $P_{NS}^{V,(1)}=P_{NS}^{-,(1)}$,
i.e.~the nonsinglets $q^{(-)}$ and $q_{NS,ik}^{(-)}$ evolve with
the same kernel, and the same does each linear combination thereof,
in particular $q_{i}^{(-)}$ for each flavor $i$. The basis of the $2n_{f}-1$
nonsinglet distributions that we choose to evolve at NLO is

\begin{enumerate}
\item $q_{i}^{(-)}$ (for each $i\leq n_{f}$), evolving with $P_{NS}^{-,(1)}$,
\item $q_{NS,1i}^{(+)}=q_{1}^{(+)}-q_{i}^{(+)}$ (for each $i$ such that
$2\leq i\leq n_{f}$), evolving with $P_{NS}^{+,(1)}$,
\end{enumerate}
and the same we do at LO, where we have in addition $P_{NS}^{+,(0)}=P_{NS}^{-,(0)}$,
being $P_{q\bar{q}}^{V,(0)}=0$.

\begin{figure}
\includegraphics[width=9.5cm,angle=-90]{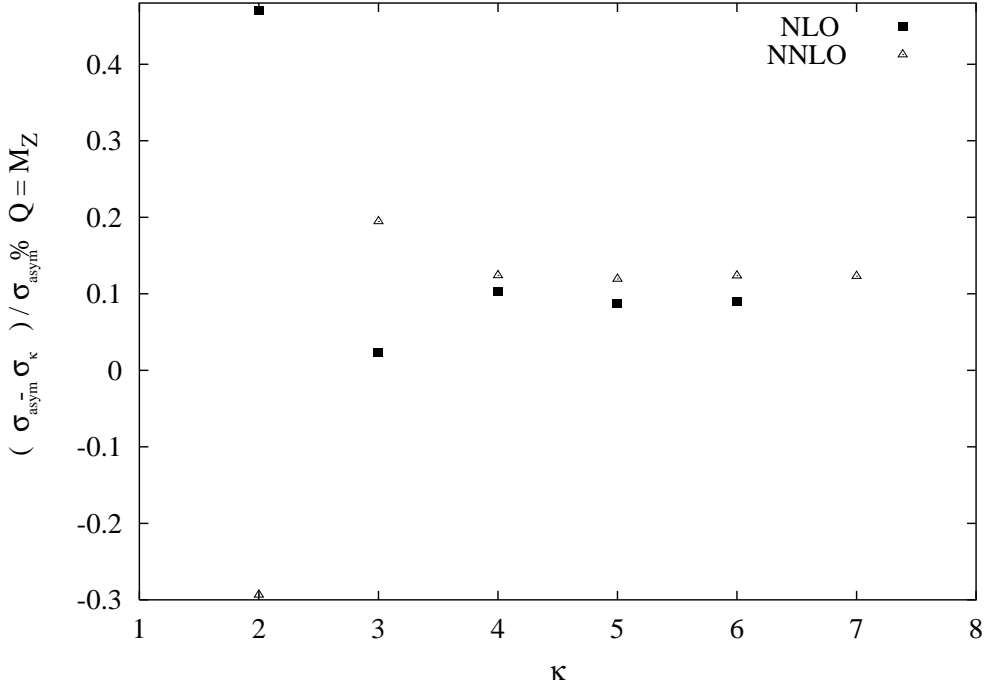}
\caption{ Plot of the percentage differences between the asymptotic Drell-Yan cross section
and those obtained using
expansions of the pdf's of a fixed order $\kappa$, shown
as a function of $\kappa$ for the NLO and NNLO cases. We have used the
MRST parametric input with $\mu_0=1$ GeV and $Q=M_{Z}$. The evolution is based on \textsc{Candia}.}
\label{TR_NNLO.ps}
\end{figure}

\section{The cross section and the parton luminosities}
Our NNLO analysis of the total cross section for lepton pair production combines
the hard scatterings of \cite{Van_Neerven1}, implemented by us in a program called 
\textsc{Candia}$_{DY}$, which combines the hard scatterings with the evolution performed by \textsc{Candia}.
We will present in a section below some results obtained by interfacing  \textsc{Vrap} and \textsc{Candia} that allow to extend some of the predictions of \cite{Anastasiou} with the inclusion of
the factorization/renormalization scale dependence not only in the hard scatterings but also in the
evolution. Here our main analysis is instead focused on the cross section for the mass distribution 
${d\sigma}/{d Q^2}$. 

Lepton pair production at low $Q$ via the Drell-Yan is sensitive to the pdf's at small-x values while in the high mass region, above the peak, is essential for the search of 
additional neutral currents. The general structure of the factorization formula for the color averaged
inclusive cross section for lepton pair production is given by \cite{Van_Neerven1}
\ba
\frac{d\sigma}{d Q^2}=\tau \sigma_V(Q^2,M_V^2)W_V(\tau,Q^2)&& \tau=Q^2/S,
\ea
where $\sigma_V$ is the point-like cross section in the case of the $\gamma,Z$
and the interference $\gamma$-$Z$. $S$ is the center of mass energy
of the incoming hadrons and $Q^2$ is the invariant mass of the di-lepton pair, respectively.
We have used the relations
\ba
&&\sigma_{\gamma}(Q^2)=\frac{4\pi\alpha_{em}^2}{3Q^4}\frac{1}{N_C}\nonumber\\
&&\sigma_{Z}(Q^2)=\frac{\pi\alpha_{em}}{4 M_Z N_C\sin^2{\theta_W}\cos^2{\theta_W} }
\frac{\Gamma_{Z\rightarrow l\bar{l}}}{(Q^2-M_Z^2)^2 + M_Z^2\Gamma^2_Z}\nonumber\\
&&\sigma_{\gamma Z}(Q^2)=\frac{\pi\alpha_{em}^2}{6} \frac{(1-4\sin^2{\theta_W})}{\sin^2{\theta_W}\cos^2{\theta_W}}
\frac{(Q^2-M_Z^2)}{N_C Q^2(Q^2-M_Z^2)^2+M_Z^2\Gamma_Z^2},\nonumber\\
\ea
where $\Gamma_{Z\rightarrow l\bar{l}}=0.0839136$ GeV, $\Gamma_Z=2.4952$ GeV,
$\sin^2{\theta_W}=0.23143$ and $\alpha_{em}(M_Z)=1/128$. These choices, performed as in
\cite{Anastasiou} are expected to account for the factorizable 
electroweak corrections, using the effective Born approximation \cite{Baur},\cite{All}. The non-factorizable contribution,
very relevant in the large invariant mass region ($Q=160$ GeV and above) are 
estimated to be much larger \cite{Baur}.

In all our studies we have fixed the 
energy of the collision to be $\sqrt{S}=14$ TeV.

The hadronic structure function $W_V(\tau,Q^2)$ is represented by a convolution product
between the parton luminosities $\Phi^{V}_{i j}(x,\mu_R^2,\mu_F^2)$ and the Wilson coefficients
$\Delta_{i j}(x,Q^2,\mu_R^2,\mu_F^2)$
\ba
W_Z(\tau,Q^2, \mu_R^2,\mu_F^2)&=&\sum_{i, j}\int_{\tau}^{1}\frac{d x}{x}\Phi_{i j}(x,\mu_R^2,\mu_F^2)
\Delta_{i j}(\frac{\tau}{x},Q^2,\mu_F^2),
\ea
where the luminosities are given by
\ba
\Phi_{i j}(x,\mu_R^2,\mu_F^2)=\int_{x}^{1}\frac{d y}{y}f_{i}(y,\mu_R^2,\mu_F^2) f_{j}\left(\frac{x}{y},\mu_R^2,\mu_F^2\right)
\equiv  \left[f_{i}\otimes f_{j}\right](x,\mu_R^2,\mu_F^2)
\ea
and the Wilson coefficients depend from both scales
\ba
\Delta_{i j}(x,Q^2,\mu_F^2)=
\sum_{n=0}^{\infty}\alpha_s^n(\mu_R^2)\Delta^{(n)}_{i j}(x,Q^2,\mu_F^2,\mu_R^2).
\ea
The explicit expressions of the hard scatterings coefficients have been taken from \cite{Van_Neerven1} and implemented in \textsc{Candia}. Moving to the parton densities, these are decomposed into their singlet ({\bf S}) and nonsinglet
({\bf NS}) contributions starting from the explicit expression
\ba
&&\left[q_{i}\otimes \bar{q}_{j}\right](x,\mu_F^2)=\frac{1}{4}
\left(q^{(+)}_{i}+q^{(-)}_{i}\right)\otimes\left(q^{(+)}_{j}-q^{(-)}_{j}\right)=
\nonumber\\
&&\hspace{2cm}\frac{1}{4 n_f^2}\left[\left(q^{(+)}+\sum_{k=1,k\neq i}^{n_{f}}q_{NS,ik}^{(+)}\right)+
\left(q^{(-)}+\sum_{k=1,k\neq i}^{n_{f}}q_{NS,ik}^{(-)}\right)\right]\otimes
\nonumber\\
&&\hspace{2cm}\left[\left(q^{(+)}+\sum_{k=1,k\neq j}^{n_{f}}q_{NS,jk}^{(+)}\right)+
\left(q^{(-)}+\sum_{k=1,k\neq j}^{n_{f}}q_{NS,jk}^{(-)}\right)\right]\,,
\ea
and after an expansion, one identifies, as usual, the convolution products
$\textbf{S}\otimes\textbf{S}$, $ \textbf{NS}\otimes\textbf{NS}$
and $ \textbf{S}\otimes\textbf{NS}$.

As we have already mentioned, in each of this sectors we are entitled
to implement evolved pdf's of different accuracy, according to the
classification presented in the previous section. Summarizing, we have, for the nonsinglet sector: 
1) exact solutions of the NNLO exact equation; 
2) exact solution of the NNLO truncated equation;
3) truncated solution of the NNLO truncated equation, while for the singlet case we 
have only the option of the $\kappa^{\prime}$-truncated solutions.

As we have already explained, we work with the equations written according to 
form-1, which has a single expansion parameter $(\kappa^{\prime})$.
This implies that the parton luminosities can be of a varying accuracy depending on the type of the solutions. The numerical analysis of
these choices is very involved for realistic distributions, as we are going to discuss next. 
We remark that there are differences between the iterated solutions of type-1 
and the {\em brute force}
solutions or the exact solutions, which are also available in the nonsinglet case. 
We have tried to answer this subtle point by showing in Fig.~\ref{TR_NNLO.ps} the results for 
the cross sections determined at NLO and at NNLO using as input the MRST conditions taken from the grid,
evolved by us using different sets of solutions. We recall that the initial condition $\mu_0^2=1$ GeV$^2$ means
that we are using the MRST parametric input \cite{MRST1}.
We have defined the ``asymptotic solution'' to be 
$\sigma_{asymp}$, built using the exact solution in the nonsinglet sector and a truncated 
solution in the singlet, with the index of truncation $\kappa$ sufficiently large so that an asymptotic value for the logarithmic expansion ($\kappa=8$) is obtained. We plot the percentage difference, normalized as shown 
in the figure, between truncated solutions of a varying $\kappa$ index and this asymptotic 
cross section. It is clear, from this analysis, that the iterates of fixed accuracy, expanded in powers of
$\log(\alpha_s)$, do not converge to the asymptotic solution but 
give cross sections that 
differ by a small but finite amount from that. This is quantified to be of the order of $0.1-0.5 \%$ at 
the energy reported in the plot. This estimate is subject to change as we vary the energy scale 
and the model of the initial conditions.
 On the basis of this result, we may reasonably assume that the sequence of truncations, respect to 
the {\em brute force} solution, or exact solution, should be of the order of a percent or so. This could be 
quantified better using a numerical code that solves the DGLAP by direct discretization, which is not available to us. From this point on, all the analysis that follows is going to be based on the implementation of $\sigma_{asymp}$, as defined above. More details concerning the difference between truncated and asymptotic solutions, a critical analysis of these results and of their 
implications for precision studies of the parton model at NNLO  will be presented below. 

\begin{figure}
\subfigure[$xu_v(x,\mu_F)$ at NLO]{\includegraphics[%
  width=6cm,
  angle=-90]{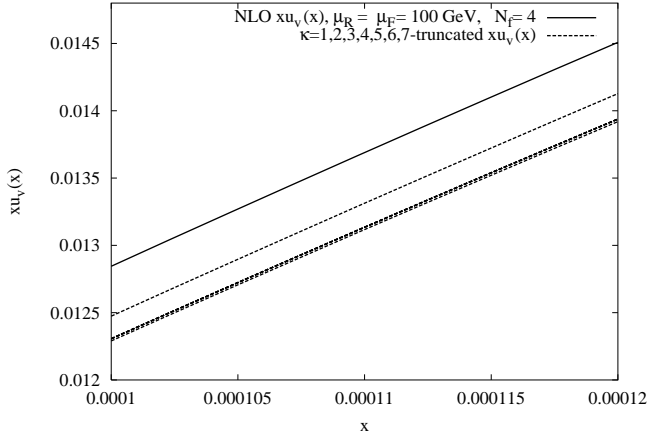}}
\subfigure[$xu_v(x,\mu_F)$ at NNLO]{\includegraphics[%
  width=6cm,
  angle=-90]{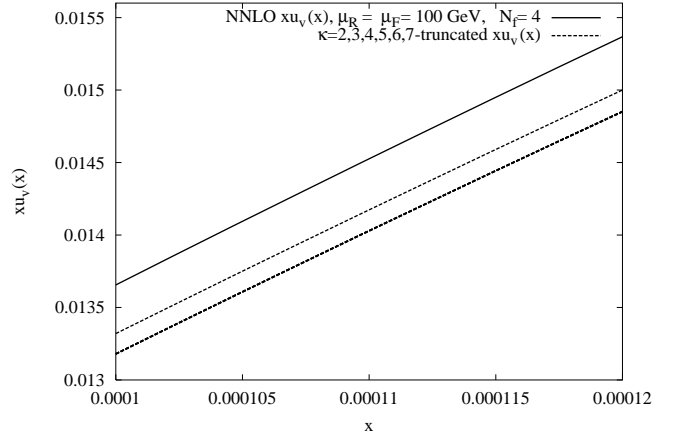}}
\subfigure[$xg(x,\mu_F)$ at NLO]{\includegraphics[%
  width=6cm,
  angle=-90]{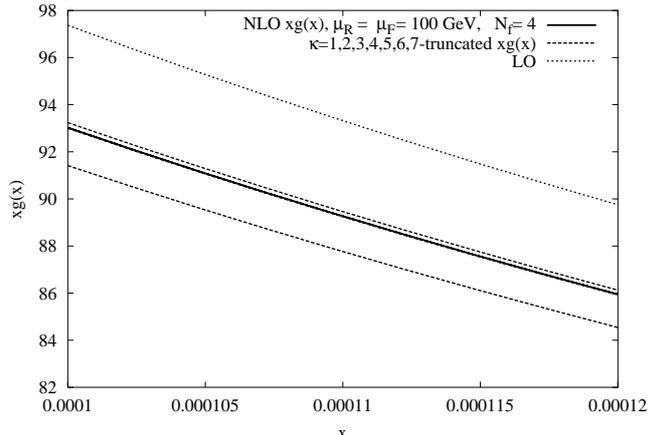}}
\subfigure[$xg(x,\mu_F)$ at NNLO]{\includegraphics[%
  width=6cm,
  angle=-90]{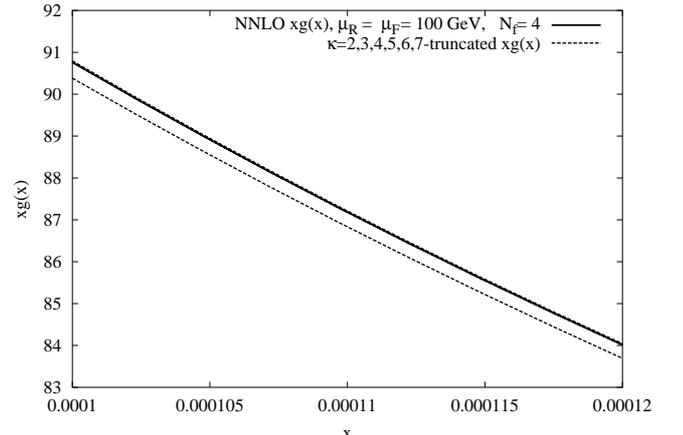}}
\caption{Asymptotic and truncated pdf's for the valence up-quark and for the gluons
at NLO and NNLO with $\mu_R=\mu_F=100$ GeV. We have selected the Les Houches input and a fixed number of flavors, $N_{f}=4$. The small 
range for $x$ has been chosen to resolve among the various predictions.}
\label{PDF1}
\end{figure}

\section{Numerical Analysis: Comparison with the Les Houches and the MRST Models}

We start presenting in this section our comparisons between the results for the evolution and the cross sections obtained using \textsc{Candia} against those of other implementations. In doing this we have made sure that the same conditions are kept 
in regard both to the treatment of the heavy flavors and of the initial conditions when running the different codes. In particular, 
the parameters of the runs have been selected so to generate either truncated solutions or asymptotic solutions, as specified above. 
\subsection{Comparisons with the Les Houches benchmarks}

We start our comparison using as initial conditions those presented in the Les Houches Model
\cite{LesHouches02}, which have been used to determine some benchmarks for the evolution.
The implementations that we compare, in this case, are those of \textsc{Candia} and \textsc{Pegasus}, the latter running with the option IMODEV$=1$. This option generates
exact solutions of the evolution equations by using a large sequence of truncates in Mellin space, with both parameters $\kappa$ and
$\kappa^{\prime}$ large, according to the $U$-ansatz (\ref{Uansatz}). The heavy quarks have been treated according to the VFN scheme. 

In the Les Houches model \cite{LesHouches02} the input distributions mimic
the CTEQ5M \cite{CTEQ5M} parameterization and are used regardless of the order of the evolution equations. They are given by
\ba
&&xu_{v}(x) =  5.107200x^{0.8}(1-x)^{3}\nonumber \\
&&xd_{v}(x)  =  3.064320x^{0.8}(1-x)^{4}\nonumber \\
&&xg(x)  =  1.700000x^{-0.1}(1-x)^{5}\nonumber \\
&&x\bar{d}(x)  =  0.1939875x^{-0.1}(1-x)^{6}\nonumber \\
&&x\bar{u}(x)  =  (1-x)x\bar{d}(x)\nonumber \\
&&xs(x)=x\bar{s}(x)  =  0.2x(\bar{u}+\bar{d})(x)\,,
\ea
and the running coupling has the value $\alpha_{s}(\mu_{R,0}^{2}=2\,\textrm{GeV}^{2})=0.35$. Our 
implementation in \textsc{Candia} of the heavy thresholds, in this case, follows exactly the one described in 
\cite{Pegasus}. To show the very good agreement between our
method of solution and \textsc{Pegasus} we detail the results for all the sectors. 
We have included both the numerical values for the pdf's and the LO,
NLO and NNLO predictions for the cross sections obtained by the two different implementations of the evolution. Tables~\ref{comp1}-
\ref{comp6} show the gluon and u-quark
distributions using the two evolutions at the various orders. In both cases we keep the ``asymptotic'' mode
(IMODEV=1 for \textsc{Pegasus}) and the asymptotic solutions in \textsc{Candia}, with the nonsinglet
treated using the exact iterated ansatz. It can be noticed that the differences are very 
small for all the densities up to NNLO. They can be read directly from the Tables~(\ref{comp1},\ref{comp3},\ref{comp5}) since
$x\delta f(x)$ are the relative differences normalized to the \textsc{Pegasus} determination, i.e. 
$x\delta f(x)\equiv (x f(x)_{\textsc{Pegasus}} - x f(x)_{\textsc{Candia}})/x f(x)_{\textsc{Pegasus}}$. 
The percentage differences for the gluon densities are 0.2 \% or smaller at NLO, $0.4 \%$ and smaller at NNLO.
In the kinematical region relevant for the LHC they stay around $0.1 \%$ at NNLO.
The valence u-quark distributions, at NNLO, reach at most $1\%$ at $x=10^{-4}$, while they are about $0.1-0.2 \%$ at $x=10^{-2}$.
Coming to the cross sections, the differences between the two determinations
are pretty small.
They essentially coincide at LO, they are about $0.6 \%$ at NLO, while they are
about $0.3 \%$ and below at NNLO (see Tables~\ref{comp2},\ref{comp4} and \ref{comp6}).

\subsection{ Truncations and asymptotic solutions}

The reader can find
in a sequence of 8 tables (see Tabs.~\ref{tab1},\ref{tab2},\ref{tab3},\ref{tab4},\ref{tab5},\ref{tab6},\ref{tab7},\ref{tab8}) 
added at the end of this work detailed numerical results for the various
truncated solutions and for the corresponding asymptotic solution in the 
Les Houches model and in a realistic model, MRST \cite{MRST1}.

We show in Figs.~(\ref{PDF1}) four plots of the valence up-quark and of the gluon distributions for various $\kappa$ values. The small range
of variability in $x$ has been chosen so to render the differences in the plots visible, since they are quite small.
The various $\kappa$ solutions converge toward the asymptotic solution as the index of 
truncation increases. We show the exact (for the valence up-quark distribution)
or the asymptotic solution (for the gluon density) and the various 
truncated solutions for several $\kappa$ values. 
In the case of the Les Houches model, table \ref{tab1} and \ref{tab2} are particularly significant, since these show
for the nonsinglet the existence of a difference between the exact solutions,
that performs a resummation of the $\log(\alpha_s)$, and the sequence of truncated solutions,
which reach saturation at $\kappa=6$. The differences for the  valence up-quark distribution 
($x u_v(x)$) at NLO vary from $1 \%$ at $(x=10^{-3})$ to $0.7 \%$ 
($x=10^{-4}$), growing larger at $x=10^{-5}$, where they reach $7 \%$. This last value is presented only for comparison, although it is not relevant at the LHC. Moving to NNLO, the differences 
are about $4 \%$ at $x=10^{-5}$, $3\%$ at $x=10^{-4}$, decreasing to $0.4 \%$ at $x=10^{-2}$. 
They become significant at large $x$ values, being around $9 \%$ at $x=0.9$. These determinations, of course, need to be tested in 
the related cross sections in order to appreciate their real impact.
As we have already shown in Fig.~\ref{TR_NNLO.ps} 
the various determinations stay below $1\%$ for $Q=M_Z$. Even if these differences are not big, they will become 
more significant as the determination of the pdf's is going to 
improve in the near future, using the large amount of 
data coming from the LHC. This will allow to reduce the errors on the pdf's and, therefore, on the cross sections. As we are going to show next, these errors remain, at the moment, larger than 
the theoretical indetermination coming from the choice of the solution, at least in the region that we have explored. In the gluon sector (see Tab.\ref{tab3}, \ref{tab4}) the situation seems to improve, and the differences 
stay below 1 $\% $ in all the x-range, but this can be misleading: asymptotic and truncated 
solutions in the singlet sector are in fact both determined by the same logarithmic ansatz.  

A similar analysis has been performed for the MRST model. 
In this case we perform the evolution using \textsc{Candia}, the MRST input and a treatment of the heavy flavors
exactly as in MRST, with the thresholds for the heavy quarks
chosen as in \cite{MRST1}. Also in this case truncated solutions and asymptotic
solutions show a small difference, both for the valence distributions and for the singlet ones. 
We show in tables \ref{tab5},\ref{tab6},\ref{tab7} and \ref{tab8} results
for the various $\kappa$-truncated (accurate) solutions.

For instance, in the case of the gluon density, if we choose $\kappa=3$ 
(3rd truncated solution), at $x=10^{-3}$ the difference in the gluon density 
respect to the asymptotic solution is about $0.01 \%$ at NNLO, which appears to be small, but can easily grow to $0.5 \% $ or so if would let 
a {\em brute force} solution replace the asymptotic determination. In fact the valence u-quark distribution, whose asymptotic value is supposed to be 
pretty close to the exact value, shows more substantial differences. For instance, at NLO, for $x=10^{-3}$  the same truncated ansatz $(\kappa=3)$ 
differs from the exact one by $2.6 \%$. At NNLO in the more relevant region 
of $x$ (0.01-0.1) is about $2 \%$ and below. The differences grow bigger 
at larger x-values, for instance they are $9 \% $ for $x=0.5$ at NNLO. 

Coming to the cross sections obtained by the various truncated
solutions, these are shown in two tables (see Tabs. \ref{sigma_1}, \ref{sigma_2}),
which summarize these studies at NLO and NNLO respectively. Using again the 
$\kappa=3$ solution, for $Q=M_Z$ the NLO determination differs by 
$0.2 \%$ compared to the asymptotic one. They tend to grow at larger 
Q-values, $0.4 \%$ at $Q=200$ GeV (NNLO). 

There are some conclusions that we can draw from this analysis. We clearly have {\em several} ways to choose the solution and by doing so we make errors which are around $1 \% $. They tend to grow as Q increases, at larger invariant mass of the lepton pair, 
where we get more sensitive to larger x-values. This theoretical 
errors may grow slightly bigger at very large Q-values, say for Q 
around 1 TeV or so, where we need specific 
studies of that kinematical region, since we could expect that extra neutral interactions be found. It is important, however, to remind that 
the DY cross section is anyhow quite sensitive to the behavior of the hard scatterings around $x=1$, as pointed out in \cite{Van_Neerven1}. This implies that various determinations may 
differ already at percent level because of the different treatment 
of the edge-point region in the Bjorken variable even for moderate Q 
values.

\begin{table}
\begin{footnotesize}
\begin{center}
\begin{tabular}{|c||c|c|c|c|c|c|}
\hline
\multicolumn{7}{|c|}{\textsc{Candia} vs \textsc{Pegasus} PDFs at LO, Les Houches input, VFN scheme, $Q=\mu_{F}=\mu_{R}=100$ GeV}
\tabularnewline
\hline
$ x $&
$xg(x)^{\textsc{Candia}}_{asymp}$     &
$xg(x)^{\textsc{Pegasus}}$&
$\delta xg(x) $         &
$xu_v(x)^{\textsc{Candia}}_{asymp}$     &
$xu_v(x)^{\textsc{Pegasus}}$&
$\delta xu_v(x) $\tabularnewline
\hline
\hline
$1e-05$&
$2.5282\cdot10^{+2}$&
$2.5282\cdot10^{+2}$&
$5.0194\cdot10^{-6}$&
$1.9006\cdot10^{-3}$&
$1.9006\cdot10^{-3}$&
$2.2551\cdot10^{-5}$
\tabularnewline
\hline
$0.0001$&
$9.6048\cdot10^{+1}$&
$9.6048\cdot10^{+1}$&
$9.8076\cdot10^{-7}$&
$1.0186\cdot10^{-2}$&
$1.0186\cdot10^{-2}$&
$1.6788\cdot10^{-5}$
\tabularnewline
\hline
$0.001$&
$3.1333\cdot10^{+1}$&
$3.1333\cdot10^{+1}$&
$5.5756\cdot10^{-6}$&
$5.0893\cdot10^{-2}$&
$5.0893\cdot10^{-2}$&
$6.7161\cdot10^{-6}$
\tabularnewline
\hline
$0.01$&
$7.7728\cdot10^{+0}$&
$7.7728\cdot10^{+0}$&
$3.4093\cdot10^{-7}$&
$2.2080\cdot10^{-1}$&
$2.2080\cdot10^{-1}$&
$7.6268\cdot10^{-6}$
\tabularnewline
\hline
$0.1$&
$8.4358\cdot10^{-1}$&
$8.4358\cdot10^{-1}$&
$4.8152\cdot10^{-6}$&
$5.7166\cdot10^{-1}$&
$5.7166\cdot10^{-1}$&
$5.8339\cdot10^{-6}$
\tabularnewline
\hline
$0.2$&
$2.3925\cdot10^{-1}$&
$2.3925\cdot10^{-1}$&
$1.0157\cdot10^{-6}$&
$5.1570\cdot10^{-1}$&
$5.1570\cdot10^{-1}$&
$2.5305\cdot10^{-6}$
\tabularnewline
\hline
$0.3$&
$7.8026\cdot10^{-2}$&
$7.8026\cdot10^{-2}$&
$4.1486\cdot10^{-6}$&
$3.7597\cdot10^{-1}$&
$3.7597\cdot10^{-1}$&
$6.3782\cdot10^{-6}$
\tabularnewline
\hline
$0.4$&
$2.5211\cdot10^{-2}$&
$2.5211\cdot10^{-2}$&
$1.7143\cdot10^{-5}$&
$2.3918\cdot10^{-1}$&
$2.3918\cdot10^{-1}$&
$6.3425\cdot10^{-6}$
\tabularnewline
\hline
$0.5$&
$7.4719\cdot10^{-3}$&
$7.4719\cdot10^{-3}$&
$6.1470\cdot10^{-6}$&
$1.3284\cdot10^{-1}$&
$1.3284\cdot10^{-1}$&
$2.7469\cdot10^{-5}$
\tabularnewline
\hline
$0.6$&
$1.8760\cdot10^{-3}$&
$1.8760\cdot10^{-3}$&
$1.1295\cdot10^{-5}$&
$6.2211\cdot10^{-2}$&
$6.2211\cdot10^{-2}$&
$6.2272\cdot10^{-6}$
\tabularnewline
\hline
$0.7$&
$3.5241\cdot10^{-4}$&
$3.5241\cdot10^{-4}$&
$1.0386\cdot10^{-6}$&
$2.2643\cdot10^{-2}$&
$2.2643\cdot10^{-2}$&
$1.1717\cdot10^{-5}$
\tabularnewline
\hline
$0.8$&
$3.8055\cdot10^{-5}$&
$3.8054\cdot10^{-5}$&
$1.9078\cdot10^{-5}$&
$5.2773\cdot10^{-3}$&
$5.2773\cdot10^{-3}$&
$4.5213\cdot10^{-6}$
\tabularnewline
\hline
$0.9$&
$1.0310\cdot10^{-6}$&
$1.0306\cdot10^{-6}$&
$3.9758\cdot10^{-4}$&
$4.2048\cdot10^{-4}$&
$4.2047\cdot10^{-4}$&
$3.0730\cdot10^{-5}$
\tabularnewline
\hline
\end{tabular}
\caption{Comparison between the pdf's obtained using \textsc{Candia} versus those obtained using \textsc{Pegasus}
and the normalized differences, ex.: $\delta xg(x)=|xg(x)^{\textsc{Candia}}-xg(x)^{\textsc{Pegasus}}|/xg(x)^{\textsc{Pegasus}}$ at LO.}
\label{comp1}
\end{center}
\end{footnotesize}
\end{table}

\begin{table}
\begin{center}
\begin{tabular}{|c||c|c|c|}
\hline
\multicolumn{4}{|c|}{$\textrm{d}\sigma_{LO}/\textrm{d}Q$ [pb/GeV]. \textsc{Candia} vs \textsc{Pegasus} with Les Houches input.}
\tabularnewline
\hline
$Q ~[\textrm{GeV}]$&
$\sigma_{LO}^{\textsc{Candia}}$&
$\sigma_{LO}^{\textsc{Pegasus}}$  &
$\delta\sigma_{LO}  $ \tabularnewline
\hline
\hline
$50.0000$&
$4.8995\cdot10^{+0}$&
$4.8995\cdot10^{+0}$&
$6.1231\cdot10^{-7}$
\tabularnewline
\hline
$60.0469$&
$3.0209\cdot10^{+0}$&
$3.0209\cdot10^{+0}$&
$1.3241\cdot10^{-6}$
\tabularnewline
\hline
$70.0938$&
$2.7805\cdot10^{+0}$&
$2.7805\cdot10^{+0}$&
$4.3157\cdot10^{-6}$
\tabularnewline
\hline
$80.1407$&
$5.7936\cdot10^{+0}$&
$5.7936\cdot10^{+0}$&
$1.7260\cdot10^{-6}$
\tabularnewline
\hline
$90.1876$&
$2.2499\cdot10^{+2}$&
$2.2499\cdot10^{+2}$&
$2.0712\cdot10^{-6}$

\tabularnewline
\hline
$91.1876$&
$3.6905\cdot10^{+2}$&
$3.6905\cdot10^{+2}$&
$3.4413\cdot10^{-6}$
\tabularnewline
\hline
$92.1876$&
$2.2475\cdot10^{+2}$&
$2.2475\cdot10^{+2}$&
$1.6907\cdot10^{-6}$
\tabularnewline
\hline
$120.0701$&
$7.2456\cdot10^{-1}$&
$7.2456\cdot10^{-1}$&
$0$
\tabularnewline
\hline
$146.0938$&
$2.0557\cdot10^{-1}$&
$2.0557\cdot10^{-1}$&
$9.7291\cdot10^{-6}$
\tabularnewline
\hline
$172.1175$&
$8.9583\cdot10^{-2}$&
$8.9584\cdot10^{-2}$&
$1.1163\cdot10^{-5}$
\tabularnewline
\hline
$200.0000$&
$4.4674\cdot10^{-2}$&
$4.4674\cdot10^{-2}$&
$0$
\tabularnewline
\hline
\end{tabular}
\caption{Comparison between the cross sections obtained using \textsc{Candia} and \textsc{Pegasus} at LO.}
\label{comp2}
\end{center}
\end{table}

\begin{table}
\begin{footnotesize}
\begin{center}
\begin{tabular}{|c||c|c|c|c|c|c|}
\hline
\multicolumn{7}{|c|}{\textsc{Candia} vs \textsc{Pegasus} PDFs at NLO, Les Houches input, VFN scheme, $Q=\mu_{F}=\mu_{R}=100$ GeV}
\tabularnewline
\hline
$ x $&
$xg(x)^{\textsc{Candia}}_{asymp}$     &
$xg(x)^{\textsc{Pegasus}}$&
$\delta xg(x) $         &
$xu_v(x)^{\textsc{Candia}}_{asymp}$     &
$xu_v(x)^{\textsc{Pegasus}}$&
$\delta xu_v(x) $\tabularnewline
\hline
\hline
$1e-05$&
$2.2804\cdot10^{+2}$&
$2.2753\cdot10^{+2}$&
$2.2623\cdot10^{-3}$&
$2.7428\cdot10^{-3}$&
$2.7419\cdot10^{-3}$&
$3.2619\cdot10^{-4}$
\tabularnewline
\hline
$0.0001$&
$8.9671\cdot10^{+1}$&
$8.9513\cdot10^{+1}$&
$1.7658\cdot10^{-3}$&
$1.3042\cdot10^{-2}$&
$1.3039\cdot10^{-2}$&
$2.5581\cdot10^{-4}$
\tabularnewline
\hline
$0.001$&
$3.0284\cdot10^{+1}$&
$3.0245\cdot10^{+1}$&
$1.2762\cdot10^{-3}$&
$5.8519\cdot10^{-2}$&
$5.8507\cdot10^{-2}$&
$2.1253\cdot10^{-4}$
\tabularnewline
\hline
$0.01$&
$7.7547\cdot10^{+0}$&
$7.7491\cdot10^{+0}$&
$7.1653\cdot10^{-4}$&
$2.3132\cdot10^{-1}$&
$2.3128\cdot10^{-1}$&
$1.5701\cdot10^{-4}$
\tabularnewline
\hline
$0.1$&
$8.5590\cdot10^{-1}$&
$8.5586\cdot10^{-1}$&
$4.3846\cdot10^{-5}$&
$5.5328\cdot10^{-1}$&
$5.5324\cdot10^{-1}$&
$8.1196\cdot10^{-5}$
\tabularnewline
\hline
$0.2$&
$2.4330\cdot10^{-1}$&
$2.4335\cdot10^{-1}$&
$2.1829\cdot10^{-4}$&
$4.8848\cdot10^{-1}$&
$4.8845\cdot10^{-1}$&
$5.5160\cdot10^{-5}$
\tabularnewline
\hline
$0.3$&
$7.9588\cdot10^{-2}$&
$7.9625\cdot10^{-2}$&
$4.5913\cdot10^{-4}$&
$3.5131\cdot10^{-1}$&
$3.5129\cdot10^{-1}$&
$4.3636\cdot10^{-5}$
\tabularnewline
\hline
$0.4$&
$2.5845\cdot10^{-2}$&
$2.5862\cdot10^{-2}$&
$6.4662\cdot10^{-4}$&
$2.2093\cdot10^{-1}$&
$2.2092\cdot10^{-1}$&
$5.2929\cdot10^{-5}$
\tabularnewline
\hline
$0.5$&
$7.7200\cdot10^{-3}$&
$7.7265\cdot10^{-3}$&
$8.4504\cdot10^{-4}$&
$1.2130\cdot10^{-1}$&
$1.2130\cdot10^{-1}$&
$4.1179\cdot10^{-5}$
\tabularnewline
\hline
$0.6$&
$1.9616\cdot10^{-3}$&
$1.9637\cdot10^{-3}$&
$1.0442\cdot10^{-3}$&
$5.6094\cdot10^{-2}$&
$5.6093\cdot10^{-2}$&
$1.8017\cdot10^{-5}$
\tabularnewline
\hline
$0.7$&
$3.7529\cdot10^{-4}$&
$3.7574\cdot10^{-4}$&
$1.1940\cdot10^{-3}$&
$2.0103\cdot10^{-2}$&
$2.0102\cdot10^{-2}$&
$3.3196\cdot10^{-5}$
\tabularnewline
\hline
$0.8$&
$4.1724\cdot10^{-5}$&
$4.1780\cdot10^{-5}$&
$1.3352\cdot10^{-3}$&
$4.5862\cdot10^{-3}$&
$4.5861\cdot10^{-3}$&
$2.0342\cdot10^{-5}$
\tabularnewline
\hline
$0.9$&
$1.1941\cdot10^{-6}$&
$1.1955\cdot10^{-6}$&
$1.1525\cdot10^{-3}$&
$3.5234\cdot10^{-4}$&
$3.5233\cdot10^{-4}$&
$1.9592\cdot10^{-5}$
\tabularnewline
\hline
\end{tabular}
\caption{Pdf's obtained in the two evolutions at NLO}
\label{comp3}
\end{center}
\end{footnotesize}
\end{table}

\begin{table}
\begin{center}
\begin{tabular}{|c||c|c|c|}
\hline
\multicolumn{4}{|c|}{$\textrm{d}\sigma_{NLO}/\textrm{d}Q$ [pb/GeV]. \textsc{Candia} vs \textsc{Pegasus} with Les Houches input.}
\tabularnewline
\hline
$Q ~[\textrm{GeV}]$&
$\sigma_{NLO}^{\textsc{Candia}}$&
$\sigma_{NLO}^{\textsc{Pegasus}}$  &
$\delta\sigma_{NLO}  $ \tabularnewline
\hline
\hline
$50.0000$&
$7.6946\cdot10^{+0}$&
$7.6419\cdot10^{+0}$&
$6.8857\cdot10^{-3}$
\tabularnewline
\hline
$60.0469$&
$4.6319\cdot10^{+0}$&
$4.6010\cdot10^{+0}$&
$6.7059\cdot10^{-3}$
\tabularnewline
\hline
$70.0938$&
$4.1787\cdot10^{+0}$&
$4.1515\cdot10^{+0}$&
$6.5564\cdot10^{-3}$
\tabularnewline
\hline
$80.1407$&
$8.5604\cdot10^{+0}$&
$8.5055\cdot10^{+0}$&
$6.4543\cdot10^{-3}$
\tabularnewline
\hline
$90.1876$&
$3.2787\cdot10^{+2}$&
$3.2581\cdot10^{+2}$&
$6.3294\cdot10^{-3}$
\tabularnewline
\hline
$91.1876$&
$5.3713\cdot10^{+2}$&
$5.3376\cdot10^{+2}$&
$6.3133\cdot10^{-3}$
\tabularnewline
\hline
$92.1876$&
$3.2672\cdot10^{+2}$&
$3.2468\cdot10^{+2}$&
$6.2844\cdot10^{-3}$
\tabularnewline
\hline
$120.0701$&
$1.0243\cdot10^{+0}$&
$1.0183\cdot10^{+0}$&
$5.8833\cdot10^{-3}$
\tabularnewline
\hline
$146.0938$&
$2.8483\cdot10^{-1}$&
$2.8325\cdot10^{-1}$&
$5.5852\cdot10^{-3}$
\tabularnewline
\hline
$172.1175$&
$1.2208\cdot10^{-1}$&
$1.2144\cdot10^{-1}$&
$5.2947\cdot10^{-3}$
\tabularnewline
\hline
$200.0000$&
$5.9997\cdot10^{-2}$&
$5.9694\cdot10^{-2}$&
$5.0759\cdot10^{-3}$
\tabularnewline
\hline
\end{tabular}
\caption{NLO cross sections obtained using \textsc{Candia} and \textsc{Pegasus} using the Les Houches model.}
\label{comp4}
\end{center}
\end{table}

\begin{table}
\begin{footnotesize}
\begin{center}
\begin{tabular}{|c||c|c|c|c|c|c|}
\hline
\multicolumn{7}{|c|}{\textsc{Candia} vs \textsc{Pegasus} PDFs at NNLO, Les Houches input, VFN scheme, $Q=\mu_{F}=\mu_{R}=100$ GeV}
\tabularnewline
\hline
$ x $&
$xg(x)^{\textsc{Candia}}_{asymp}$     &
$xg(x)^{\textsc{Pegasus}}$&
$\delta xg(x) $         &
$xu_v(x)^{\textsc{Candia}}_{asymp}$     &
$xu_v(x)^{\textsc{Pegasus}}$&
$\delta xu_v(x) $\tabularnewline
\hline
\hline
$1e-05$&
$2.1922\cdot10^{+2}$&
$2.2012\cdot10^{+2}$&
$4.1108\cdot10^{-3}$&
$3.0823\cdot10^{-3}$&
$3.1907\cdot10^{-3}$&
$3.3962\cdot10^{-2}$
\tabularnewline
\hline
$0.0001$&
$8.8486\cdot10^{+1}$&
$8.8804\cdot10^{+1}$&
$3.5856\cdot10^{-3}$&
$1.3871\cdot10^{-2}$&
$1.4023\cdot10^{-2}$&
$1.0811\cdot10^{-2}$
\tabularnewline
\hline
$0.001$&
$3.0319\cdot10^{+1}$&
$3.0404\cdot10^{+1}$&
$2.8106\cdot10^{-3}$&
$6.0060\cdot10^{-2}$&
$6.0019\cdot10^{-2}$&
$6.9117\cdot10^{-4}$
\tabularnewline
\hline
$0.01$&
$7.7785\cdot10^{+0}$&
$7.7912\cdot10^{+0}$&
$1.6326\cdot10^{-3}$&
$2.3287\cdot10^{-1}$&
$2.3244\cdot10^{-1}$&
$1.8584\cdot10^{-3}$
\tabularnewline
\hline
$0.1$&
$8.5284\cdot10^{-1}$&
$8.5266\cdot10^{-1}$&
$2.1595\cdot10^{-4}$&
$5.4977\cdot10^{-1}$&
$5.4993\cdot10^{-1}$&
$2.9526\cdot10^{-4}$
\tabularnewline
\hline
$0.2$&
$2.4183\cdot10^{-1}$&
$2.4161\cdot10^{-1}$&
$9.1195\cdot10^{-4}$&
$4.8313\cdot10^{-1}$&
$4.8323\cdot10^{-1}$&
$2.0148\cdot10^{-4}$
\tabularnewline
\hline
$0.3$&
$7.9005\cdot10^{-2}$&
$7.8898\cdot10^{-2}$&
$1.3515\cdot10^{-3}$&
$3.4629\cdot10^{-1}$&
$3.4622\cdot10^{-1}$&
$1.9857\cdot10^{-4}$
\tabularnewline
\hline
$0.4$&
$2.5636\cdot10^{-2}$&
$2.5594\cdot10^{-2}$&
$1.6452\cdot10^{-3}$&
$2.1711\cdot10^{-1}$&
$2.1696\cdot10^{-1}$&
$6.7488\cdot10^{-4}$
\tabularnewline
\hline
$0.5$&
$7.6538\cdot10^{-3}$&
$7.6398\cdot10^{-3}$&
$1.8314\cdot10^{-3}$&
$1.1883\cdot10^{-1}$&
$1.1868\cdot10^{-1}$&
$1.2434\cdot10^{-3}$
\tabularnewline
\hline
$0.6$&
$1.9439\cdot10^{-3}$&
$1.9401\cdot10^{-3}$&
$1.9844\cdot10^{-3}$&
$5.4753\cdot10^{-2}$&
$5.4652\cdot10^{-2}$&
$1.8520\cdot10^{-3}$
\tabularnewline
\hline
$0.7$&
$3.7162\cdot10^{-4}$&
$3.7080\cdot10^{-4}$&
$2.2059\cdot10^{-3}$&
$1.9537\cdot10^{-2}$&
$1.9486\cdot10^{-2}$&
$2.6105\cdot10^{-3}$
\tabularnewline
\hline
$0.8$&
$4.1248\cdot10^{-5}$&
$4.1141\cdot10^{-5}$&
$2.5990\cdot10^{-3}$&
$4.4306\cdot10^{-3}$&
$4.4148\cdot10^{-3}$&
$3.5750\cdot10^{-3}$
\tabularnewline
\hline
$0.9$&
$1.1766\cdot10^{-6}$&
$1.1722\cdot10^{-6}$&
$3.7723\cdot10^{-3}$&
$3.3696\cdot10^{-4}$&
$3.3522\cdot10^{-4}$&
$5.1816\cdot10^{-3}$
\tabularnewline
\hline
\end{tabular}
\end{center}
\caption{NNLO pdf's determined with \textsc{Candia} and \textsc{Pegasus} using the Les Houches model.}
\label{comp5}
\end{footnotesize}
\end{table}

\begin{table}
\begin{center}
\begin{tabular}{|c||c|c|c|}
\hline
\multicolumn{4}{|c|}{$\textrm{d}\sigma_{NNLO}/\textrm{d}Q$ [pb/GeV]. \textsc{Candia} vs \textsc{Pegasus} with Les Houches input.}
\tabularnewline
\hline
$Q ~[\textrm{GeV}]$&
$\sigma_{NNLO}^{\textsc{Candia}}$&
$\sigma_{NNLO}^{\textsc{Pegasus}}$  &
$\delta\sigma_{NNLO}  $ \tabularnewline
\hline
\hline
$50.0000$&
$8.0734\cdot10^{+0}$&
$8.1044\cdot10^{+0}$&
$3.8288\cdot10^{-3}$
\tabularnewline
\hline
$60.0469$&
$4.8771\cdot10^{+0}$&
$4.8948\cdot10^{+0}$&
$3.6106\cdot10^{-3}$
\tabularnewline
\hline
$70.0938$&
$4.4033\cdot10^{+0}$&
$4.4184\cdot10^{+0}$&
$3.4110\cdot10^{-3}$
\tabularnewline
\hline
$80.1407$&
$8.9241\cdot10^{+0}$&
$8.9527\cdot10^{+0}$&
$3.1936\cdot10^{-3}$
\tabularnewline
\hline
$90.1876$&
$3.3570\cdot10^{+2}$&
$3.3669\cdot10^{+2}$&
$2.9388\cdot10^{-3}$
\tabularnewline
\hline
$91.1876$&
$5.4905\cdot10^{+2}$&
$5.5067\cdot10^{+2}$&
$2.9299\cdot10^{-3}$
\tabularnewline
\hline
$92.1876$&
$3.3344\cdot10^{+2}$&
$3.3441\cdot10^{+2}$&
$2.8919\cdot10^{-3}$
\tabularnewline
\hline
$120.0701$&
$1.0249\cdot10^{+0}$&
$1.0274\cdot10^{+0}$&
$2.4285\cdot10^{-3}$
\tabularnewline
\hline
$146.0938$&
$2.8527\cdot10^{-1}$&
$2.8590\cdot10^{-1}$&
$2.1826\cdot10^{-3}$
\tabularnewline
\hline
$172.1175$&
$1.2295\cdot10^{-1}$&
$1.2319\cdot10^{-1}$&
$1.9887\cdot10^{-3}$
\tabularnewline
\hline
$200.0000$&
$6.0923\cdot10^{-2}$&
$6.1029\cdot10^{-2}$&
$1.7369\cdot10^{-3}$
\tabularnewline
\hline
\end{tabular}
\caption{NNLO cross sections in the two evolution methods.}
\label{comp6}
\end{center}
\end{table}

\section{Other comparisons with the MRST evolution}
Now we perform a comparison between the various cross sections
obtained using the pdf's evolved by MRST and the same distributions, taken at their starting value, but evolved by us using \textsc{Candia}. These studies are 
performed using in \textsc{Candia} the asymptotic solutions in the singlet 
and non singlet sectors. We use the MRST input in a grid form
with an initial scale $\mu_0^2=1.25$ GeV$^2$, $\sqrt{S}=14$ TeV and with
$\mu_F^2=\mu_R^2=Q^2$. The choices for the thresholds of the heavy 
flavors have been chosen in \textsc{Candia} to coincide with those reported by 
MRST. For this reason we have used for comparison the asymptotic solution and the VFN scheme. The 
relative variations are computed respect to the MRST value and are indicated 
in the columns labeled as $\delta\sigma$. Also 
in this case we show in 3 tables (\ref{comp7}-\ref{comp9})
the results for the LO, NLO and NNLO cross sections. 
The differences between our prediction and the MRST result for the total cross sections are 
around 1 per cent or below at LO, vary from  $0.02 \%$ to $0.3 \%$ 
at NLO and are $2.6 \%$ and below at NNLO. In this case the maximum 
difference has been found for $Q=50$ GeV. These differences, clearly, affect the values
of the $K$-factors, as we are going to discuss below, which in our evolution are
larger compared to those of MRST.

We perform some more tests using  \textsc{Vrap} \cite{Anastasiou} and compare the results against those of \textsc{Candia}$_{{\rm DY}}$ for the calculations of the hard scattering piece. In the results given below $\hat{\sigma}_{ \textsc{Vrap}}$ refers to the 
hard scatterings for the invariant mass distributions computed using  \textsc{Vrap}, 
while $\Phi_{MRST_{\textsc{Candia}}}$ refers to the luminosities using one of the 
MRST inputs evolved using \textsc{Candia}. Similarly, $\Phi_{MRST_{evol}}$ denotes the luminosities predicted by MRST with their evolution. In this case the original scale is not indicated, but 
the grid scale $\mu_0=1.25$ GeV$^2$ is the first available point, at which the evolution with 
\textsc{Candia} is also interfaced.

The NLO total cross sections in [pb/GeV] at the peak $Q=M_Z$ and at 
$\sqrt{S}=14$ TeV using the MRST inputs with 
$\mu_0^2=1$ GeV$^2$ and $\mu_0^2=1.25$ GeV$^2$, in the various cases, are given by 

\ba
&&\hat{\sigma}^{NLO}_{Vrap}\otimes\Phi_{MRST_{evol}}=501.96\,,\nonumber\\
&&\hat{\sigma}^{NLO}_{Vrap}\otimes\Phi_{MRST_{\textsc{Candia}}}=505.87 \hspace{0.5cm} {\rm from }\,\,\mu_0^2=1\,\, {\rm GeV}^2\,,\nonumber\\
&&\hat{\sigma}^{NLO}_{Vrap}\otimes\Phi_{MRST_{\textsc{Candia}}}=502.65  \hspace{0.5cm} {\rm from }\,\,\mu_0^2=1.25\,\, {\rm GeV}^2\,,\nonumber\\
&&\hat{\sigma}^{NLO}_{\textsc{Candia}DY}\otimes\Phi_{MRST_{evol}}=501.72\,,\nonumber\\
&&\hat{\sigma}^{NLO}_{\textsc{Candia}DY}\otimes\Phi_{MRST_{\textsc{Candia}}}=505.82 \hspace{0.5cm} {\rm from }\,\,\mu_0^2=1 \,\, {\rm GeV}^2\,,\nonumber\\
&&\hat{\sigma}^{NLO}_{\textsc{Candia}DY}\otimes\Phi_{MRST_{\textsc{Candia}}}=502.42 \hspace{0.5cm} {\rm from }\,\,\mu_0^2=1.25\,\, {\rm GeV}^2\,
\ea
with differences that stay well below $1 \%$, while at NNLO we obtain

\ba
&&\hat{\sigma}^{NNLO}_{Vrap}\otimes\Phi_{MRST_{evol}}=490.51\,,\nonumber\\
&&\hat{\sigma}^{NNLO}_{Vrap}\otimes\Phi_{MRST_{\textsc{Candia}}}=479.60 \hspace{0.5cm} {\rm from }\,\,\mu_0^2=1\,\, {\rm GeV}^2\,,\nonumber\\
&&\hat{\sigma}^{NNLO}_{Vrap}\otimes\Phi_{MRST_{\textsc{Candia}}}=482.63  \hspace{0.5cm} {\rm from }\,\,\mu_0^2=1.25\,\, {\rm GeV}^2\,,\nonumber\\
&&\hat{\sigma}^{NNLO}_{\textsc{Candia}DY}\otimes\Phi_{MRST_{evol}}=488.22\,,\nonumber\\
&&\hat{\sigma}^{NNLO}_{\textsc{Candia}DY}\otimes\Phi_{MRST_{\textsc{Candia}}}=477.81 \hspace{0.5cm} {\rm from }\,\,\mu_0^2=1 \,\, {\rm GeV}^2\,,\nonumber\\
&&\hat{\sigma}^{NNLO}_{\textsc{Candia}DY}\otimes\Phi_{MRST_{\textsc{Candia}}}=480.27 \hspace{0.5cm} {\rm from }\,\,\mu_0^2=1.25\,\, {\rm GeV}^2\,
\ea
cross sections that differ approximately by $2 \%$. The reduction of the cross section in 
\textsc{Candia} is more remarked compared to MRST and is due to the evolution.


\begin{table}
\begin{center}
\begin{tabular}{|c||c|c|c|}
\hline
\multicolumn{4}{|c|}{$\textrm{d}\sigma_{LO}/\textrm{d}Q$ [pb/GeV]. \textsc{Candia} vs MRST evol. with MRST input, $\mu_0^2=1.25$ GeV$^2$}
\tabularnewline
\hline
$Q ~[\textrm{GeV}]$&
$\sigma_{LO}^{\textsc{Candia}}$&
$\sigma_{LO}^{MRST}$  &
$\delta\sigma_{LO}  $ \tabularnewline
\hline
\hline
$50.0000$&
$5.6629\cdot10^{+0}$&
$5.7110\cdot10^{+0}$&
$8.4230\cdot10^{-3}$
\tabularnewline
\hline
$60.0469$&
$3.4301\cdot10^{+0}$&
$3.4692\cdot10^{+0}$&
$1.1274\cdot10^{-2}$
\tabularnewline
\hline
$70.0938$&
$3.1248\cdot10^{+0}$&
$3.1646\cdot10^{+0}$&
$1.2583\cdot10^{-2}$
\tabularnewline
\hline
$80.1407$&
$6.4675\cdot10^{+0}$&
$6.5540\cdot10^{+0}$&
$1.3191\cdot10^{-2}$
\tabularnewline
\hline
$90.1876$&
$2.4859\cdot10^{+2}$&
$2.5189\cdot10^{+2}$&
$1.3086\cdot10^{-2}$
\tabularnewline
\hline
$91.1876$&
$4.0723\cdot10^{+2}$&
$4.1261\cdot10^{+2}$&
$1.3059\cdot10^{-2}$
\tabularnewline
\hline
$92.1876$&
$2.4767\cdot10^{+2}$&
$2.5094\cdot10^{+2}$&
$1.3033\cdot10^{-2}$
\tabularnewline
\hline
$120.0701$&
$7.6837\cdot10^{-1}$&
$7.7755\cdot10^{-1}$&
$1.1796\cdot10^{-2}$
\tabularnewline
\hline
$146.0938$&
$2.1196\cdot10^{-1}$&
$2.1415\cdot10^{-1}$&
$1.0240\cdot10^{-2}$
\tabularnewline
\hline
$172.1175$&

$9.0345\cdot10^{-2}$&
$9.1149\cdot10^{-2}$&
$8.8207\cdot10^{-3}$
\tabularnewline
\hline
$200.0000$&
$4.4185\cdot10^{-2}$&
$4.4504\cdot10^{-2}$&
$7.1679\cdot10^{-3}$
\tabularnewline
\hline
\end{tabular}
\caption{LO cross section for Drell-Yan obtained by \textsc{Candia} using the MRST input and the evolved MRST pdf's}
\label{comp7}
\end{center}
\end{table}

\begin{table}
\begin{center}
\begin{tabular}{|c||c|c|c|}

\hline
\multicolumn{4}{|c|}{$\textrm{d}\sigma_{NLO}/\textrm{d}Q$ [pb/GeV]. \textsc{Candia} vs MRST evol. with MRST input, $\mu_0^2=1.25$ GeV$^2$}
\tabularnewline
\hline
$Q ~[\textrm{GeV}]$&
$\sigma_{NLO}^{\textsc{Candia}}$&
$\sigma_{NLO}^{MRST}$  &
$\delta\sigma_{NLO}  $ \tabularnewline
\hline
\hline
$50.0000$&
$6.8119\cdot10^{+0}$&
$6.8100\cdot10^{+0}$&
$2.7680\cdot10^{-4}$
\tabularnewline
\hline
$60.0469$&
$4.1552\cdot10^{+0}$&
$4.1521\cdot10^{+0}$&
$7.5793\cdot10^{-4}$
\tabularnewline
\hline
$70.0938$&
$3.8110\cdot10^{+0}$&
$3.8080\cdot10^{+0}$&
$8.1120\cdot10^{-4}$
\tabularnewline
\hline
$80.1407$&
$7.9371\cdot10^{+0}$&
$7.9287\cdot10^{+0}$&
$1.0526\cdot10^{-3}$
\tabularnewline
\hline
$90.1876$&
$3.0657\cdot10^{+2}$&
$3.0615\cdot10^{+2}$&
$1.3656\cdot10^{-3}$
\tabularnewline
\hline
$91.1876$&
$5.0242\cdot10^{+2}$&
$5.0172\cdot10^{+2}$&
$1.3903\cdot10^{-3}$
\tabularnewline
\hline
$92.1876$&
$3.0569\cdot10^{+2}$&
$3.0526\cdot10^{+2}$&
$1.4133\cdot10^{-3}$
\tabularnewline
\hline
$120.0701$&
$9.5677\cdot10^{-1}$&
$9.5496\cdot10^{-1}$&
$1.8964\cdot10^{-3}$
\tabularnewline
\hline
$146.0938$&
$2.6562\cdot10^{-1}$&
$2.6504\cdot10^{-1}$&
$2.1997\cdot10^{-3}$
\tabularnewline
\hline
$172.1175$&
$1.1382\cdot10^{-1}$&
$1.1356\cdot10^{-1}$&
$2.2278\cdot10^{-3}$
\tabularnewline
\hline
$200.0000$&
$5.5940\cdot10^{-2}$&
$5.5778\cdot10^{-2}$&
$2.9044\cdot10^{-3}$
\tabularnewline
\hline
\end{tabular}
\caption{NLO cross section for Drell-Yan obtained by \textsc{Candia} using the MRST input and the evolved MRST pdf's}
\label{comp8}
\end{center}
\end{table}

\begin{table}
\begin{center}
\begin{tabular}{|c||c|c|c|}
\hline
\multicolumn{4}{|c|}{$\textrm{d}\sigma_{NNLO}/\textrm{d}Q$ [pb/GeV]. \textsc{Candia} vs MRST evolution with MRST input, $\mu_0^2=1.25$ GeV$^2$}
\tabularnewline
\hline
$Q ~[\textrm{GeV}]$&
$\sigma_{NNLO}^{\textsc{Candia}}$&
$\sigma_{NNLO}^{MRST}$  &
$\delta\sigma_{NNLO}  $ \tabularnewline
\hline
\hline
$50.0000$&
$6.4935\cdot10^{+0}$&
$6.6707\cdot10^{+0}$&
$2.6560\cdot10^{-2}$
\tabularnewline
\hline
$60.0469$&
$3.9997\cdot10^{+0}$&
$4.0961\cdot10^{+0}$&
$2.3534\cdot10^{-2}$
\tabularnewline
\hline
$70.0938$&
$3.6962\cdot10^{+0}$&
$3.7743\cdot10^{+0}$&
$2.0678\cdot10^{-2}$
\tabularnewline
\hline
$80.1407$&
$7.6755\cdot10^{+0}$&
$7.8198\cdot10^{+0}$&
$1.8455\cdot10^{-2}$
\tabularnewline
\hline
$90.1876$&
$2.9325\cdot10^{+2}$&
$2.9827\cdot10^{+2}$&
$1.6834\cdot10^{-2}$
\tabularnewline
\hline
$91.1876$&
$4.8006\cdot10^{+2}$&
$4.8822\cdot10^{+2}$&
$1.6702\cdot10^{-2}$
\tabularnewline
\hline
$92.1876$&
$2.9179\cdot10^{+2}$&
$2.9671\cdot10^{+2}$&
$1.6575\cdot10^{-2}$
\tabularnewline
\hline
$120.0701$&
$9.0411\cdot10^{-1}$&
$9.1687\cdot10^{-1}$&
$1.3918\cdot10^{-2}$
\tabularnewline
\hline
$146.0938$&
$2.5267\cdot10^{-1}$&
$2.5567\cdot10^{-1}$&
$1.1714\cdot10^{-2}$
\tabularnewline
\hline
$172.1175$&
$1.0938\cdot10^{-1}$&
$1.1049\cdot10^{-1}$&
$1.0028\cdot10^{-2}$
\tabularnewline
\hline
$200.0000$&
$5.4431\cdot10^{-2}$&
$5.4876\cdot10^{-2}$&
$8.1092\cdot10^{-3}$
\tabularnewline
\hline
\end{tabular}
\caption{NNLO cross section for Drell-Yan obtained by \textsc{Candia} using the MRST input and the evolved MRST pdf's}
\label{comp9}
\end{center}
\end{table}

\begin{figure}
\subfigure[$K=\sigma_{NNLO}/\sigma_{NLO}$]{\includegraphics[%
  width=6cm,
  angle=-90]{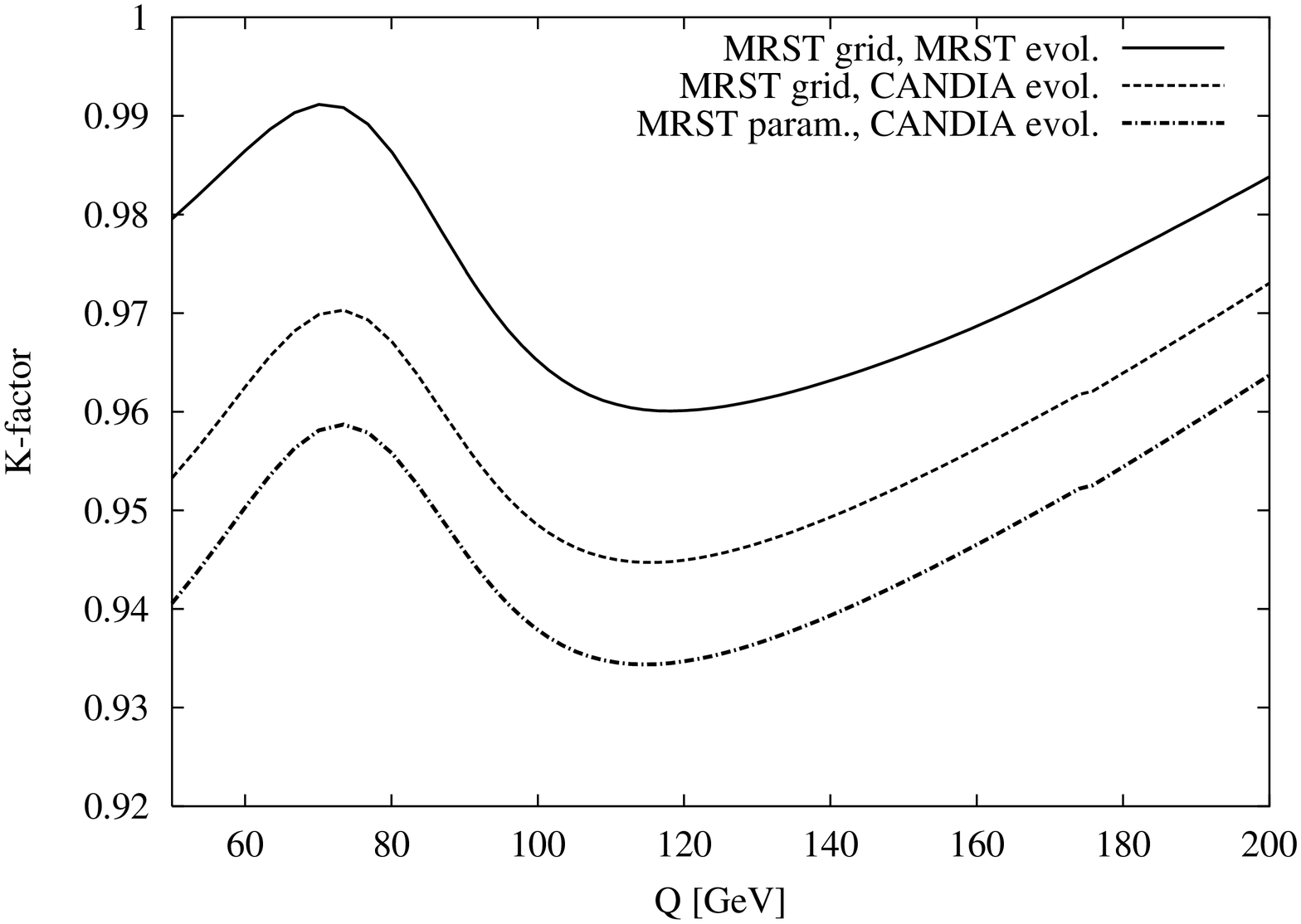}}
\subfigure[$K_1=\sigma_{NLO}/\sigma_{LO}$  and $K_2=\sigma_{NNLO}/\sigma_{LO}$]{\includegraphics[%
  width=6cm,
  angle=-90]{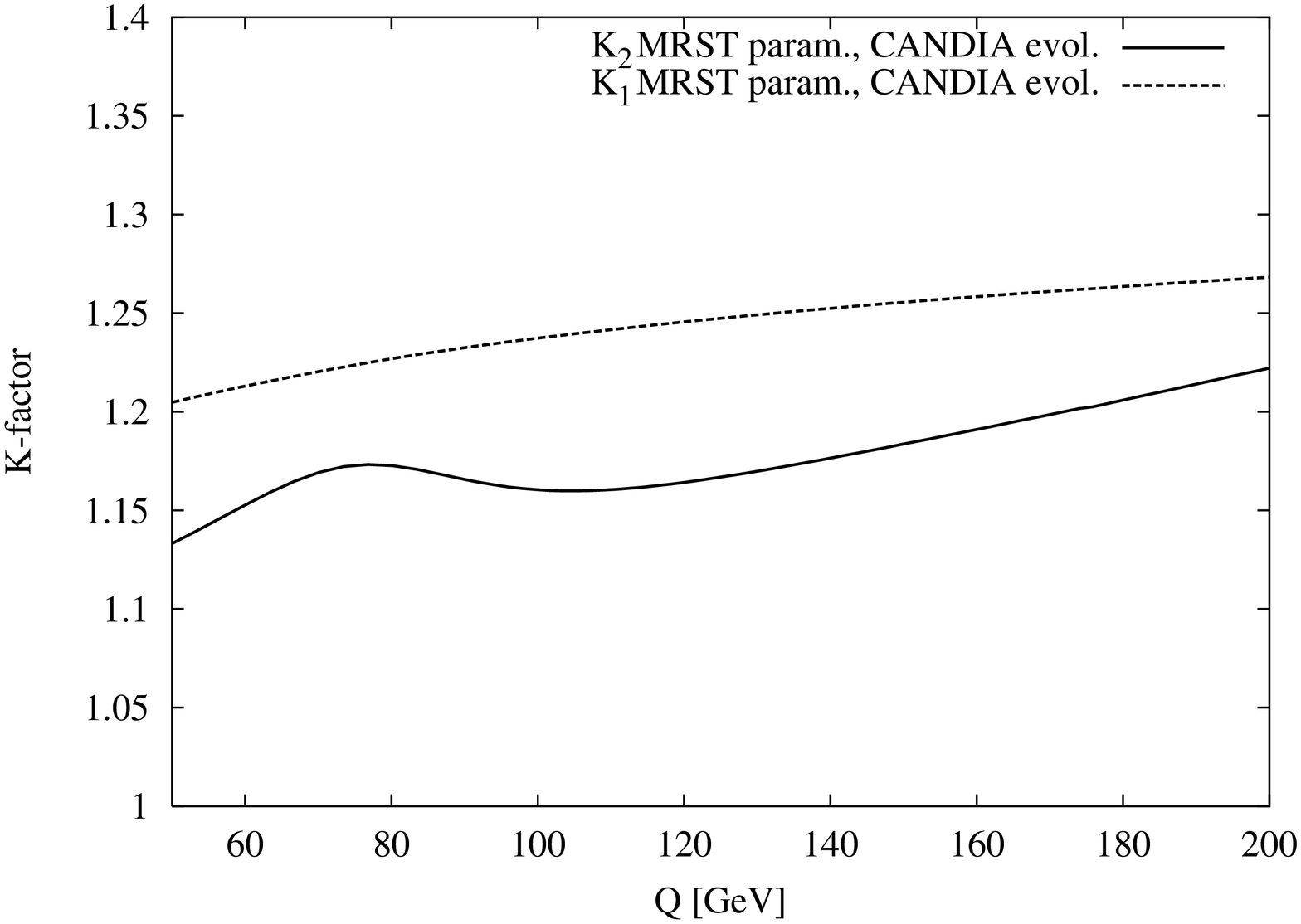}}
\subfigure[$K_1=\sigma_{NLO}/\sigma_{LO}$  and $K_2=\sigma_{NNLO}/\sigma_{LO}$]{\includegraphics[%
  width=6cm,
  angle=-90]{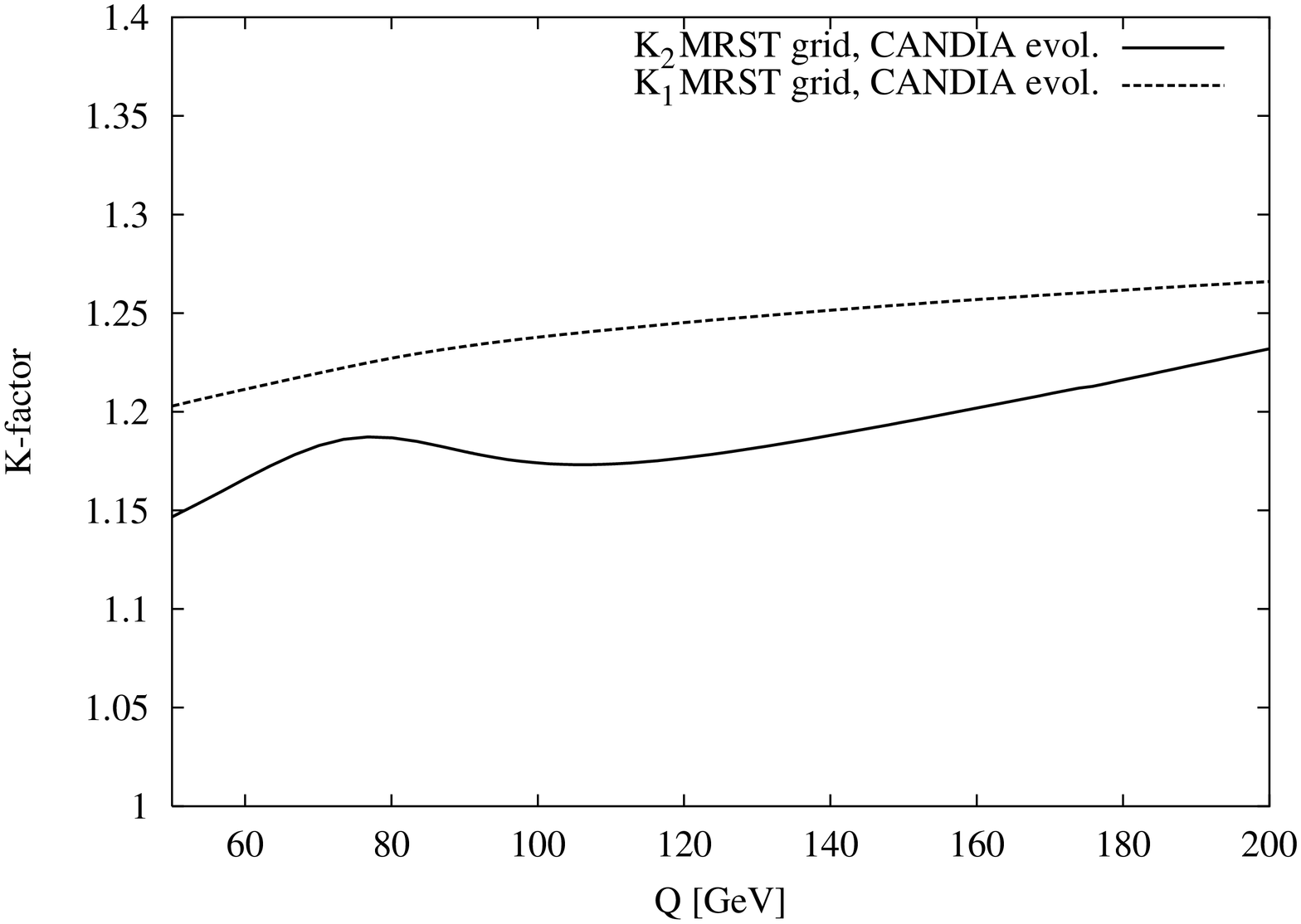}}
\subfigure[$K_1=\sigma_{NLO}/\sigma_{LO}$ and $K_2=\sigma_{NNLO}/\sigma_{LO}$]{\includegraphics[%
  width=6cm,
  angle=-90]{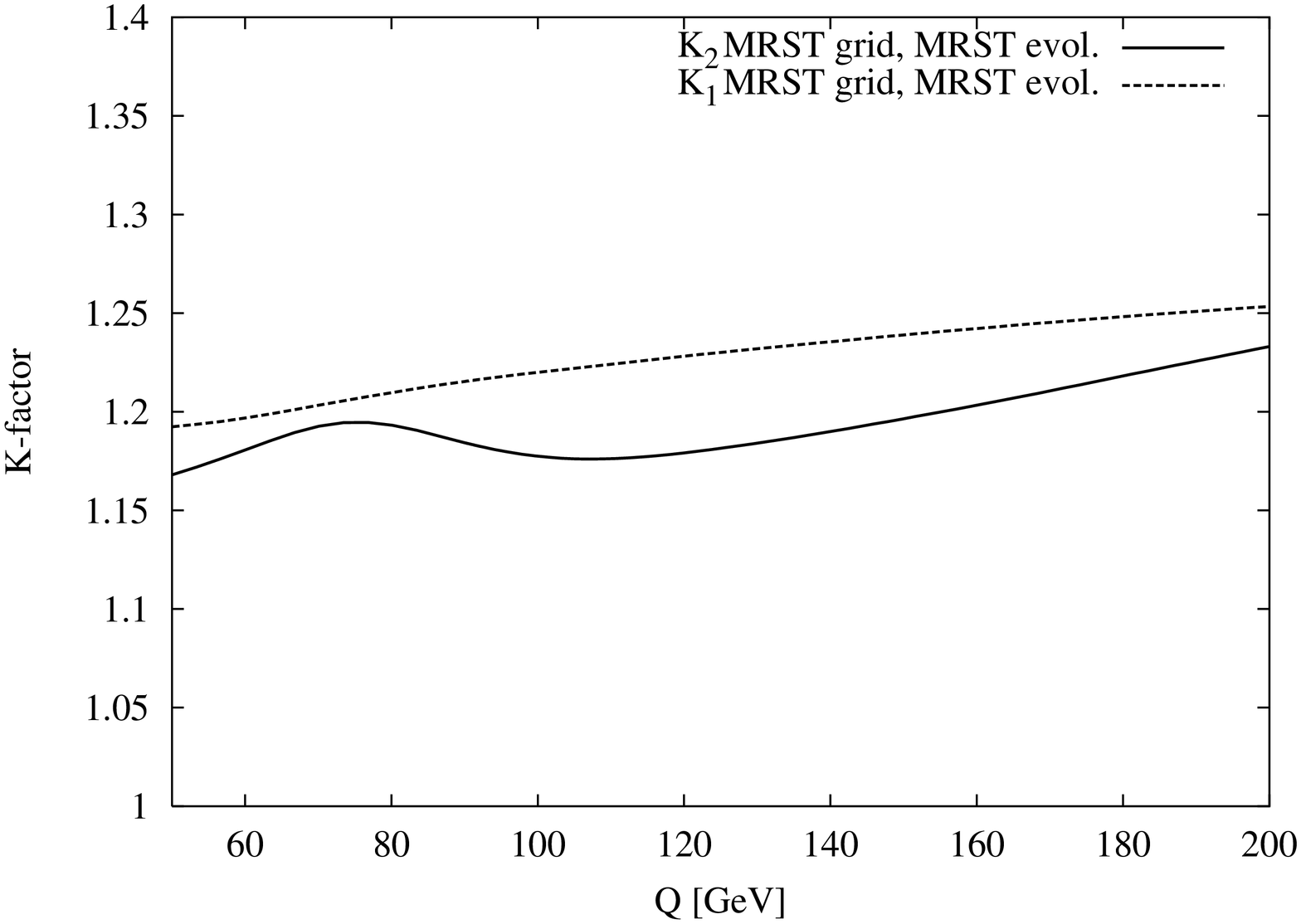}}
\caption{Various $K$-factors obtained with the evolution performed by \textsc{Candia} and MRST.}
\label{Kfact}
\end{figure}


\subsection{The renormalization/factorization scale dependence of the cross section}

An interesting aspect of the prediction of the QCD observables is their factorization
and renormalization scale dependence. We will denote by $\mu_F$ and
$\mu_R$ the two scales. The dependence is important and appears 
both in the hard scatterings and in the evolved pdf's, using the modified NNLO kernels defined 
above. The optimal choices for these scales are identified in the region of stability of the cross section,
usually a small plateau in a multi-parameter space, which can be searched numerically. 
In the case of the Higgs, for instance, a rather general analysis of
the structure of these surfaces for specific observables (such as
the total cross section for Higgs production and the corresponding $K$-factors)
through NNLO has been given. There one can show, but the result is quite general, that the 
concavity of the bidimensional surfaces describing the cross sections, plotted in terms of the two independent scales, changes sign as we move from leading to next-to-leading order \cite{CCG2}. 
We show in Figs.~\ref{Cross1} and \ref{Cross2} global plots of the DY cross section near the Z peak 
for the two models Alekhin and MRST, evolved by the same authors, and zooms of the peak region (Fig.\ref{Cross2}), in which the LO, NLO and NNLO contributions are resolved in great detail. Here we have set the factorization scale to be $Q$ $(\mu_F=Q)$. In two following plots, Fig.~\ref{Cross3} and \ref{Cross4}, 
we show instead the variation of the same cross section using an evolution provided by \textsc{Candia} at LO, NLO and NNLO of the MRST input from the grids ($\mu_0^2=1.25$ GeV$^2$)and we have varied $\mu_F$ and $\mu_R$.
As we have already mentioned, our analysis includes all the $\mu_R$ dependence
(see Tabs. \ref{tab1_muR_muF},\ref{tab2_muR_muF},\ref{tab3_muR_muF},\ref{tab4_muR_muF},\ref{tab5_muR_muF}),
coming both from the pdf's and from the hard scatterings. The first is usually not reported
in the standard parameterizations such as MRST and Alekhin.
The variation has been performed setting, for each (fixed) value of $Q$, $\mu_F=Q$ and studying
the variation of the renormalization scale $\mu_R$ in the ratio $k_F\equiv$ $\mu_R^2/
\mu_F^2$,
which has been taken to vary between 1/2 and 2. The decreased dependence of the result
on the spurious scales of the process as we move toward the NNLO predictions from the LO ones are quite visible. 
This is particularly easy to see from Fig.~\ref{Cross4}. 
The options $\mu_R=\mu_F$ and $Q \neq\mu_F$ are shown both for
Alekhin's and the MRST inputs (Figs.~\ref{Cross5} and \ref{Cross6}) where
a zoom of the region above the Z peak and of the tail of the cross
section are presented.  
The region covered is quite small (100-110 GeV) so to allow to discern between the various results. 
The bands of variations of the LO, NLO and NNLO results can be identified by a close look at these figures. 
One can see immediately the reduced sizes of these bands as we increase the perturbative 
order of accuracy. The same bands are shown right on the peak of the Z in Fig.~\ref{Cross6}. One can immediately notice that the the NNLO variations take place right inside the NLO error band for the Alekhin model (Fig.~\ref{Cross6} (a)),
while they overlap at the edge in the MRST model (Fig.~\ref{Cross6} (b)). 
Regarding the precise size of these variations, 
these can be inferred from the corresponding tables. The range explored in our 
analysis $(1/2 <k_F < 2)$ is somehow smaller than that explored in \cite{Anastasiou}, 
but includes the entire dependence on the renormalization scale of the 
pdf's. Being the evolution rather important in the determination of the NNLO total cross section, it is clear that also 
the $\mu_R$ dependence on the evolution is not negligible. 
The two cases $\mu_R < \mu_F$ and $\mu_R > \mu_F$  are 
characterized by substantially different excursions in range. 
In the first case the variations, at LO, are from 25 $\% $ at 50 GeV 
down to 10 $\%$ for $Q=200$ GeV, while for $k_F=2$
 they are more moderate (from 17 $\%$ down to 8 $\%$). At NLO 
the excursions are approximately from 11$\%$ down to $6 \%$ in the same range of Q, 
for both cases of $k_F$. The variations at NNLO can be found in 
\ref{tab3_muR_muF}, and are in the range of 1-3 $\% $. We have also 
shown in tab.~\ref{tab4_muR_muF} and \ref{tab5_muR_muF} results for the scale 
dependence when we remove $\mu_R$ in the pdf's, by 
equating $\mu_R$ to $\mu_F$ and keep them separate only in the 
hard scatterings. The range of variation are sensibly reduced especially at 
lower values of $Q$, with a drastic reduction especially around the peak. The reduction 
in the variation is by a factor of 10 less: from about $10 \%$ down to less than $1 \%$.
On the peak the NNLO variations are between 0.1 and 0.03 $\%$. 
It is clear from this results 
that the $\mu_R$ scale dependence coming from the evolution is pretty relevant and, in a complete analysis of the stability of the NNLO corrections can't be forgotten.

\begin{table}
\begin{footnotesize}
\begin{center}
\begin{tabular}{|c||c|c|c|c|c|}
\hline
\multicolumn{6}{|c|}{$\textrm{d}\sigma_{LO}(Q,\mu_F,\mu_R)/\textrm{d}Q$ [pb/GeV]. \textsc{Candia} evolution with MRST input, $\mu_0^2=1.25$ GeV$^2$}
\tabularnewline
\hline
$Q ~[\textrm{GeV}]$       &
$\sigma^{LO}(Q)~ k_F=2$ &
$\sigma^{LO}(Q)~ k_F=1$   &
$\sigma^{LO}(Q)~ k_F=1/2$   &
$\delta \sigma_{k_F=1/2} \% $ &
$\delta \sigma_{k_F=2}\% $ \tabularnewline
\hline
\hline
$50.0000$&
$6.0789\cdot10^{+0}$&
$5.6629\cdot10^{+0}$&
$7.0877\cdot10^{+0}$&
$2.5159\cdot10^{+1}$&
$1.7291\cdot10^{+1}$
\tabularnewline
\hline
$60.0469$&
$2.8718\cdot10^{+0}$&
$3.4301\cdot10^{+0}$&
$4.2241\cdot10^{+0}$&
$2.3148\cdot10^{+1}$&
$1.6276\cdot10^{+1}$
\tabularnewline
\hline
$70.0938$&
$2.6438\cdot10^{+0}$&
$3.1248\cdot10^{+0}$&
$3.7952\cdot10^{+0}$&
$2.1454\cdot10^{+1}$&
$1.5394\cdot10^{+1}$
\tabularnewline
\hline
$80.1407$&
$5.5230\cdot10^{+0}$&
$6.4675\cdot10^{+0}$&
$7.7598\cdot10^{+0}$&
$1.9981\cdot10^{+1}$&
$1.4604\cdot10^{+1}$
\tabularnewline
\hline
$90.1876$&
$2.1411\cdot10^{+2}$&
$2.4859\cdot10^{+2}$&
$2.9498\cdot10^{+2}$&
$1.8662\cdot10^{+1}$&
$1.3869\cdot10^{+1}$
\tabularnewline
\hline
$91.1876$&
$3.5104\cdot10^{+2}$&
$4.0723\cdot10^{+2}$&
$4.8272\cdot10^{+2}$&
$1.8538\cdot10^{+1}$&
$1.3798\cdot10^{+1}$
\tabularnewline
\hline
$120.0701$&
$6.7639\cdot10^{-1}$&
$7.6837\cdot10^{-1}$&
$8.8706\cdot10^{-1}$&
$1.5446\cdot10^{+1}$&
$1.1972\cdot10^{+1}$
\tabularnewline
\hline
$146.0938$&
$1.8945\cdot10^{-1}$&
$2.1196\cdot10^{-1}$&
$2.4008\cdot10^{-1}$&
$1.3267\cdot10^{+1}$&
$1.0620\cdot10^{+1}$
\tabularnewline
\hline
$172.1175$&
$8.1799\cdot10^{-2}$&
$9.0345\cdot10^{-2}$&
$1.0070\cdot10^{-1}$&
$1.1463\cdot10^{+1}$&
$9.4593\cdot10^{+0}$
\tabularnewline
\hline
$200.0000$&
$4.0486\cdot10^{-2}$&
$4.4185\cdot10^{-2}$&
$4.8524\cdot10^{-2}$&
$9.8201\cdot10^{+0}$&
$8.3716\cdot10^{+0}$
\tabularnewline
\hline
\end{tabular}
\caption{Study of the variation of the LO cross sections with respect to $k_F=\mu_R^2/\mu_F^2$.
Here we choose $Q=\mu_F$ and the $\mu_R^2/\mu_F^2$ variation is also included in the pdf's evolved
with \textsc{Candia}.}
\label{tab1_muR_muF}
\end{center}
\end{footnotesize}
\end{table}

\begin{table}
\begin{footnotesize}
\begin{center}
\begin{tabular}{|c||c|c|c|c|c|}
\hline
\multicolumn{6}{|c|}{$\textrm{d}\sigma_{NLO}(Q,\mu_F,\mu_R)/\textrm{d}Q$ [pb/GeV]. \textsc{Candia} evol. with MRST input, $\mu_0^2=1.25$ GeV$^2$}
\tabularnewline
\hline
$Q ~[\textrm{GeV}]$       &
$\sigma^{NLO}(Q)~ k_F=2$ &
$\sigma^{NLO}(Q)~ k_F=1$   &
$\sigma^{NLO}(Q)~ k_F=1/2$   &
$\delta \sigma_{k_F=1/2} \% $ &
$\delta \sigma_{k_F=2}\% $ \tabularnewline
\hline
\hline
$50.0000$&
$6.0789\cdot10^{+0}$&
$6.8121\cdot10^{+0}$&
$7.5694\cdot10^{+0}$&
$1.1116\cdot10^{+1}$&
$1.0763\cdot10^{+1}$
\tabularnewline
\hline
$60.0469$&
$3.7343\cdot10^{+0}$&
$4.1554\cdot10^{+0}$&
$4.5906\cdot10^{+0}$&
$1.0473\cdot10^{+1}$&
$1.0134\cdot10^{+1}$
\tabularnewline
\hline
$70.0938$&
$3.4443\cdot10^{+0}$&
$3.8112\cdot10^{+0}$&
$4.1920\cdot10^{+0}$&
$9.9929\cdot10^{+0}$&
$9.6262\cdot10^{+0}$
\tabularnewline
\hline
$80.1407$&
$7.2077\cdot10^{+0}$&
$7.9374\cdot10^{+0}$&
$8.6992\cdot10^{+0}$&
$9.5986\cdot10^{+0}$&
$9.1922\cdot10^{+0}$
\tabularnewline
\hline
$90.1876$&
$2.7965\cdot10^{+2}$&
$3.0658\cdot10^{+2}$&
$3.3480\cdot10^{+2}$&
$9.2047\cdot10^{+0}$&
$8.7856\cdot10^{+0}$
\tabularnewline
\hline
$91.1876$&
$4.5849\cdot10^{+2}$&
$5.0243\cdot10^{+2}$&
$5.4848\cdot10^{+2}$&
$9.1649\cdot10^{+0}$&
$8.7461\cdot10^{+0}$
\tabularnewline
\hline
$120.0701$&
$8.8285\cdot10^{-1}$&
$9.5681\cdot10^{-1}$&
$1.0345\cdot10^{+0}$&
$8.1197\cdot10^{+0}$&
$7.7303\cdot10^{+0}$
\tabularnewline
\hline
$146.0938$&
$2.4702\cdot10^{-1}$&
$2.6563\cdot10^{-1}$&
$2.8526\cdot10^{-1}$&
$7.3907\cdot10^{+0}$&
$7.0048\cdot10^{+0}$
\tabularnewline
\hline
$172.1175$&
$1.0654\cdot10^{-1}$&
$1.1382\cdot10^{-1}$&
$1.2155\cdot10^{-1}$&
$6.7940\cdot10^{+0}$&
$6.3978\cdot10^{+0}$
\tabularnewline
\hline
$200.0000$&
$5.2673\cdot10^{-2}$&
$5.5942\cdot10^{-2}$&
$5.9441\cdot10^{-2}$&
$6.2547\cdot10^{+0}$&
$5.8436\cdot10^{+0}$
\tabularnewline
\hline
\end{tabular}
\caption{Study of the variation of the NLO cross sections with respect to $k_F=\mu_R^2/\mu_F^2$.
Here we choose $Q=\mu_F$ and the $\mu_R^2/\mu_F^2$ variation is also included in the pdf's evolved
with \textsc{Candia}.}
\label{tab2_muR_muF}
\end{center}
\end{footnotesize}
\end{table}

\begin{table}
\begin{footnotesize}
\begin{center}
\begin{tabular}{|c||c|c|c|c|c|}
\hline
\multicolumn{6}{|c|}{$\textrm{d}\sigma_{NNLO}(Q,\mu_F,\mu_R)/\textrm{d}Q$ [pb/GeV]. \textsc{Candia} evolution with MRST input, $\mu_0^2=1.25$ GeV$^2$}
\tabularnewline
\hline
$Q ~[\textrm{GeV}]$       &
$\sigma^{NNLO}(Q)~ k_F=2$ &
$\sigma^{NNLO}(Q)~ k_F=1$   &
$\sigma^{NNLO}(Q)~ k_F=1/2$   &
$\delta \sigma_{k_F=1/2} \% $ &
$\delta \sigma_{k_F=2}\% $ \tabularnewline
\hline
\hline
$50.0000$&
$6.2855\cdot10^{+0}$&
$6.4940\cdot10^{+0}$&
$6.5465\cdot10^{+0}$&
$8.0790\cdot10^{-1}$&
$3.2107\cdot10^{+0}$
\tabularnewline
\hline
$60.0469$&
$3.8626\cdot10^{+0}$&
$3.9989\cdot10^{+0}$&
$4.0503\cdot10^{+0}$&
$1.2850\cdot10^{+0}$&
$3.4090\cdot10^{+0}$
\tabularnewline
\hline
$70.0938$&
$3.5635\cdot10^{+0}$&
$3.6948\cdot10^{+0}$&
$3.7557\cdot10^{+0}$&
$1.6481\cdot10^{+0}$&
$3.5528\cdot10^{+0}$
\tabularnewline
\hline
$80.1407$&
$7.4312\cdot10^{+0}$&
$7.6740\cdot10^{+0}$&
$7.7729\cdot10^{+0}$&
$1.2890\cdot10^{+0}$&
$3.1640\cdot10^{+0}$
\tabularnewline
\hline
$90.1876$&
$2.8672\cdot10^{+2}$&
$2.9335\cdot10^{+2}$&
$2.9409\cdot10^{+2}$&
$2.5092\cdot10^{-1}$&
$2.2615\cdot10^{+0}$
\tabularnewline
\hline
$91.1876$&
$4.6986\cdot10^{+2}$&
$4.8027\cdot10^{+2}$&
$4.8094\cdot10^{+2}$&
$1.4037\cdot10^{-1}$&
$2.1663\cdot10^{+0}$
\tabularnewline
\hline
$120.0701$&
$9.0247\cdot10^{-1}$&
$9.0552\cdot10^{-1}$&
$8.8841\cdot10^{-1}$&
$-1.8901\cdot10^{+0}$&
$3.3748\cdot10^{-1}$
\tabularnewline
\hline
$146.0938$&
$2.5416\cdot10^{-1}$&
$2.5318\cdot10^{-1}$&
$2.4664\cdot10^{-1}$&
$-2.5808\cdot10^{+0}$&
$-3.8826\cdot10^{-1}$
\tabularnewline
\hline
$172.1175$&
$1.1060\cdot10^{-1}$&
$1.0963\cdot10^{-1}$&
$1.0633\cdot10^{-1}$&
$-3.0129\cdot10^{+0}$&
$-8.8480\cdot10^{-1}$
\tabularnewline
\hline
$200.0000$&
$5.5290\cdot10^{-2}$&
$5.4572\cdot10^{-2}$&
$5.2727\cdot10^{-2}$&
$-3.3809\cdot10^{+0}$&
$-1.3157\cdot10^{+0}$
\tabularnewline
\hline
\end{tabular}
\caption{Study of the variation of the NNLO cross sections with respect to $k_F=\mu_R^2/\mu_F^2$.
Here we choose $Q=\mu_F$ and the $\mu_R^2/\mu_F^2$ variation is also included in the pdf's evolved
with \textsc{Candia}.}
\label{tab3_muR_muF}
\end{center}
\end{footnotesize}
\end{table}

\begin{table}
\begin{footnotesize}
\begin{center}
\begin{tabular}{|c||c|c|c|c|c|}
\hline
\multicolumn{6}{|c|}{$\textrm{d}\sigma_{NLO}(Q,\mu_F,\mu_R)/\textrm{d}Q$ [pb/GeV]. \textsc{Candia} evolution with MRST input, $\mu_0^2=1.25$ GeV$^2$ $\hat{\sigma}(k_F)\otimes\Phi(\mu_F)$}
\tabularnewline
\hline
$Q ~[\textrm{GeV}]$       &
$\sigma^{NLO}(Q)~ k_F=2$ &
$\sigma^{NLO}(Q)~ k_F=1$   &
$\sigma^{NLO}(Q)~ k_F=1/2$   &
$\delta \sigma_{k_F=1/2} \% $ &
$\delta \sigma_{k_F=2}\% $ \tabularnewline
\hline
\hline
$50.0000$&
$6.7636\cdot10^{+0}$&
$6.8121\cdot10^{+0}$&
$6.8667\cdot10^{+0}$&
$8.0201\cdot10^{-1}$&
$7.1156\cdot10^{-1}$
\tabularnewline
\hline
$60.0469$&
$4.1271\cdot10^{+0}$&
$4.1554\cdot10^{+0}$&
$4.1871\cdot10^{+0}$&
$7.6402\cdot10^{-1}$&
$6.8044\cdot10^{-1}$
\tabularnewline
\hline
$70.0938$&
$3.7863\cdot10^{+0}$&
$3.8112\cdot10^{+0}$&
$3.8390\cdot10^{+0}$&
$7.3124\cdot10^{-1}$&
$6.5324\cdot10^{-1}$
\tabularnewline
\hline
$80.1407$&
$7.8873\cdot10^{+0}$&
$7.9374\cdot10^{+0}$&
$7.9933\cdot10^{+0}$&
$7.0434\cdot10^{-1}$&
$6.3080\cdot10^{-1}$
\tabularnewline
\hline
$90.1876$&
$3.0469\cdot10^{+2}$&
$3.0658\cdot10^{+2}$&
$3.0869\cdot10^{+2}$&
$6.8790\cdot10^{-1}$&
$6.1737\cdot10^{-1}$
\tabularnewline
\hline
$91.1876$&
$4.9934\cdot10^{+2}$&
$5.0243\cdot10^{+2}$&
$5.0589\cdot10^{+2}$&
$6.8677\cdot10^{-1}$&
$6.1649\cdot10^{-1}$
\tabularnewline
\hline
$120.0701$&
$9.5104\cdot10^{-1}$&
$9.5681\cdot10^{-1}$&
$9.6321\cdot10^{-1}$&
$6.6868\cdot10^{-1}$&
$6.0294\cdot10^{-1}$
\tabularnewline
\hline
$146.0938$&
$2.6405\cdot10^{-1}$&
$2.6563\cdot10^{-1}$&
$2.6738\cdot10^{-1}$&
$6.5806\cdot10^{-1}$&
$5.9519\cdot10^{-1}$
\tabularnewline
\hline
$172.1175$&
$1.1315\cdot10^{-1}$&
$1.1382\cdot10^{-1}$&
$1.1456\cdot10^{-1}$&
$6.5014\cdot10^{-1}$&
$5.8952\cdot10^{-1}$
\tabularnewline
\hline
$200.0000$&
$5.5615\cdot10^{-2}$&
$5.5942\cdot10^{-2}$&
$5.6303\cdot10^{-2}$&
$6.4531\cdot10^{-1}$&
$5.8453\cdot10^{-1}$
\tabularnewline
\hline
\end{tabular}
\caption{Study of the variation of the NLO cross sections with respect to $k_F=\mu_R^2/\mu_F^2$.
Here we choose $Q=\mu_F$ and the $\mu_R^2/\mu_F^2$ variation is only included in the hard scattering
piece.}
\label{tab4_muR_muF}
\end{center}
\end{footnotesize}
\end{table}

\begin{table}
\begin{footnotesize}
\begin{center}
\begin{tabular}{|c||c|c|c|c|c|}
\hline
\multicolumn{6}{|c|}{$\textrm{d}\sigma_{NNLO}(Q,\mu_F,\mu_R)/\textrm{d}Q$ [pb/GeV]. \textsc{Candia} evolution with MRST input, $\mu_0^2=1.25$ GeV$^2$ $\hat{\sigma}(k_F)\otimes\Phi(\mu_F)$}
\tabularnewline
\hline
$Q ~[\textrm{GeV}]$       &
$\sigma^{NNLO}(Q)~ k_F=2$ &
$\sigma^{NNLO}(Q)~ k_F=1$   &
$\sigma^{NNLO}(Q)~ k_F=1/2$   &
$\delta \sigma_{k_F=1/2} \% $ &
$\delta \sigma_{k_F=2}\% $ \tabularnewline
\hline
\hline
$50.0000$&
$6.4534\cdot10^{+0}$&
$6.4940\cdot10^{+0}$&
$6.5389\cdot10^{+0}$&
$6.9133\cdot10^{-1}$&
$6.2544\cdot10^{-1}$
\tabularnewline
\hline
$60.0469$&
$3.9563\cdot10^{+0}$&
$3.9989\cdot10^{+0}$&
$4.0484\cdot10^{+0}$&
$1.2371\cdot10^{+0}$&
$1.0648\cdot10^{+0}$
\tabularnewline
\hline
$70.0938$&
$3.6442\cdot10^{+0}$&
$3.6948\cdot10^{+0}$&
$3.7544\cdot10^{+0}$&
$1.6123\cdot10^{+0}$&
$1.3704\cdot10^{+0}$
\tabularnewline
\hline
$80.1407$&
$7.5963\cdot10^{+0}$&
$7.6740\cdot10^{+0}$&
$7.7638\cdot10^{+0}$&
$1.1711\cdot10^{+0}$&
$1.0122\cdot10^{+0}$
\tabularnewline
\hline
$90.1876$&
$2.9315\cdot10^{+2}$&
$2.9335\cdot10^{+2}$&
$2.9340\cdot10^{+2}$&
$1.6344\cdot10^{-2}$&
$6.7867\cdot10^{-2}$
\tabularnewline
\hline
$91.1876$&
$4.8042\cdot10^{+2}$&
$4.8027\cdot10^{+2}$&
$4.7977\cdot10^{+2}$&
$-1.0389\cdot10^{-1}$&
$-3.0745\cdot10^{-2}$
\tabularnewline
\hline
$120.0701$&
$9.2166\cdot10^{-1}$&
$9.0552\cdot10^{-1}$&
$8.8539\cdot10^{-1}$&
$-2.2235\cdot10^{+0}$&
$-1.7820\cdot10^{+0}$
\tabularnewline
\hline
$146.0938$&
$2.5909\cdot10^{-1}$&
$2.5318\cdot10^{-1}$&
$2.4588\cdot10^{-1}$&
$-2.8814\cdot10^{+0}$&
$-2.3367\cdot10^{+0}$
\tabularnewline
\hline
$172.1175$&
$1.1256\cdot10^{-1}$&
$1.0963\cdot10^{-1}$&
$1.0604\cdot10^{-1}$&
$-3.2701\cdot10^{+0}$&
$-2.6699\cdot10^{+0}$
\tabularnewline
\hline
$200.0000$&
$5.6183\cdot10^{-2}$&
$5.4572\cdot10^{-2}$&
$5.2608\cdot10^{-2}$&
$-3.5989\cdot10^{+0}$&
$-2.9521\cdot10^{+0}$
\tabularnewline
\hline
\end{tabular}
\caption{Study of the variation of the NNLO cross sections with respect to $k_F=\mu_R^2/\mu_F^2$.
Here we choose $Q=\mu_F$ and the $\mu_R^2/\mu_F^2$ variation is only included in the hard scattering
piece.}
\label{tab5_muR_muF}
\end{center}
\end{footnotesize}
\end{table}

\subsection{The K factors}

We have summarized in Fig.~\ref{Kfact} four plots of the behavior of the 3 $K$-factors
$K=\sigma_{NNLO}/\sigma_{NLO}$, $K_1=\sigma_{NLO}/\sigma_{LO}$ and $K_2=\sigma_{NNLO}/\sigma_{LO}$
obtained using \textsc{Candia} and the MRST evolution.
These are shown as a function of $Q$, and evaluated at the center of mass energy of $\sqrt{S}=14$
TeV. The dependence of the results on
the evolution is significant. In fact, from Fig.~\ref{Kfact} it is evident that while the shapes of the
plots of the $K$-factors are similar, there are
variations of the order $2\%$, in the results using the two different evolutions.
Both in the evolution performed with \textsc{Candia} and in the MRST evolution we use the same
MRST input, choosing the initial scale $\mu_0^2=1.25$ GeV$^2$, and the same treatment of the heavy flavors. On the Z resonance we get
\ba
&&K(M_Z)=(\hat{\sigma}_{NNLO}\otimes\Phi^{NNLO}_{MRST})/(\hat{\sigma}_{NLO}\otimes\Phi^{NLO}_{MRST})=0.97\nonumber\\
&&K(M_Z)=(\hat{\sigma}_{NNLO}\otimes\Phi^{NNLO}_{\textsc{Candia}})/(\hat{\sigma}_{NLO}\otimes\Phi^{NLO}_{\textsc{Candia}})=0.95 \nonumber \\
&& K(M_Z)=(\hat{\sigma}_{NNLO}\otimes\Phi^{NNLO}_{Alekhin})/(\hat{\sigma}_{NLO}\otimes\Phi^{NLO}_{Alekhin})=0.98
\ea
which corresponds to a reduction by $2.7 \%$ of the NNLO cross section compared to the NLO result, 
(MRST evolution) and larger for the \textsc{Candia} evolution, 
$4.4 \% $, while for Alekhin is $1.5 \%$. From the analysis of the errors on the pdf's to NNLO, for instance for the Alekhin's set, the differences among these determinations are still 
compatible, being the variations on the $K$-factors of the order of $4 \%$. 
We will get back to this point in the next sections.
Similar $K$-factors can be introduced to study the variations from LO to NNLO. We obtain
\ba
&&K_2(M_Z)=(\hat{\sigma}_{NNLO}\otimes\Phi^{NNLO}_{MRST})/(\hat{\sigma}_{LO}\otimes\Phi^{LO}_{MRST})=1.18\nonumber\\
&&K_1(M_Z)=(\hat{\sigma}_{NLO}\otimes\Phi^{NLO}_{MRST})/(\hat{\sigma}_{LO}\otimes\Phi^{LO}_{MRST})=1.21\nonumber\\
&&K_2(M_Z)=(\hat{\sigma}_{NNLO}\otimes\Phi^{NNLO}_{\textsc{Candia}})/(\hat{\sigma}_{LO}\otimes\Phi^{LO}_{\textsc{Candia}})=1.17\nonumber\\
&&K_1(M_Z)=(\hat{\sigma}_{NLO}\otimes\Phi^{NLO}_{\textsc{Candia}})/(\hat{\sigma}_{LO}\otimes\Phi^{LO}_{\textsc{Candia}})=1.23\,,\nonumber \\
&&K_1(M_Z)=(\hat{\sigma}_{NLO}\otimes\Phi^{NLO}_{Alekhin})/(\hat{\sigma}_{LO}\otimes\Phi^{LO}_{Alekhin})=1.23\,,\nonumber \\
&&K_2(M_Z)=(\hat{\sigma}_{NNLO}\otimes\Phi^{NLO}_{Alekhin})/(\hat{\sigma}_{LO}\otimes\Phi^{LO}_{Alekhin})=1.21.\,\nonumber
\ea
corresponding to a growth around $17-23 \%$.

\subsection{The rapidity distributions}

Another cross section of relevance 
is the calculation of the rapidity distributions of the lepton pair in the final state 
at the resonance of the Z at NNLO.
Since the number of events expected from Drell-Yan cross at LHC is large,
the study of these distributions will be very important for partonometry.
As we have already mentioned, the analysis presented here is going to be rather short, 
and we hope to return to this point in a separate work. We perform a numerical calculation of the differential
cross section interfacing  \textsc{Vrap} \cite{Anastasiou}, which computes the hard scatterings, 
with \textsc{Candia}. At this point we recall that the rapidity of the vector boson Z is defined as
\begin{eqnarray}
Y=\frac{1}{2}\log\left(\frac{E+p_z}{E-p_z}\right),
\end{eqnarray}
where $E$ and $p_z$ are respectively the energy and the longitudinal
momentum of Z in the center of mass frame of the colliding hadrons.
Integrating over this variable one obtains the total cross section as
\begin{eqnarray}
&&\sigma^{Z}=\int_{(1/2)\ln\tau}^{(1/2)\ln 1/\tau}dY \frac{d\sigma}{dY}
\eeqa
where
\beqa
&&\frac{d\sigma^{Z}}{dY}=\sum_{ab}\int_{\sqrt{\tau}e^{Y}}^{1}
\int_{\sqrt{\tau}e^{-Y}}^{1}dx_1 dx_2
f_a^{h_1}(x_1,Q^2/\mu_F^2,\mu_R^2/\mu_F^2)f_b^{h_2}(x_2,Q^2/\mu_F^2,\mu_R^2/\mu_F^2)
\times \nonumber\\
&&\frac{d\sigma^{Z}_{ab}}{dY}(x_1,x_2,Q^2/\mu_F^2,\mu_R^2/\mu_F^2).
\end{eqnarray}
Notice that the evolution implemented in \textsc{Candia} allows to analyze the
renormalization/factorization scale dependence also in the evolution, 
which is not present in the MRST parameterizations. 

If we set the scales to be equal, $\mu_F=\mu_R$ and vary $\mu_F$ in the interval $1/2 Q \leq\mu_F\leq 2 Q$
we obtain the results in Fig.~(\ref{rapmuf}), which differ from
those obtained in \cite{Anastasiou} by $2\%$ due to the different implementation of the 
evolution.  Using \textsc{Candia} and as initial condition the 
MRST grid input with $\mu_0^2=1.25$ GeV$^2$ the NNLO band and the NLO one are resolved
separately. From Fig.~(\ref{rapmur}) it is clear that including the $\mu_R^2/\mu_F^2$ effects in the pdf's evolution, the dependence on $\mu_R$ is quite sizeable at NLO, but is reduced at NNLO. 

We show in Fig.~(\ref{Rapex}) the plots of the
variations of the rapidity distributions at the three orders and the corresponding pdf's errors for Alekhin's model and for MRST for $Q=M_Z$. In both cases the reduction of the variation of the cross sections
as we move toward higher orders is quite evident. We report also the errors
on these distributions obtained in both models, which also get systematically smaller as the accuracy of the calculation increases.

\begin{figure}
\subfigure[Alekhin Evolution]{\includegraphics[%
  width=8.5cm,
  angle=-90]{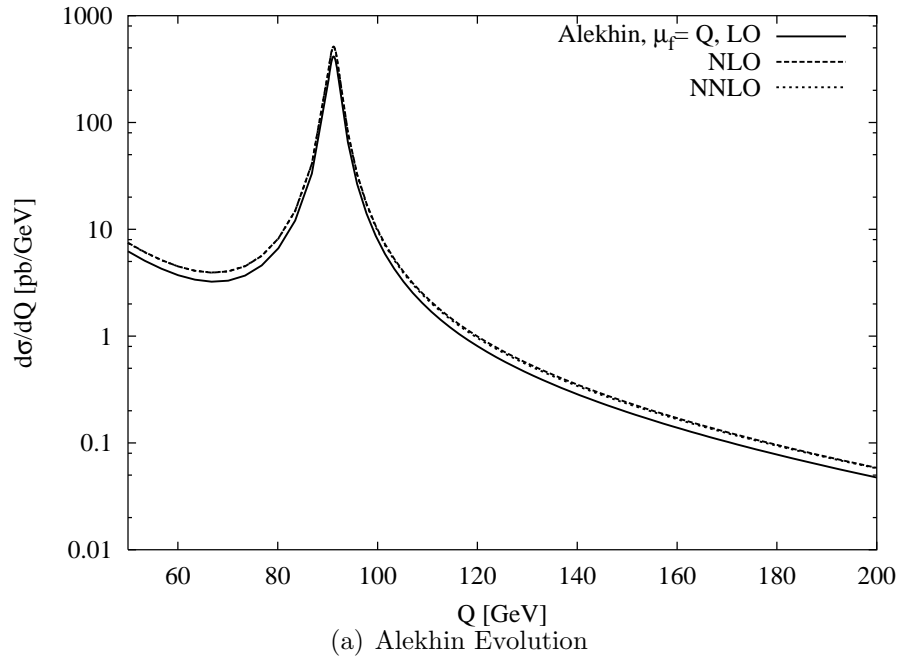}}
\subfigure[MRST Evolution]{\includegraphics[%
  width=8.5cm,
  angle=-90]{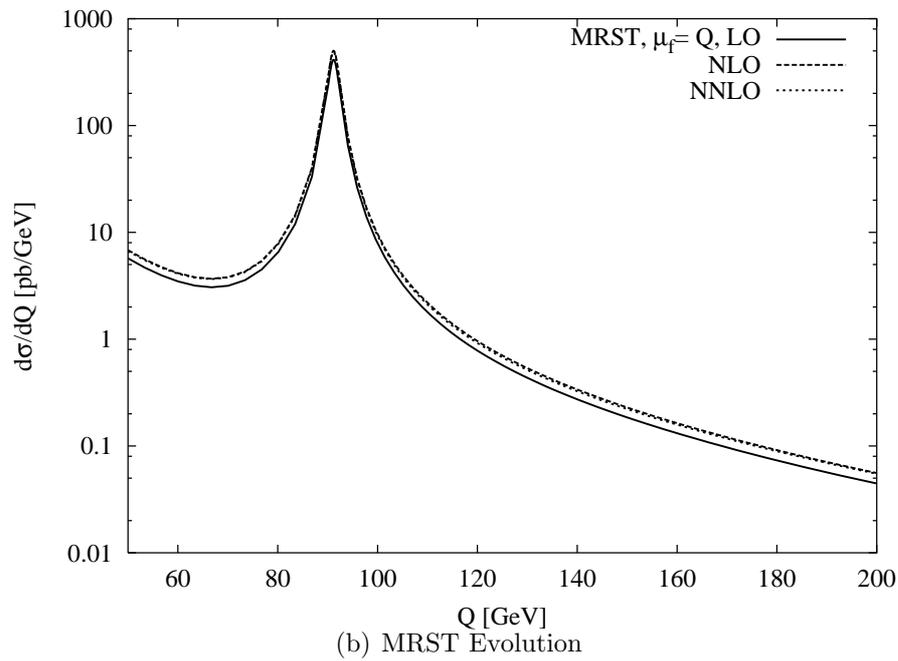}}
\caption{Cross Sections in the region of the peak of the Z
boson at LO, NLO, and NNLO obtained using the
luminosities evolved respectively by Alekhin and MRST}
\label{Cross1}
\end{figure}
\begin{figure}
\subfigure[Alekhin Evolution]{\includegraphics[%
  width=8.5cm,
  angle=-90]{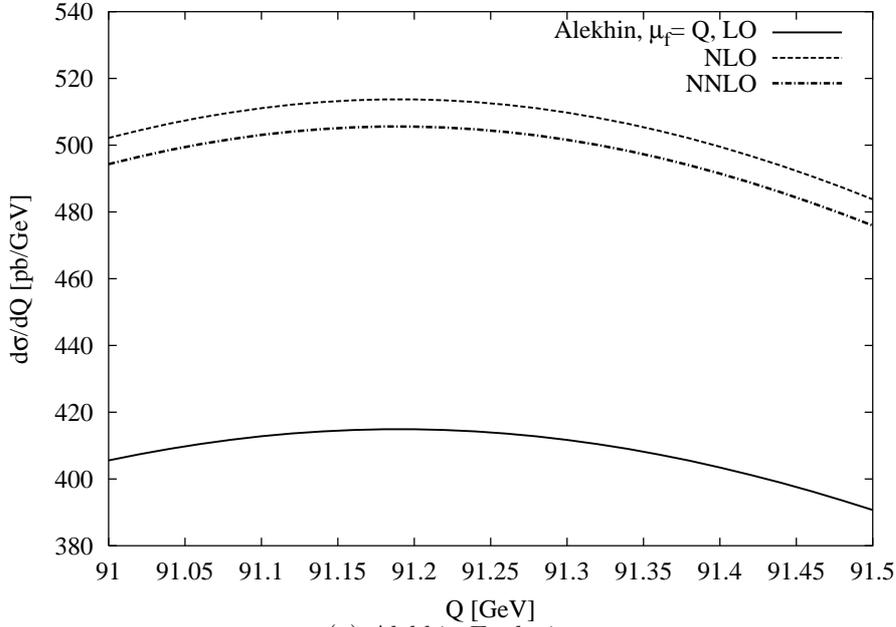}}
\subfigure[MRST Evolution]{\includegraphics[%
  width=8.5cm,
  angle=-90]{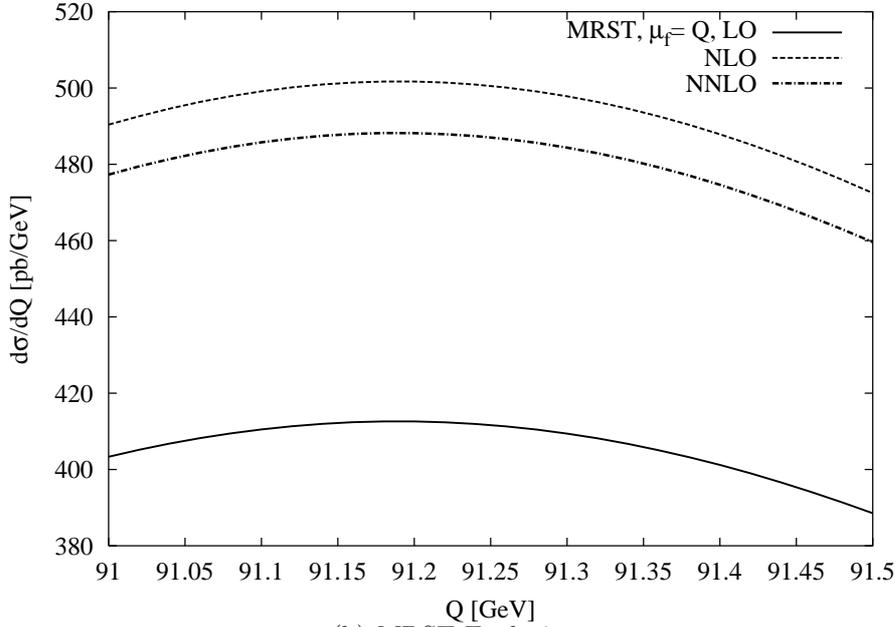}}
\caption{Cross Sections in the region of the Z
with a zoom in the peak region.}
\label{Cross2}
\end{figure}

\begin{figure}
\subfigure[\textsc{Candia} LO evolution for MRST parametric input.]{\includegraphics[%
  width=8cm,
  angle=-90]{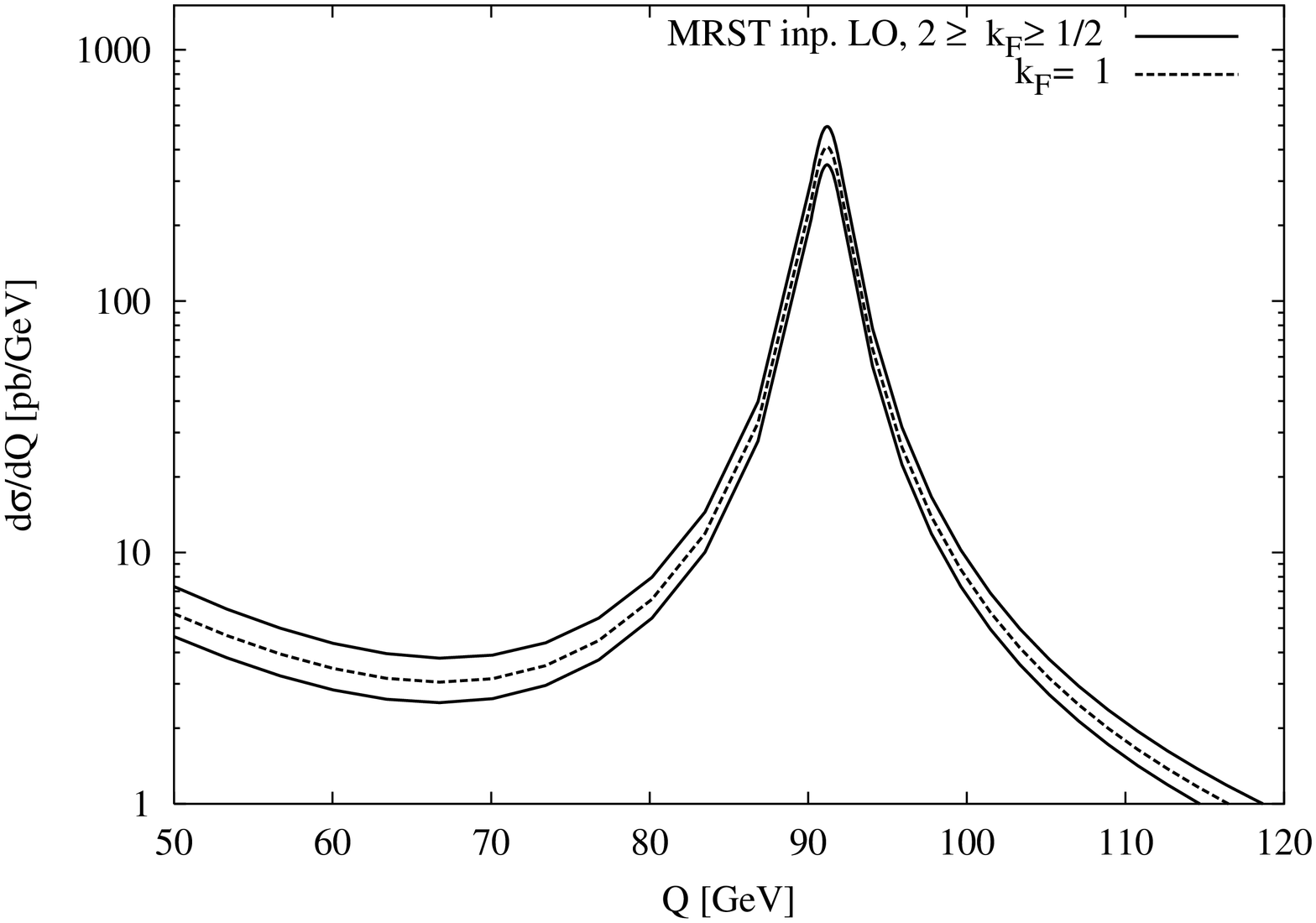}}
\subfigure[\textsc{Candia} NLO evolution for MRST parametric input.]{\includegraphics[%
  width=8cm,
  angle=-90]{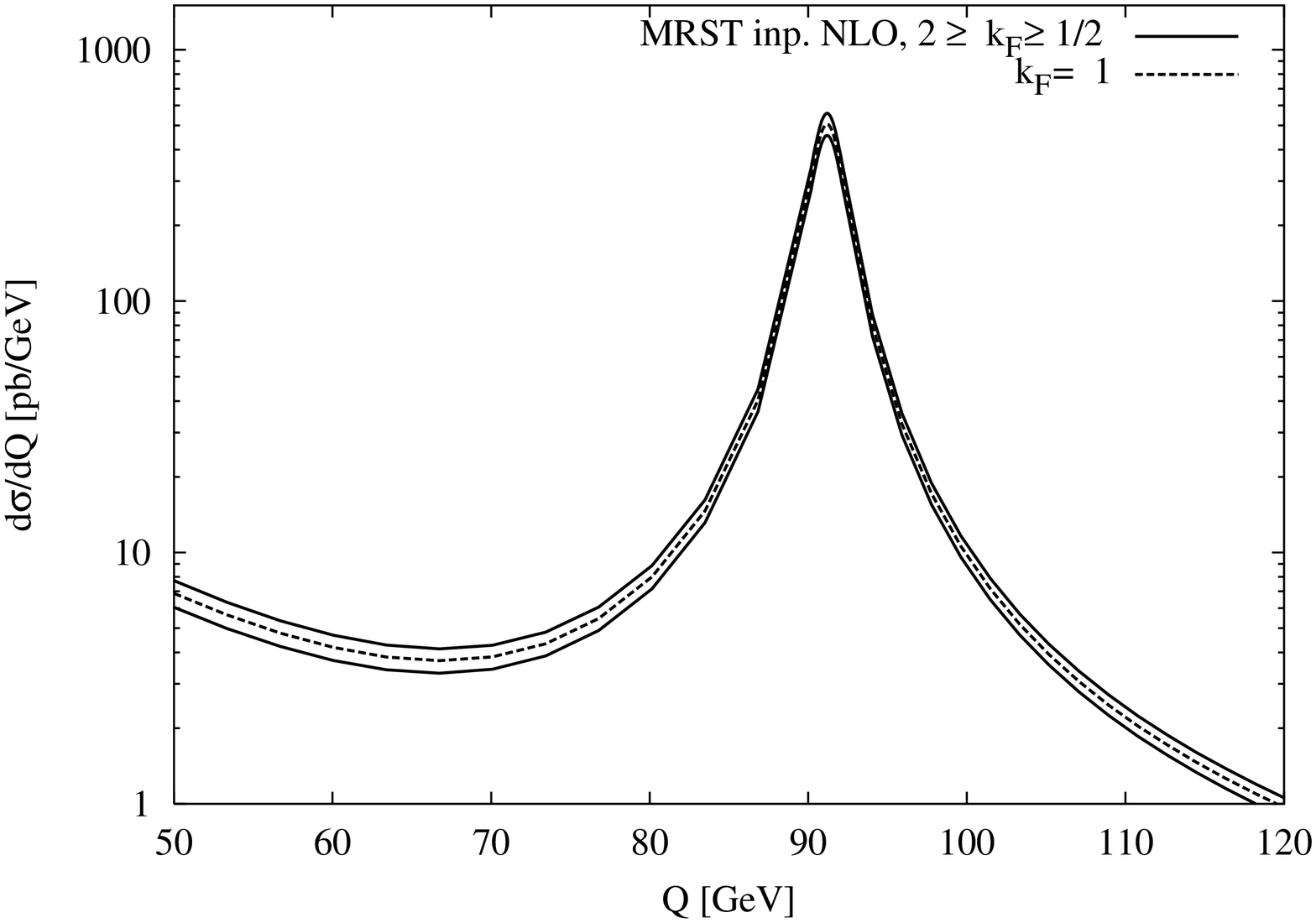}}
\caption{Factorization vs Renormalization scale dependence of
the cross section at LO, NLO with $\sqrt{S}=14$ TeV.
The pdf's have been evolved by using the MRST parametric input at $\mu_0^2=1$ GeV$^2$}
\label{Cross3}
\end{figure}

\begin{figure}
\includegraphics[width=9cm,angle=-90]{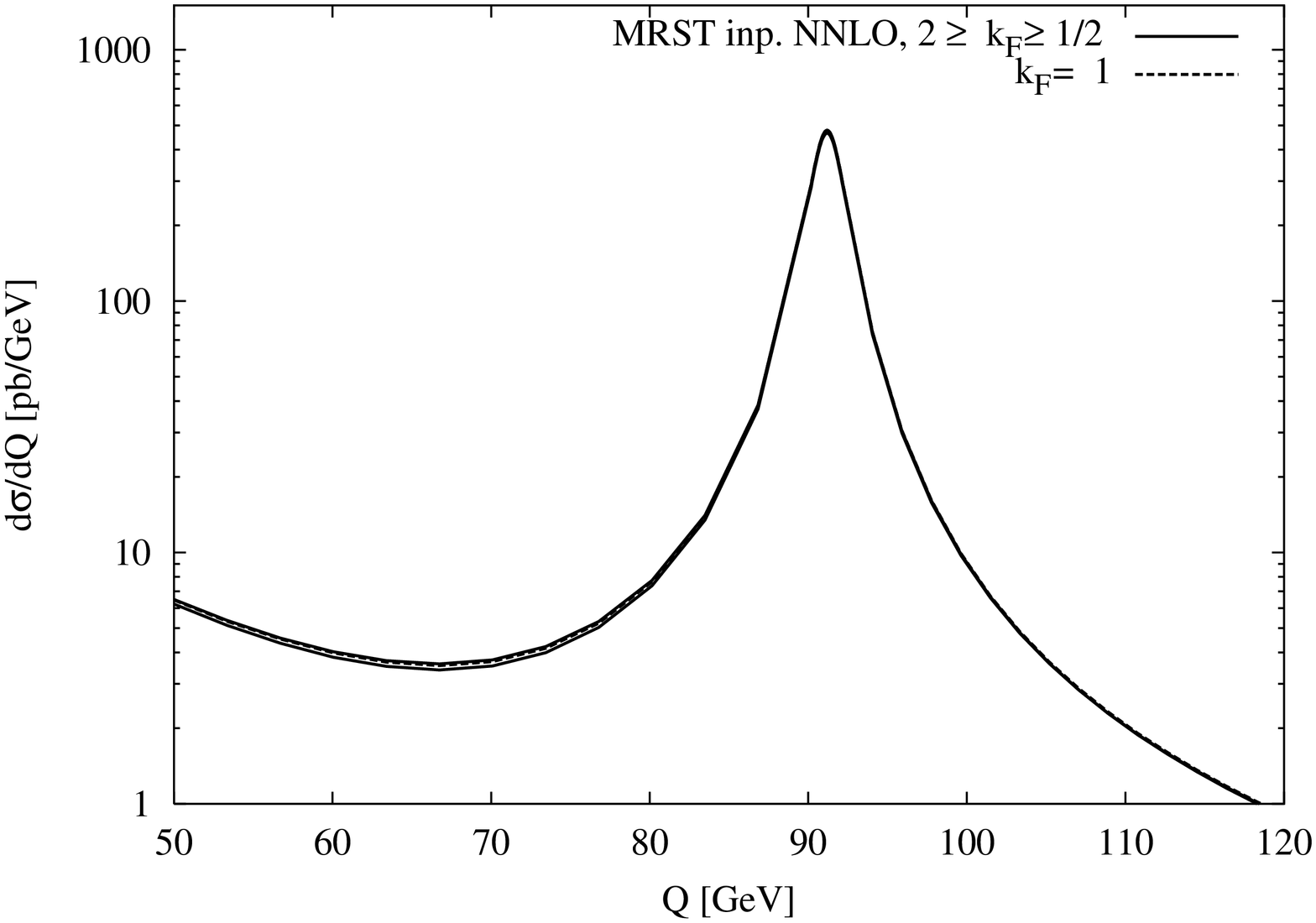}
\caption{Factorization vs renormalization scale dependence of
the cross section at NNLO at $\sqrt{S}=14$ TeV.
The pdf's have been evolved by \textsc{Candia} using the MRST parametric input at $\mu_0^2=1$ GeV$^2$.}
\label{Cross4}
\end{figure}

\begin{figure}
\subfigure[Zoom of the tail of the DY cross section above the peak of the
Z using the evolution provided by Alekhin.]{\includegraphics[%
  width=8cm,
  angle=-90]{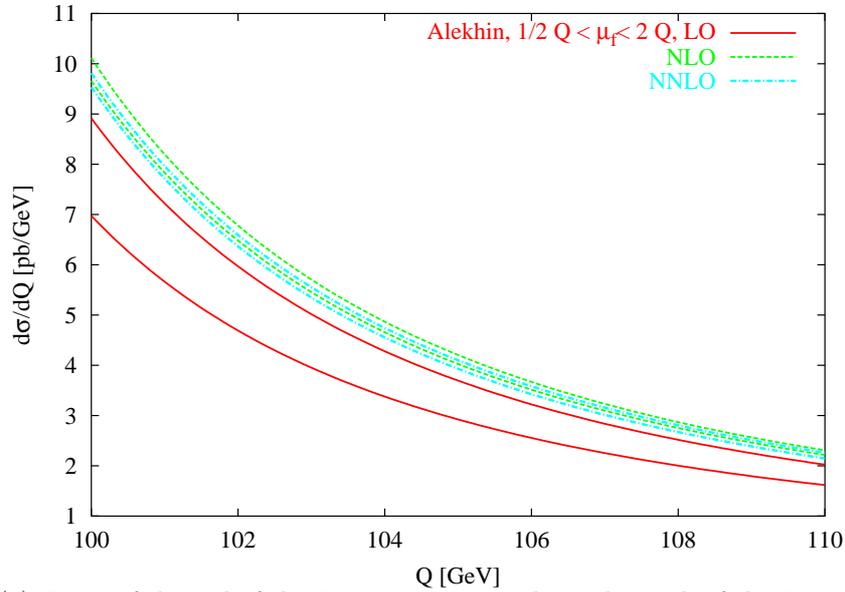}}
\subfigure[As above, but for the MRST parameterizations, with input energy $\mu_{0}^2=1.25$ GeV$^2$.]{\includegraphics[%
  width=8cm,
  angle=-90]{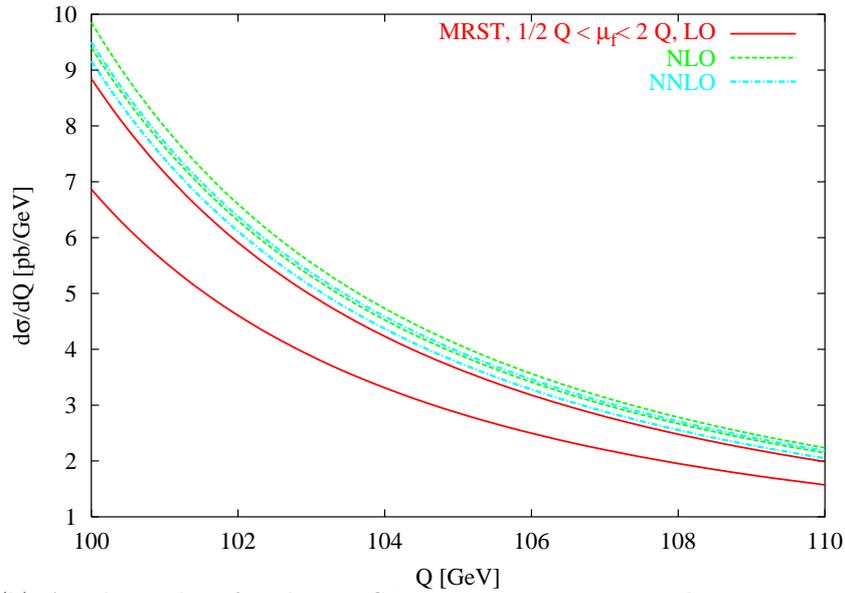}}
\caption{Factorization scale dependence of the cross section at LO NLO and NNLO with $\sqrt{S}=14$ TeV for two models.}
\label{Cross5}
\end{figure}

\begin{figure}
\subfigure[Alekhin's evolution.]{\includegraphics[%
  width=9cm,
  angle=-90]{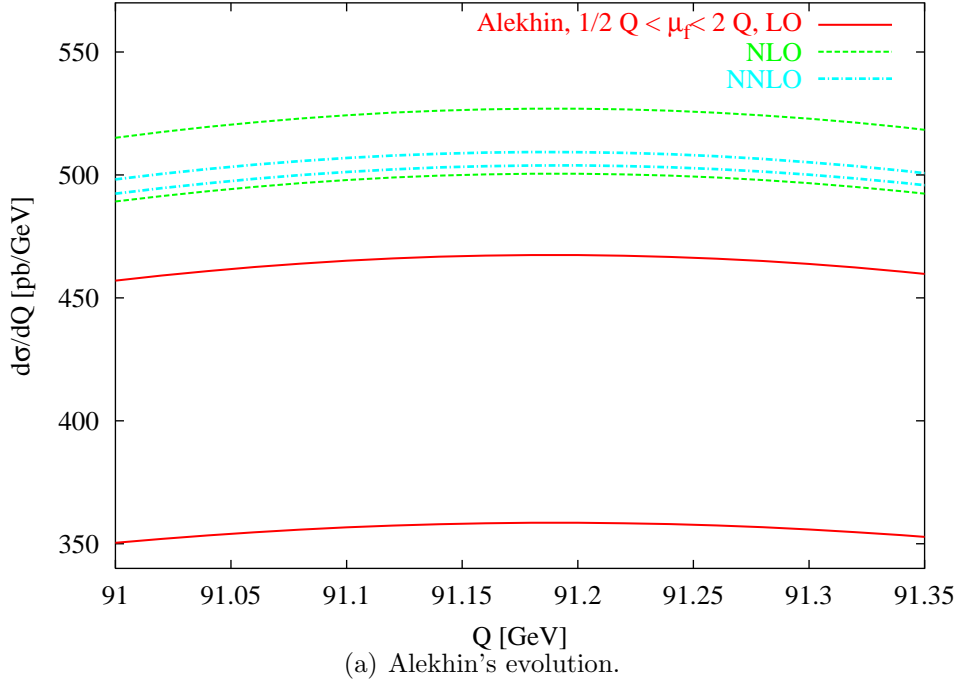}}
\subfigure[MRST evolution with input energy $\mu_{0}^2=1.25$ GeV$^2$.]{\includegraphics[%
  width=9cm,
  angle=-90]{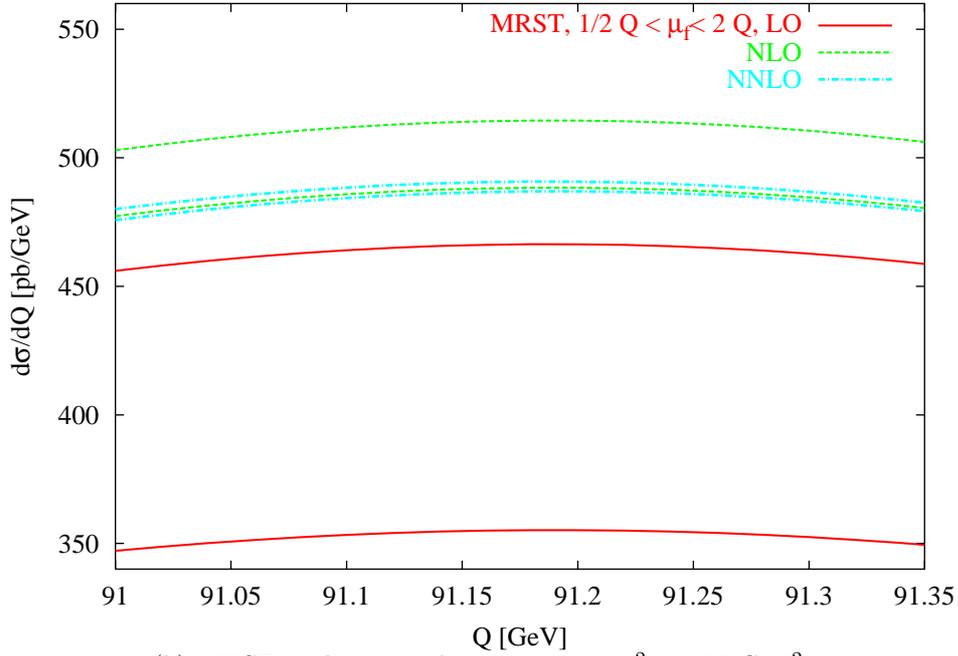}}
\caption{Factorization scale dependence of the cross section at LO NLO and NNLO with $\sqrt{S}=14$ TeV.
Zoom in the peak region.}
\label{Cross6}
\end{figure}

\begin{figure}
\includegraphics[width=9.5cm,angle=-90]{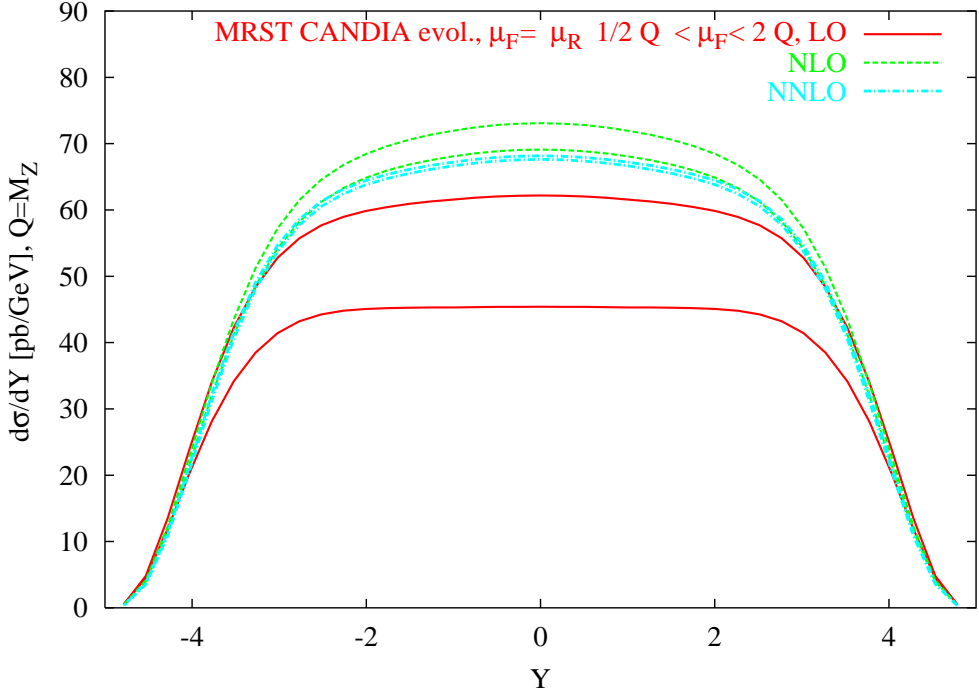}
\caption{Rapidity distributions obtained by changing $1/2 Q\leq \mu_F\leq 2 Q$.
Here we choose $Q=M_{Z}=\mu_F$. The evolution is based on \textsc{Candia} using
MRST grid input with $\mu_0^2=1.25$ GeV$^2$, while we used  \textsc{Vrap} for
the calculation of the hard scattering pieces }
\label{rapmuf}
\end{figure}

\begin{figure}
\includegraphics[width=9.5cm,angle=-90]{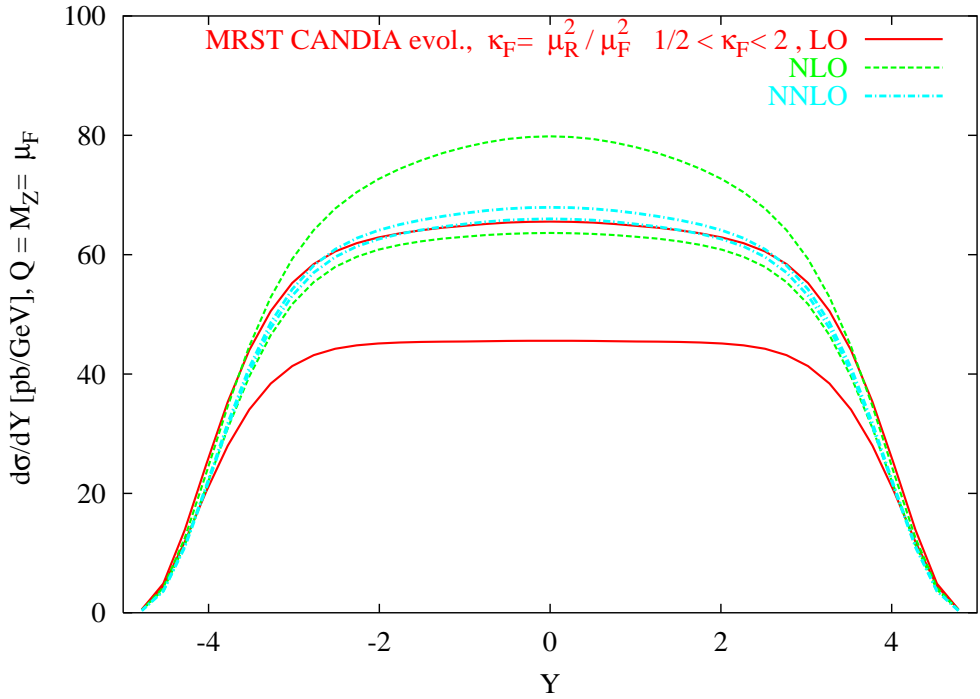}
\caption{Rapidity distributions obtained by changing $1/2 \leq k_F\leq 2$
where $k_F=\mu_R^2/\mu_F^2 $. Here we choose $Q=M_{Z}=\mu_F$.
The evolution is based on \textsc{Candia} using
MRST grid input with $\mu_0^2=1.25$ GeV$^2$. As before  \textsc{Vrap} has been used
for the calculation of the hard scattering}
\label{rapmur}
\end{figure}

\begin{figure}
\subfigure[Alekhin's model]{\includegraphics[%
  width=8.2cm,
  angle=-90]{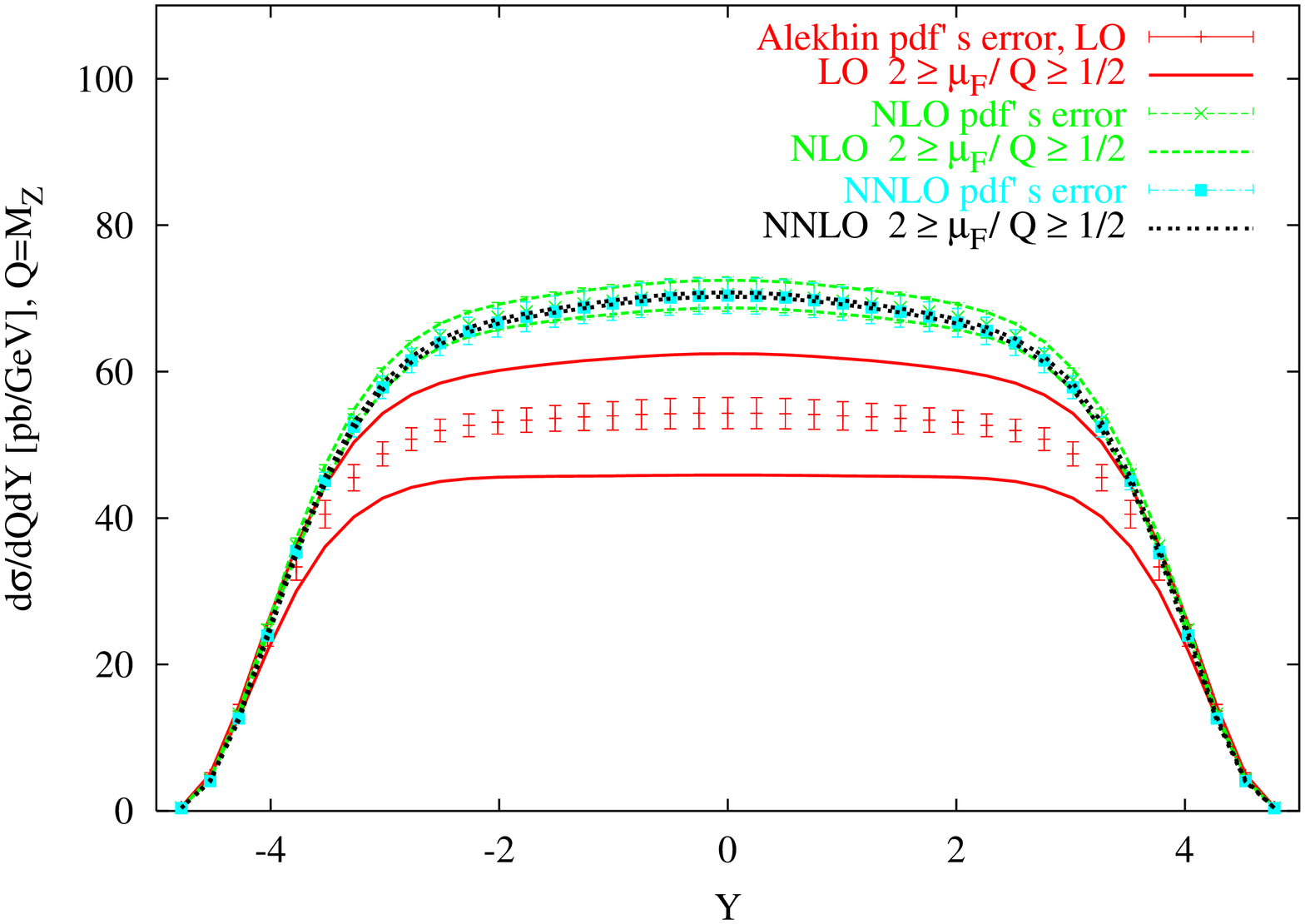}}
\subfigure[MRST model]{\includegraphics[%
  width=8.2cm,
  angle=-90]{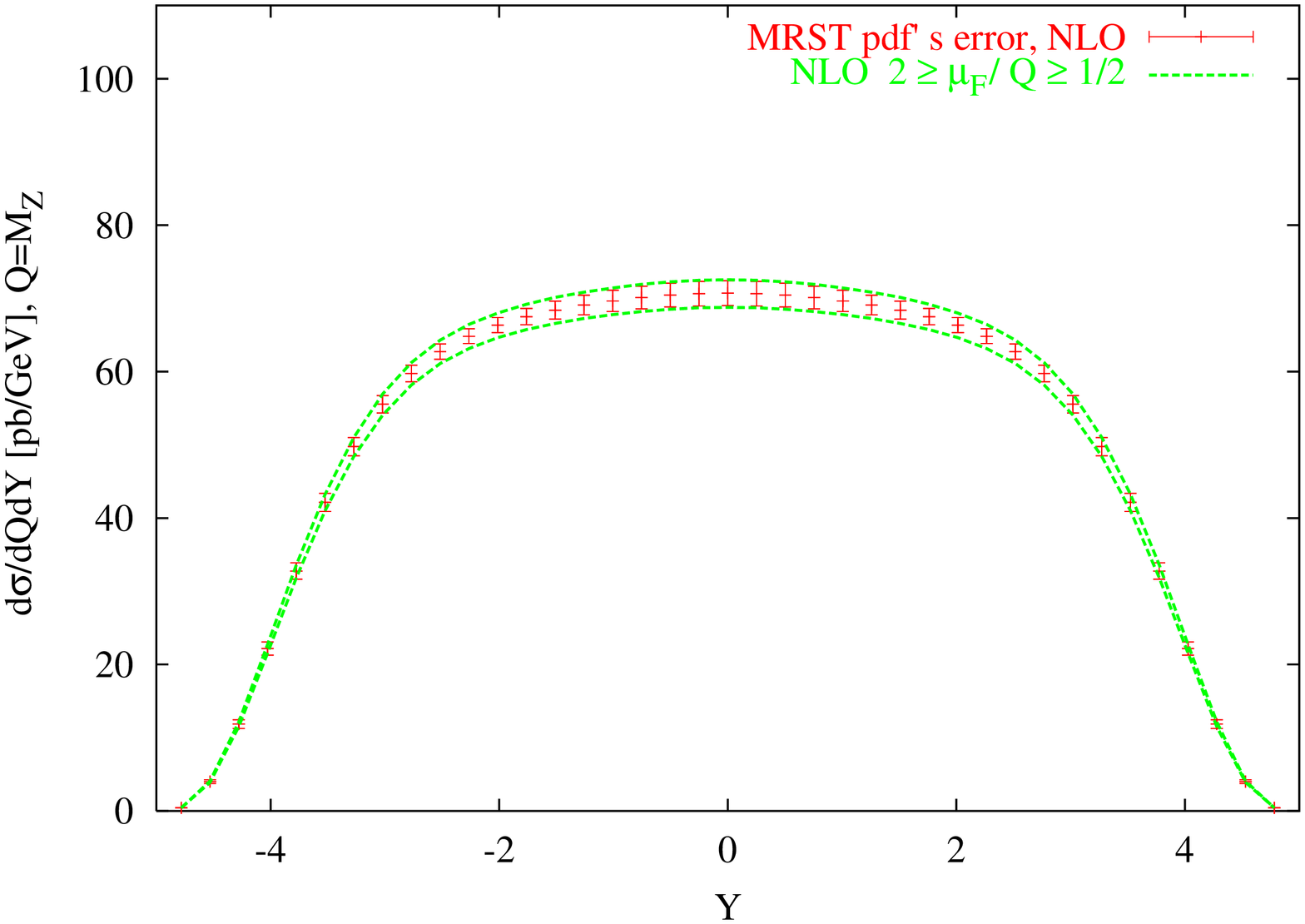}}
\caption{Plot of the rapidity distributions at LO, NLO and NNLO for Alekhin's model and MRST.
Shown are also the bands due to the variation of the $\mu_F$ scale,
and the errors on the cross sections at the corresponding orders.}
\label{Rapex}
\end{figure}

\section{The Cross Sections and the Errors}
We are now going to quantify the errors coming from the pdf's on the
differential cross section in the peak region of the Z by setting the condition $\mu_R=\mu_F$.
The numerical determination of the errors
is computationally very intensive. We perform the analysis at LO, NLO and NNLO for the case
of the Alekhin's pdf's, and only at NLO for the case of the MRST model,
since the error analysis in the latter case is
not available at LO and at NNLO.
We present our results in Figs. (\ref{error1}), (\ref{error2}),
(\ref{error3}) and (\ref{error4}) at typical LHC energy ($\sqrt{S}=14$ TeV).

Fig.~\ref{error1} (a)
shows that the 2 error bands at NLO and NNLO intersect, though the average NNLO
cross section is located outside the area covered by the error band in the fast fall off region.
This feature of the result is shown more clearly in Figs.~\ref{error2} and
\ref{error3}. A zoom of the same region is shown in Fig. \ref{error4}.

The calculation of the error bands has been done following the usual
theory of the linear propagation of the errors.
Starting from the errors on the pdf's known in the
literature (see \cite{MRST1},\cite{MRST2},\cite{Alekhin}), we have generated different
sets of cross sections. Then, the error on the cross section has been
calculated using the formula
\ba
\Delta \sigma=\frac{1}{2}\sqrt{\sum_{k=1}^{N}\left[\sigma_{2k-1}-\sigma_{2 k}\right]^{2}},
\ea
where $\sigma_{k}$ is the $k$-th cross section belonging to a certain set,
and $N$ is the number of free parameters, which is 15 for MRST and 17 for Alekhin.

We show in Table \ref{alek1} the values for the cross section with values obtained by the best fits and the errors at
the corresponding orders. It is evident that the relevance of the NNLO corrections is reduced at
lower $Q$, given the actual quantification of the pdf's errors, since the NNLO corrections
are not outside the error bands. The situation, however, changes beyond the resonance (120 GeV and above) , where it is clear that
the cross section of best the fit at NNLO lays outside the error band,
on the tail of the region that we analyze. The errors induced on the $K$-factors 
( $(K(Q)-1)\% $), as one can easily figure out, are of the order of $4 \%$ 
on the peak ($K(M_Z)$ for the Alekhin set) from their best fit value,
widening quite sharply that determination ($K(M_Z)=0.98 \pm 0.04$).
For $Q=50$ GeV the percentile variation of the cross section in moving from NLO to NNLO is about $1.5\pm 4\%$. 
As a last example, for $Q=146$ GeV we obtain a rate of variation of $3\%$ ($K(146)=0.97\pm 0.03$). Regarding the size
of the errors at NNLO at various Q values, in the Alekhin set these equal - for $Q=50$ GeV - $2.8 \% $ of the best fit value, raising to almost $3 \%$ at $M_Z$ and decreasing to $1.7 \%$ at $200$ GeV.
For the MRST set we have determined the error on the NLO cross sections in Tab.~\ref{mrst1}. They are about $2.8 \%$ of the best fit value at $Q=50$ GeV, decrease to $1.6 \%$ on the peak and decrease moving toward the tail, equating $ 1.2 \%$ at 200 GeV. 

A more complete view of the role played both by the errors at each perturbative order and the corresponding 
best-fit values can be obtained from Figs.~\ref{error1}-\ref{error4}. In Fig.~\ref{error1} 
we have zoomed on the region of invariant mass of the lepton  pair around 100-102 GeV and presented plots of the Alekhin model with the relative errors. The NLO and NNLO predictions show 
overlapping error bands, while the NLO error band for the MRST set (Fig.~\ref{error2}) 
lays slightly below the Alekhin's result at the corresponding order. We have also tried 
to provide an overall view of the tail of the distribution in Fig.~\ref{error3}, where 
we show the best-fit result at NNLO and the NLO error band. The best-fit value lays outside 
this band. A similar result holds for the MRST set and is shown in Fig.~\ref{error3}. 
Finally, in Fig.~\ref{error4} we have zoomed over the region of the resonance, where the best-fit 
value is shown to lay inside the error band.      

\section{Estimating the size of the QCD corrections due to the evolution}

To estimate the role played by the evolution in determining the full NNLO prediction, we show in three tables results for some approximations of 
the NNLO DY cross sections obtained by varying either the hard 
scattering or the order of the evolved pdf's in the factorization 
formula. These approximations may serve as possible ways 
to estimate the contribution coming from the evolution from 
that of the hard scatterings, and may provide some partial 
information on their role in the final result. Tables 
\ref{limit1} and  \ref{limit} show that the error made by neglecting the NNLO 
corrections in the hard scatterings - while keeping the entire NNLO evolution -  
is around 2-3 $\%$, and the correct NNLO cross section is both underestimated and overestimated, while a slight bigger error is made if we neglect the NNLO corrections to the evolution ($4 \%$). The overall decrease of the total cross section appears only after the inclusion of the NNLO evolution. In a final table (Tab.~\ref{limit2}) we 
repeat the trick at NLO, by keeping the hard scatterings at NLO and 
convoluting with the NNLO evolution. Also in this case the errors are 
around $4 \%$ and below in the region of Q that we have studied. 

 There are some features which are quite 
evident from this analysis. The first is that both at NLO (tab.~\ref{limit2}) and at NNLO (tab.~\ref{limit1}) the role of the NNLO terms is to reduce the 
contribution to the cross section. A second 
piece of information can be extracted by comparing all the tables, 
and extracting the differences $\Delta_0\equiv \sigma_{NNLO}\otimes \Phi_{NNLO}- \sigma_{NLO}\otimes \Phi_{NNLO} $ over the entire range of variability of $Q $ and comparing them with the canonical NLO cross section, $\sigma_{NLO}\otimes \Phi_{NLO}$. One can easily come to the conclusion that by combining the NNLO 
evolution with the NLO hard scatterings this ``improved'' NLO cross section is much closer to the true NNLO result than the canonical NLO approximation obtained using the NLO pdf's. For instance for Q=50 GeV the improved NLO result 
differs by $0.3 \%$ from the correct determination, while the ordinary 
NLO prediction differs from it by $4 \%$. On the Z peak 
the improved NLO result differs by $1 \%$ respect to the correct NNLO 
prediction, while the standard NLO cross section is $4 \% $ away. The pattern 
is quite general. It would be interesting to test the same approach 
on other NNLO computations and check whether on a more general basis 
the NLO ``improved'' cross section can be used also for other 
processes as a better estimate of the NNLO result when this is not 
available. We have seen that using these types of approaches, one can estimate the role played by the evolution, which in DY dominates over the NNLO corrections to the hard scatterings.

It is clear from the results of these studied that the role played by the NNLO QCD corrections 
in the $K$-factors at NNLO on the Z peak is relevant, corresponding to variations that can be 
reasonably assumed  around the few percent level. 
We recall that 
with $10^{-1}$ fb of integrated luminosity the statistical error expected on the Z peak is 
around $0.05 \%$ at the LHC. As we have mentioned, suitable choices of the 
electroweak parameters allow to take into account the bulk of the electroweak effects, while the 
non-factorizable contributions are not included in this approach \cite{Baur}. It is then clear that, 
given the size of the QCD NNLO corrections, 
we need to worry about these additional effects, which are clearly dominant especially if we are interested in having 
a robust determination of all the contributions to this process. Searching for heavy extra 
Z' is going to be critically linked to the correct quantification of these additional 
corrections \cite{Baur}.

\begin{table}
\begin{center}
\begin{tabular}{|c||c|c|c|c|}
\hline
\multicolumn{4}{|c|}{$\textrm{d}\sigma/\textrm{d}Q$ in [pb/GeV] for Alekhin with $Q^2=\mu_{F}^2=\mu_{R}^2$, $\sqrt{S}=14$ TeV}
\tabularnewline
\hline
$Q ~[\textrm{GeV}]$&
$\sigma_{LO}$&
$\sigma_{NLO}$&
$\sigma_{NNLO}$\tabularnewline
\hline
\hline
$        50$&
$         6.22$ $\pm$ $         0.27$&
$         7.48$ $\pm$ $         0.24$&
$         7.43$ $\pm$ $         0.21$
\tabularnewline
\hline
$        60.04$&
$         3.72$ $\pm$ $         0.15$&
$         4.50$ $\pm$ $         0.13$&
$         4.49$ $\pm$ $         0.12$
\tabularnewline
\hline
$        70.1$&
$         3.30$ $\pm$ $         0.12$&
$         4.03$ $\pm$ $         0.11$&
$         4.05$ $\pm$ $         0.10$
\tabularnewline
\hline
$        80.1$&
$         6.65$ $\pm$ $         0.24$&
$         8.20$ $\pm$ $         0.24$&
$         8.19$ $\pm$ $         0.23$
\tabularnewline
\hline
$        90.19$&
$       253$ $\pm$ $         8$&
$       313$ $\pm$ $         9$&
$       309$ $\pm$ $         8$
\tabularnewline
\hline
$        91.19$&
$       415$ $\pm$ $        14$&
$       514$ $\pm$ $        15$&
$       506$ $\pm$ $        15$
\tabularnewline
\hline
$       120.07$&
$         0.80$ $\pm$ $         0.02$&
$         0.99$ $\pm$ $         0.02$&
$         0.96$ $\pm$ $         0.03$
\tabularnewline
\hline
$       146.1$&
$         0.225$ $\pm$ $         0.006$&
$         0.277$ $\pm$ $         0.007$&
$         0.269$ $\pm$ $         0.007$
\tabularnewline
\hline
$       172.1$&
$         0.097$ $\pm$ $         0.002$&
$         0.119$ $\pm$ $         0.003$&
$         0.117$ $\pm$ $         0.003$
\tabularnewline
\hline
$        200$&
$         0.047$ $\pm$ $         0.001$&
$         0.058$ $\pm$ $         0.001$&
$         0.058$ $\pm$ $         0.001$
\tabularnewline
\hline
\end{tabular}
\caption{Cross sections derived from the best fits for the 3 orders with their errors for the
set by Alekhin.}
\label{alek1}
\end{center}
\end{table}

\begin{table}
\begin{center}
\begin{tabular}{|c||c|c|}
\hline
\multicolumn{2}{|c|}{$\textrm{d}\sigma/\textrm{d}Q$ in [pb/GeV] for MRST with $Q^2=\mu_{F}^2=\mu_{R}^2$, $\sqrt{S}=14$ TeV}
\tabularnewline
\hline
$Q ~[\textrm{GeV}]$&
$\sigma_{NLO}$\tabularnewline
\hline
\hline
$        50$&
$         6.77$ $\pm$ $         0.19$
\tabularnewline
\hline
$        60.04$&
$         4.13$ $\pm$ $         0.10$
\tabularnewline
\hline
$        70.1$&
$         3.79$ $\pm$ $         0.08$
\tabularnewline
\hline
$        80.1$&
$         7.90$ $\pm$ $         0.14$
\tabularnewline
\hline
$        90.19$&
$       305$ $\pm$ $         5$
\tabularnewline
\hline
$        91.19$&
$       499$ $\pm$ $         8$
\tabularnewline
\hline
$       120.1$&
$         0.952$ $\pm$ $         0.014$
\tabularnewline
\hline
$       146.1$&
$         0.264$ $\pm$ $         0.003$
\tabularnewline
\hline
$       172.1$&
$         0.113$ $\pm$ $         0.001$
\tabularnewline
\hline
$       200$&
$         0.0556$ $\pm$ $         0.0007$
\tabularnewline
\hline
\end{tabular}
\caption{Cross sections derived from the best fits at NLO with the errors for the
MRST set.}
\end{center}
\label{mrst1}
\end{table}


\begin{figure}
\subfigure[Pdf's Error bands in the Alekhin model with $Q=\mu_F=\mu_R$ and $\sqrt{S}=14$ TeV]{\includegraphics[%
  width=8.5cm,
  angle=-90]{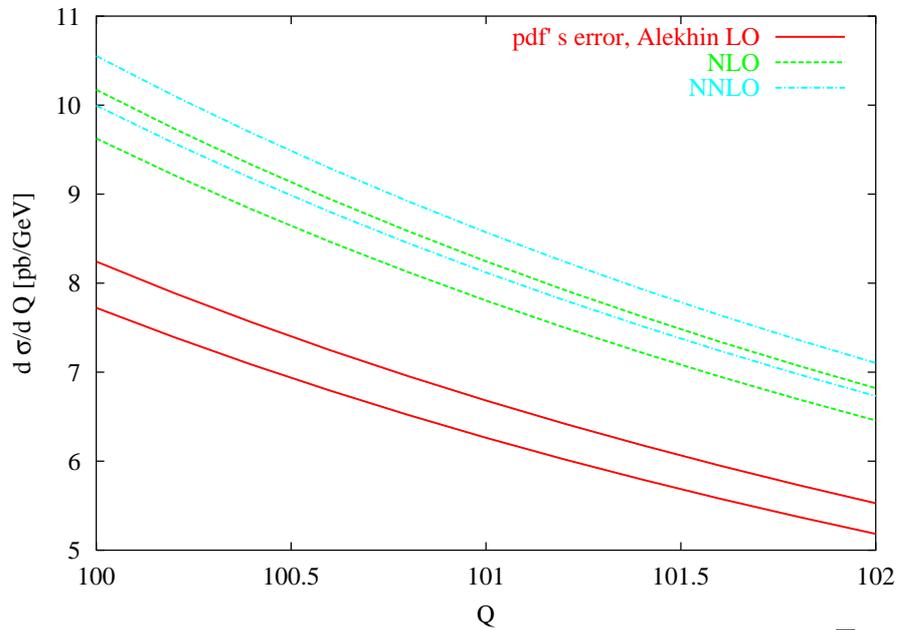}}
\subfigure[MRST error bands with $Q=\mu_F=\mu_R$ and $\sqrt{S}=14$ TeV]{\includegraphics[%
  width=8.5cm,
  angle=-90]{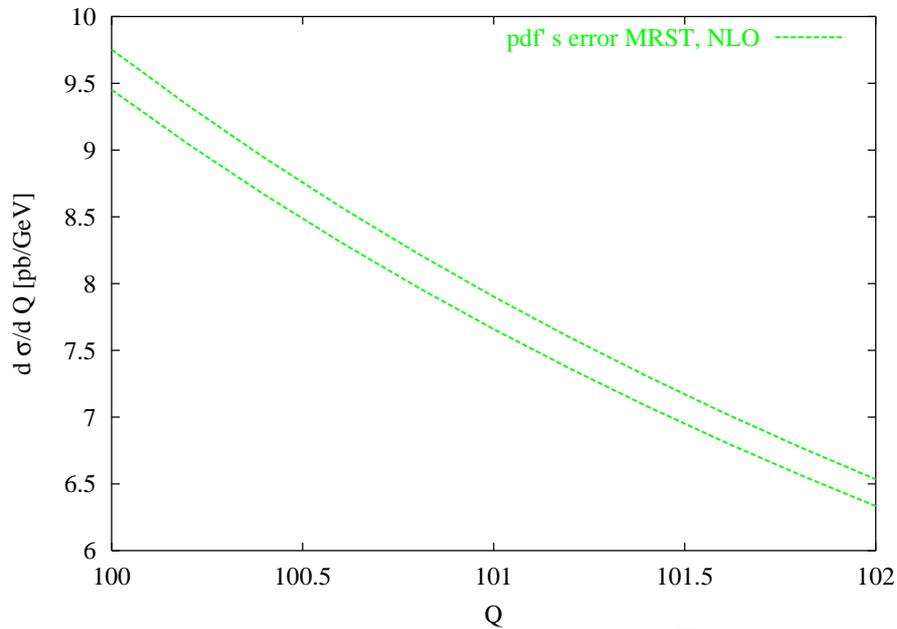}}
\caption{Errors of the pdf's on the cross sections at LHC. Zoom in the region of $100$ GeV.}
\label{error1}
\end{figure}

\begin{figure}
\includegraphics[%
  width=11cm,
  angle=-90]{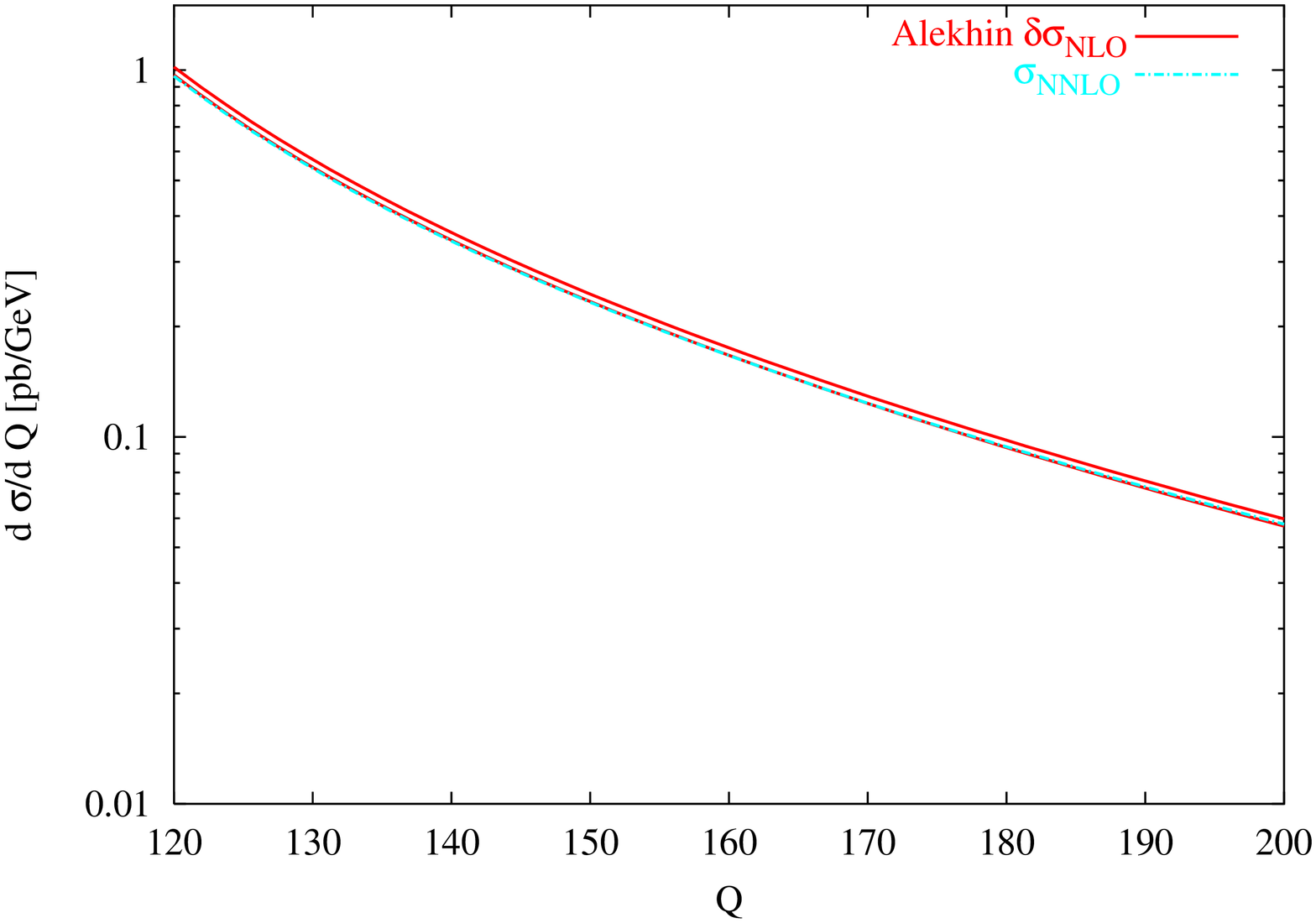}
\caption{NNLO cross section for Alekhin with the respective pdf's errors
at NLO in the $100$ GeV region, with $Q=\mu_F=\mu_R$ and $\sqrt{S}=14$ TeV}
\label{error2}
\end{figure}
\begin{figure}
\includegraphics[%
  width=11cm,
  angle=-90]{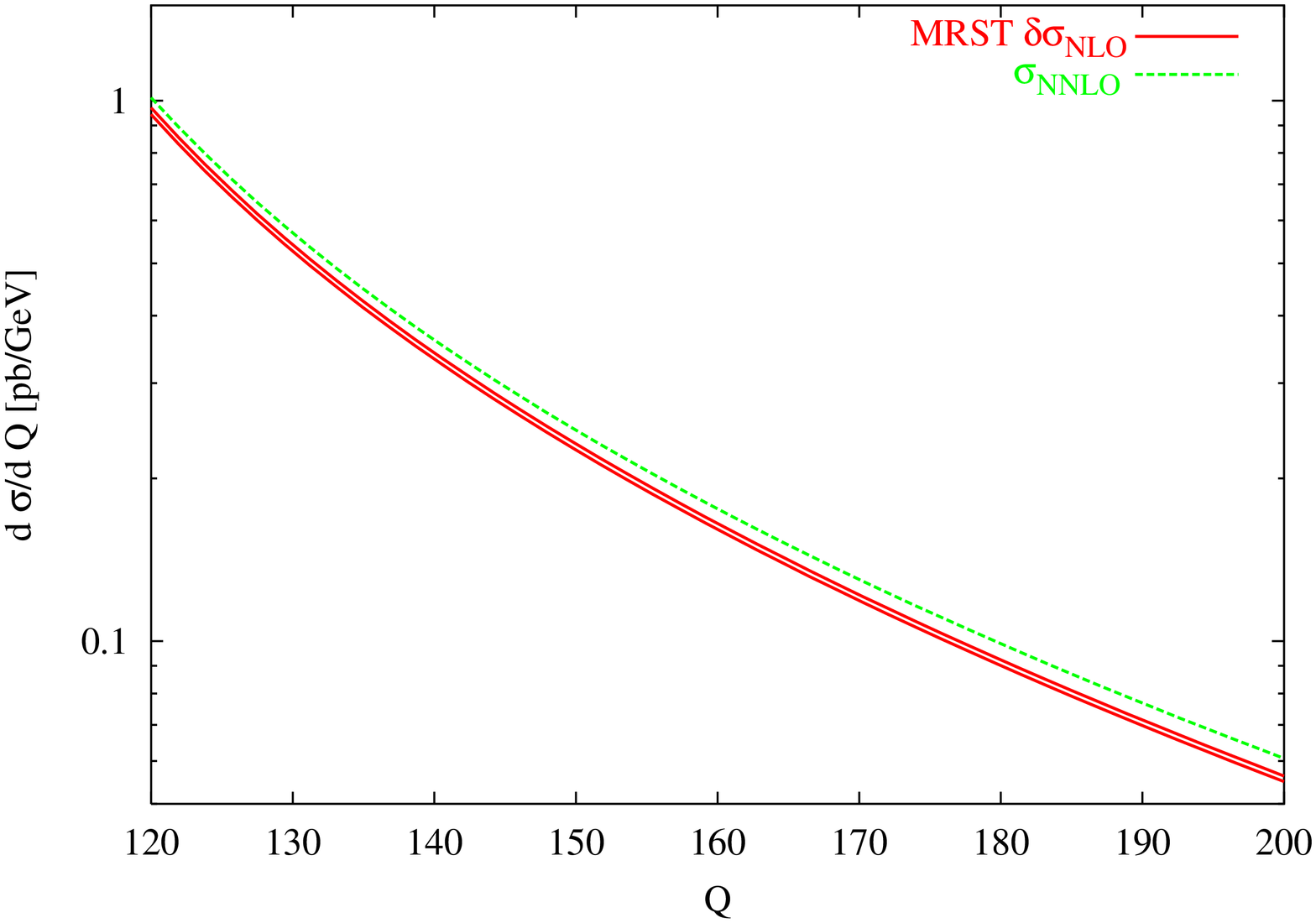}
\caption{NNLO cross section for MRST with the respective pdf's errors
at NLO in the $100$ GeV region, with $Q=\mu_F=\mu_R$ and $\sqrt{S}=14$ TeV}
\label{error3}
\end{figure}

\section{Conclusions}

We have presented a comparative study of the NNLO predictions for lepton pair production
and discussed their robustness. We have presented results
concerning $K$-factors, renormalization/factorization scale dependence and errors on the
cross sections - induced by errors on the pdf's - following different approaches.
For this reason we have put under close scrutiny the theory of the logarithmic expansions,
which we have shown to give results which are compatible with other approaches, and allows to address 
the issue of accuracy in the context of the QCD evolution.
Our estimate of the difference between the different approaches is slightly above the
level of $1 \%$. The $K$-factors found using these new methods appears to be slightly larger than
those coming from the MRST and Alekhin evolved parton distributions, but compatible with them, given the 
actual errors on the pdf's. Clearly, 
with the advent of the LHC, these analysis should be rendered even more accurate, 
especially at large values of the mass distributions, where a detailed analysis of the electroweak
effects should be included. This is particularly important in the search of extra
gauge interactions using this channel. On the Z resonance these effects are smaller, at the percent
level, but are important for calibration and partonometry.
These and other related issues will be left for future work.

\vspace{1cm}

\centerline{\bf Note Added}
The extended analysis presented in this work has been performed within
the 2006 Monte Carlo workshop held in Frascati under the sponsorship of
INFN of Italy. Detailed tables/plots are provided only for reference in this version for the arXiv
since they may be of practical use. \textsc{Candia}, \textsc{Candia}$_{DY}$ and their interface with  \textsc{Vrap} will be released and described in forthcoming work.

\centerline{\bf Acknowledgments}
We thank Lance Dixon and Andreas Vogt for discussions and correspondence.
 We thank Simone Morelli for help in the numerical implementations and for discussions. C.C. thanks Theodore Tomaras and Nikos Irges for discussions, Alon Faraggi and the Theory Group at Crete and at the University of Liverpool for hospitality. The work of A.C. is partly supported by 
the grant MTKD-CT-2004-014319 and by the EU grant MRTN-CT-2004-512194 with partial support from the INTERREG IIIA  Greece - Cyprus program. 
The work of C.C. is partly supported by the Marie Curie Research and Training network ``Universenet'' (MRTN-CT-2006-035863) and by the INTERREG IIIA Greece - Cyprus program. The numerical 
analysis has been performed on the INFN cluster at the University of Salento. The work of M.G.
is partly supported by MIUR and by INFN. We thank the Participants of the INFN 2006 Monte Carlo workshop in Frascati, and the Organizers, in particular Barbara Mele and Paolo Nason for the effort with the organization and for discussions. 

\begin{figure}
\subfigure[Alekhin with $Q=\mu_F=\mu_R$ and $\sqrt{S}=14$ TeV]{\includegraphics[%
  width=8.5cm,
  angle=-90]{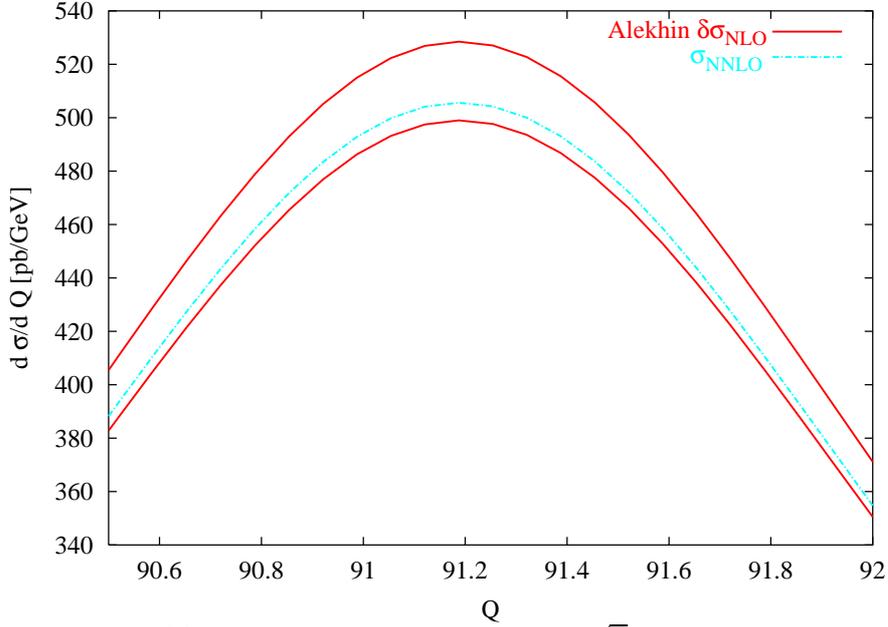}}
\subfigure[MRST with $Q=\mu_F=\mu_R$ and $\sqrt{S}=14$ TeV]{\includegraphics[%
  width=8.5cm,
  angle=-90]{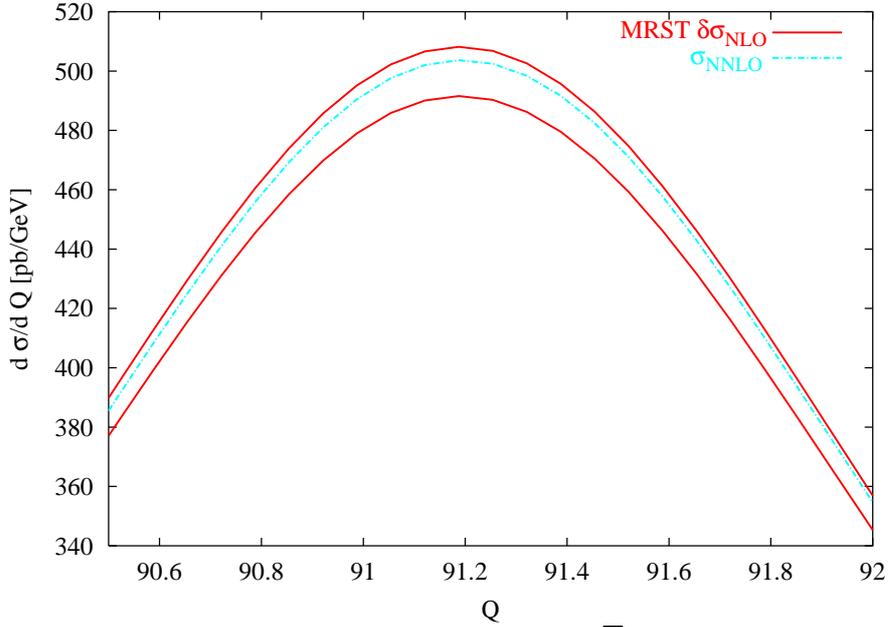}}
\caption{NNLO cross section for Alekhin and MRST with the respective pdf's errors
at NLO in the Z peak region. Zoom in the region of $90$ GeV.}
\label{error4}
\end{figure}

\begin{table}
\begin{footnotesize}
\begin{center}
\begin{tabular}{|c||c|c|c|}
\hline
\multicolumn{4}{|c|}{\textsc{Candia} evolution with MRST input, $\mu_0^2=1.25$ GeV$^2$}
\tabularnewline
\hline
$Q ~[\textrm{GeV}]$&
$\sigma_{NNLO}\otimes \Phi_{NNLO}$&
$\sigma_{NLO}\otimes \Phi_{NNLO}$  &
$\delta\sigma$ \tabularnewline
\hline
\hline
$50.0000$&
$6.4935\cdot10^{+0}$&
$6.5164\cdot10^{+0}$&
$3.5195\cdot10^{-3}$
\tabularnewline
\hline
$60.0469$&
$3.9997\cdot10^{+0}$&
$3.9864\cdot10^{+0}$&
$3.3040\cdot10^{-3}$
\tabularnewline
\hline
$70.0938$&
$3.6962\cdot10^{+0}$&
$3.6683\cdot10^{+0}$&
$7.5645\cdot10^{-3}$
\tabularnewline
\hline
$80.1407$&
$7.6755\cdot10^{+0}$&
$7.6639\cdot10^{+0}$&
$1.5087\cdot10^{-3}$
\tabularnewline
\hline
$90.1876$&
$2.9325\cdot10^{+2}$&
$2.9676\cdot10^{+2}$&
$1.1988\cdot10^{-2}$
\tabularnewline
\hline
$91.1876$&
$4.8006\cdot10^{+2}$&
$4.8644\cdot10^{+2}$&
$1.3293\cdot10^{-2}$
\tabularnewline
\hline
$92.1876$&
$2.9179\cdot10^{+2}$&
$2.9604\cdot10^{+2}$&
$1.4556\cdot10^{-2}$
\tabularnewline
\hline
$120.0701$&
$9.0411\cdot10^{-1}$&
$9.3152\cdot10^{-1}$&
$3.0318\cdot10^{-2}$
\tabularnewline
\hline
$146.0938$&
$2.5267\cdot10^{-1}$&
$2.5981\cdot10^{-1}$&
$2.8222\cdot10^{-2}$
\tabularnewline
\hline
$172.1175$&
$1.0938\cdot10^{-1}$&
$1.1179\cdot10^{-1}$&
$2.2061\cdot10^{-2}$
\tabularnewline
\hline
$200.0000$&
$5.4431\cdot10^{-2}$&
$5.5145\cdot10^{-2}$&
$1.3118\cdot10^{-2}$
\tabularnewline
\hline
\end{tabular}
\end{center}
\caption{$\sigma_{NNLO}\otimes \Phi_{NNLO}$ vs $\sigma_{NLO}\otimes \Phi_{NNLO}$ in [pb/GeV].
Comparison between NNLO and NLO cross sections obtained by the convolution of NNLO pdf's.
$\delta \sigma$ is defined as
$|\sigma_{NNLO}\otimes \Phi_{NNLO}-\sigma_{NLO}\otimes \Phi_{NNLO}|/\sigma_{NNLO}\otimes \Phi_{NNLO}$.}
\label{limit1}
\end{footnotesize}
\end{table}

\begin{table}
\begin{footnotesize}
\begin{center}
\begin{tabular}{|c||c|c|c|}
\hline
\multicolumn{4}{|c|}{\textsc{Candia} evolution with MRST input, $\mu_0^2=1.25$ GeV$^2$}
\tabularnewline
\hline
$Q ~[\textrm{GeV}]$&
$\sigma_{NNLO}\otimes \Phi_{NNLO}$&
$\sigma_{NNLO}\otimes \Phi_{NLO}$  &
$\delta\sigma$ \tabularnewline
\hline
\hline
$50.0000$&
$6.4935\cdot10^{+0}$&
$6.7853\cdot10^{+0}$&
$4.4938\cdot10^{-2}$
\tabularnewline
\hline
$60.0469$&
$3.9997\cdot10^{+0}$&
$4.1805\cdot10^{+0}$&
$4.5206\cdot10^{-2}$
\tabularnewline
\hline
$70.0938$&
$3.6962\cdot10^{+0}$&
$3.8571\cdot10^{+0}$&
$4.3521\cdot10^{-2}$
\tabularnewline
\hline
$80.1407$&
$7.6755\cdot10^{+0}$&
$7.9669\cdot10^{+0}$&
$3.7967\cdot10^{-2}$
\tabularnewline
\hline
$90.1876$&
$2.9325\cdot10^{+2}$&
$3.0219\cdot10^{+2}$&
$3.0498\cdot10^{-2}$
\tabularnewline
\hline
$91.1876$&
$4.8006\cdot10^{+2}$&
$4.9437\cdot10^{+2}$&
$2.9809\cdot10^{-2}$
\tabularnewline
\hline
$92.1876$&
$2.9179\cdot10^{+2}$&
$3.0029\cdot10^{+2}$&
$2.9141\cdot10^{-2}$
\tabularnewline
\hline
$120.0701$&
$9.0411\cdot10^{-1}$&
$9.2025\cdot10^{-1}$&
$1.7858\cdot10^{-2}$
\tabularnewline
\hline
$146.0938$&
$2.5267\cdot10^{-1}$&
$2.5593\cdot10^{-1}$&
$1.2890\cdot10^{-2}$
\tabularnewline
\hline
$172.1175$&
$1.0938\cdot10^{-1}$&
$1.1039\cdot10^{-1}$&
$9.2157\cdot10^{-3}$
\tabularnewline
\hline
$200.0000$&
$5.4431\cdot10^{-2}$&
$5.4781\cdot10^{-2}$&
$6.4302\cdot10^{-3}$
\tabularnewline
\hline
\end{tabular}
\end{center}
\caption{$\sigma_{NNLO}\otimes \Phi_{NNLO}$ vs $\sigma_{NNLO}\otimes \Phi_{NLO}$ in [pb/GeV].
Upper bound on the NNLO cross sections obtained by the convolution of NNLO and NLO pdf's.
$\delta \sigma$ is defined as
$|\sigma_{NNLO}\otimes \Phi_{NNLO}-\sigma_{NNLO}\otimes \Phi_{NLO}|/\sigma_{NNLO}\otimes \Phi_{NNLO}$.}
\label{limit}
\end{footnotesize}
\end{table}

\begin{table}
\begin{footnotesize}
\begin{center}
\begin{tabular}{|c||c|c|c|}
\hline
\multicolumn{4}{|c|}{\textsc{Candia} evolution with MRST input, $\mu_0^2=1.25$ GeV$^2$}
\tabularnewline
\hline
$Q ~[\textrm{GeV}]$&
$\sigma_{NLO}\otimes \Phi_{NLO}$&
$\sigma_{NLO}\otimes \Phi_{NNLO}$  &
$\delta\sigma$ \tabularnewline
\hline
\hline
$50.0000$&
$6.8119\cdot10^{+0}$&
$6.5164\cdot10^{+0}$&
$4.3376\cdot10^{-2}$
\tabularnewline
\hline
$60.0469$&
$4.1552\cdot10^{+0}$&
$3.9864\cdot10^{+0}$&
$4.0623\cdot10^{-2}$
\tabularnewline
\hline
$70.0938$&
$3.8110\cdot10^{+0}$&
$3.6683\cdot10^{+0}$&
$3.7465\cdot10^{-2}$
\tabularnewline
\hline
$80.1407$&
$7.9371\cdot10^{+0}$&
$7.6639\cdot10^{+0}$&
$3.4420\cdot10^{-2}$
\tabularnewline
\hline
$90.1876$&
$3.0657\cdot10^{+2}$&
$2.9676\cdot10^{+2}$&
$3.2000\cdot10^{-2}$
\tabularnewline
\hline
$91.1876$&
$5.0242\cdot10^{+2}$&
$4.8644\cdot10^{+2}$&
$3.1790\cdot10^{-2}$
\tabularnewline
\hline
$92.1876$&
$3.0569\cdot10^{+2}$&
$2.9604\cdot10^{+2}$&
$3.1584\cdot10^{-2}$
\tabularnewline
\hline
$120.0701$&
$9.5677\cdot10^{-1}$&
$9.3152\cdot10^{-1}$&
$2.6396\cdot10^{-2}$
\tabularnewline
\hline
$146.0938$&
$2.6562\cdot10^{-1}$&
$2.5981\cdot10^{-1}$&
$2.1896\cdot10^{-2}$
\tabularnewline
\hline
$172.1175$&
$1.1382\cdot10^{-1}$&
$1.1179\cdot10^{-1}$&
$1.7801\cdot10^{-2}$
\tabularnewline
\hline
$200.0000$&
$5.5940\cdot10^{-2}$&
$5.5145\cdot10^{-2}$&
$1.4212\cdot10^{-2}$
\tabularnewline
\hline
\end{tabular}
\end{center}
\caption{$\sigma_{NLO}\otimes \Phi_{NLO}$ vs $\sigma_{NLO}\otimes \Phi_{NNLO}$ in [pb/GeV].
Lower bound on the NLO cross sections obtained by the convolution of NLO and NNLO pdf's.
$\delta\sigma$ is defined as
$|\sigma_{NLO}\otimes \Phi_{NLO}-\sigma_{NLO}\otimes \Phi_{NNLO}|/\sigma_{NLO}\otimes \Phi_{NLO}$.}
\label{limit2}
\end{footnotesize}
\end{table}


\begin{table}
\begin{center}
\begin{tabular}{|c||c|c|c|}
\hline
\multicolumn{4}{|c|}{$\alpha_{s}(Q^2)$ for MRST at NNLO. Brute force vs $\Lambda_{QCD}$ parameterization. }
\tabularnewline
\hline
$Q ~[\textrm{GeV}]$&
$\alpha_{s}^{brute}(Q^2)$&
$\alpha_{s}^{\Lambda}(Q^2)$ &
$\delta\alpha_{s}^{NNLO}(Q^2) \% $ \tabularnewline
\hline
\hline
$        50.0000$&
$         0.1268$&
$         0.1251$&
$         1.3702$
\tabularnewline
\hline
$        83.4897$&
$         0.1170$&
$         0.1156$&
$         1.2570$
\tabularnewline
\hline
$        90.7209$&
$         0.1156$&
$         0.1142$& 
$         1.2405$ 
\tabularnewline
\hline
$        91.1876$&
$         0.1155$&
$         0.1141$& 
$         1.2395$ 
\tabularnewline
\hline
$        92.0543$&
$         0.1153$&
$         0.1139$& 
$         1.2377$ 
\tabularnewline
\hline
$       107.0583$&
$         0.1128$&
$         0.1115$& 
$         1.2090$ 
\tabularnewline
\hline
$       125.6466$&
$         0.1103$&
$         0.1090$&
$         1.1801$ 
\tabularnewline
\hline
$       144.2350$&
$         0.1082$&
$         0.1069$& 
$         1.1564$ 
\tabularnewline
\hline
$       162.8233$&
$         0.1064$&
$         0.1052$& 
$         1.1363$ 
\tabularnewline
\hline
$       181.4117$&
$         0.1049$&
$         0.1037$& 
$         1.1191$ 
\tabularnewline
\hline
$       200.0000$&
$         0.1035$&
$         0.1024$& 
$         1.1040$
\tabularnewline
\hline
\end{tabular}
\end{center}
\caption{NNLO running of the coupling determined using the {\em brute force} solution of the
renormalization group equations versus the asymptotic expansions in terms $\Lambda_{QCD}$. The
percentage differences are normalized respect to $\alpha_{brute}$.}
\label{running}
\end{table}

\begin{table}
\begin{footnotesize}
\hspace{-1cm}
\begin{tabular}{|c||c|c|c|c|c|c|c|}
\hline
\multicolumn{8}{|c|}{$xu_v(x)$ \textsc{Candia} evolution at NLO, Les Houches input, $N_f=4$, $Q=\mu_{F}=\mu_{R}=100$ GeV}
\tabularnewline
\hline
$ x $                       &
$x{u_v(x)}^{\textsc{Candia}}_{asymp}$     &
$\kappa=1$                 &
$\kappa=2$                 &
$\kappa=3$                 &
$\kappa=4$                 &
$\kappa=5$                 &
$\kappa=6$  \tabularnewline
\hline
\hline
$1e-05$&
$2.6878\cdot10^{-3}$&
$2.5121\cdot10^{-3}$&
$2.4829\cdot10^{-3}$&
$2.4828\cdot10^{-3}$&
$2.4833\cdot10^{-3}$&
$2.4831\cdot10^{-3}$&
$2.4832\cdot10^{-3}$
\tabularnewline
\hline
$0.0001$&
$1.2844\cdot10^{-2}$&
$1.2474\cdot10^{-2}$&
$1.2290\cdot10^{-2}$&
$1.2307\cdot10^{-2}$&
$1.2306\cdot10^{-2}$&
$1.2306\cdot10^{-2}$&
$1.2306\cdot10^{-2}$
\tabularnewline
\hline
$0.001$&
$5.7937\cdot10^{-2}$&
$5.7893\cdot10^{-2}$&
$5.7161\cdot10^{-2}$&
$5.7260\cdot10^{-2}$&
$5.7246\cdot10^{-2}$&
$5.7248\cdot10^{-2}$&
$5.7248\cdot10^{-2}$
\tabularnewline
\hline
$0.01$&
$2.3029\cdot10^{-1}$&
$2.3340\cdot10^{-1}$&
$2.3188\cdot10^{-1}$&
$2.3213\cdot10^{-1}$&
$2.3209\cdot10^{-1}$&
$2.3209\cdot10^{-1}$&
$2.3209\cdot10^{-1}$
\tabularnewline
\hline
$0.1$&
$5.5456\cdot10^{-1}$&
$5.5556\cdot10^{-1}$&
$5.5738\cdot10^{-1}$&
$5.5712\cdot10^{-1}$&
$5.5716\cdot10^{-1}$&
$5.5715\cdot10^{-1}$&
$5.5715\cdot10^{-1}$
\tabularnewline
\hline
$0.2$&
$4.9105\cdot10^{-1}$&
$4.8494\cdot10^{-1}$&
$4.8836\cdot10^{-1}$&
$4.8784\cdot10^{-1}$&
$4.8792\cdot10^{-1}$&
$4.8790\cdot10^{-1}$&
$4.8791\cdot10^{-1}$
\tabularnewline
\hline
$0.3$&
$3.5395\cdot10^{-1}$&
$3.4503\cdot10^{-1}$&
$3.4837\cdot10^{-1}$&
$3.4785\cdot10^{-1}$&
$3.4793\cdot10^{-1}$&
$3.4792\cdot10^{-1}$&
$3.4792\cdot10^{-1}$
\tabularnewline
\hline
$0.4$&
$2.2304\cdot10^{-1}$&
$2.1470\cdot10^{-1}$&
$2.1729\cdot10^{-1}$&
$2.1689\cdot10^{-1}$&
$2.1695\cdot10^{-1}$&
$2.1694\cdot10^{-1}$&
$2.1694\cdot10^{-1}$
\tabularnewline
\hline
$0.5$&
$1.2271\cdot10^{-1}$&
$1.1661\cdot10^{-1}$&
$1.1830\cdot10^{-1}$&
$1.1803\cdot10^{-1}$&
$1.1808\cdot10^{-1}$&
$1.1807\cdot10^{-1}$&
$1.1807\cdot10^{-1}$
\tabularnewline
\hline
$0.6$&
$5.6866\cdot10^{-2}$&
$5.3292\cdot10^{-2}$&
$5.4212\cdot10^{-2}$&
$5.4067\cdot10^{-2}$&
$5.4090\cdot10^{-2}$&
$5.4086\cdot10^{-2}$&
$5.4087\cdot10^{-2}$
\tabularnewline
\hline
$0.7$&
$2.0429\cdot10^{-2}$&
$1.8840\cdot10^{-2}$&
$1.9232\cdot10^{-2}$&
$1.9169\cdot10^{-2}$&
$1.9179\cdot10^{-2}$&
$1.9178\cdot10^{-2}$&
$1.9178\cdot10^{-2}$
\tabularnewline
\hline
$0.8$&
$4.6754\cdot10^{-3}$&
$4.2230\cdot10^{-3}$&
$4.3329\cdot10^{-3}$&
$4.3152\cdot10^{-3}$&
$4.3180\cdot10^{-3}$&
$4.3175\cdot10^{-3}$&
$4.3176\cdot10^{-3}$
\tabularnewline
\hline
$0.9$&
$3.6098\cdot10^{-4}$&
$3.1538\cdot10^{-4}$&
$3.2674\cdot10^{-4}$&
$3.2486\cdot10^{-4}$&
$3.2516\cdot10^{-4}$&
$3.2511\cdot10^{-4}$&
$3.2512\cdot10^{-4}$
\tabularnewline
\hline
\end{tabular}
\caption{NLO valence distribution of the up quark with the Les Houches benchmark
model}
\label{tab1}
\end{footnotesize}
\end{table}

\begin{table}
\begin{footnotesize}
\hspace{-1cm}
\begin{tabular}{|c||c|c|c|c|c|c|c|}
\hline
\multicolumn{8}{|c|}{$xu_v(x)$ \textsc{Candia} evolution at NNLO, Les Houches input, $N_f=4$, $Q=\mu_{F}=\mu_{R}=100$ GeV}
\tabularnewline
\hline
$ x $                       &
$x{u_{v}(x)}^{\textsc{Candia}}_{asymp}$     &
$\kappa=2$                 &
$\kappa=3$                 &
$\kappa=4$                 &
$\kappa=5$                 &
$\kappa=6$                 &
$\kappa=7$  \tabularnewline
\hline
\hline
$1e-05$&
$3.0260\cdot10^{-3}$&
$2.9464\cdot10^{-3}$&
$2.8972\cdot10^{-3}$&
$2.8928\cdot10^{-3}$&
$2.8945\cdot10^{-3}$&
$2.8944\cdot10^{-3}$&
$2.8944\cdot10^{-3}$
\tabularnewline
\hline
$0.0001$&
$1.3656\cdot10^{-2}$&
$1.3320\cdot10^{-2}$&
$1.3179\cdot10^{-2}$&
$1.3177\cdot10^{-2}$&
$1.3181\cdot10^{-2}$&
$1.3181\cdot10^{-2}$&
$1.3181\cdot10^{-2}$
\tabularnewline
\hline
$0.001$&
$5.9360\cdot10^{-2}$&
$5.8657\cdot10^{-2}$&
$5.8425\cdot10^{-2}$&
$5.8448\cdot10^{-2}$&
$5.8451\cdot10^{-2}$&
$5.8450\cdot10^{-2}$&
$5.8450\cdot10^{-2}$
\tabularnewline
\hline
$0.01$&
$2.3139\cdot10^{-1}$&
$2.3254\cdot10^{-1}$&
$2.3248\cdot10^{-1}$&
$2.3253\cdot10^{-1}$&
$2.3252\cdot10^{-1}$&
$2.3252\cdot10^{-1}$&
$2.3252\cdot10^{-1}$
\tabularnewline
\hline
$0.1$&
$5.5125\cdot10^{-1}$&
$5.5406\cdot10^{-1}$&
$5.5451\cdot10^{-1}$&
$5.5446\cdot10^{-1}$&
$5.5446\cdot10^{-1}$&
$5.5446\cdot10^{-1}$&
$5.5446\cdot10^{-1}$
\tabularnewline
\hline
$0.2$&
$4.8672\cdot10^{-1}$&
$4.8416\cdot10^{-1}$&
$4.8463\cdot10^{-1}$&
$4.8454\cdot10^{-1}$&
$4.8454\cdot10^{-1}$&
$4.8454\cdot10^{-1}$&
$4.8454\cdot10^{-1}$
\tabularnewline
\hline
$0.3$&
$3.5017\cdot10^{-1}$&
$3.4470\cdot10^{-1}$&
$3.4507\cdot10^{-1}$&
$3.4497\cdot10^{-1}$&
$3.4498\cdot10^{-1}$&
$3.4498\cdot10^{-1}$&
$3.4498\cdot10^{-1}$
\tabularnewline
\hline
$0.4$&
$2.2030\cdot10^{-1}$&
$2.1460\cdot10^{-1}$&
$2.1486\cdot10^{-1}$&
$2.1478\cdot10^{-1}$&
$2.1479\cdot10^{-1}$&
$2.1479\cdot10^{-1}$&
$2.1479\cdot10^{-1}$
\tabularnewline
\hline
$0.5$&
$1.2099\cdot10^{-1}$&
$1.1660\cdot10^{-1}$&
$1.1676\cdot10^{-1}$&
$1.1671\cdot10^{-1}$&
$1.1671\cdot10^{-1}$&
$1.1671\cdot10^{-1}$&
$1.1671\cdot10^{-1}$
\tabularnewline
\hline
$0.6$&
$5.5957\cdot10^{-2}$&
$5.3309\cdot10^{-2}$&
$5.3392\cdot10^{-2}$&
$5.3364\cdot10^{-2}$&
$5.3366\cdot10^{-2}$&
$5.3367\cdot10^{-2}$&
$5.3367\cdot10^{-2}$
\tabularnewline
\hline
$0.7$&
$2.0052\cdot10^{-2}$&
$1.8854\cdot10^{-2}$&
$1.8888\cdot10^{-2}$&
$1.8877\cdot10^{-2}$&
$1.8877\cdot10^{-2}$&
$1.8878\cdot10^{-2}$&
$1.8878\cdot10^{-2}$
\tabularnewline
\hline
$0.8$&
$4.5726\cdot10^{-3}$&
$4.2288\cdot10^{-3}$&
$4.2387\cdot10^{-3}$&
$4.2352\cdot10^{-3}$&
$4.2355\cdot10^{-3}$&
$4.2356\cdot10^{-3}$&
$4.2356\cdot10^{-3}$
\tabularnewline
\hline
$0.9$&
$3.5079\cdot10^{-4}$&
$3.1629\cdot10^{-4}$&
$3.1736\cdot10^{-4}$&
$3.1698\cdot10^{-4}$&
$3.1701\cdot10^{-4}$&
$3.1702\cdot10^{-4}$&
$3.1702\cdot10^{-4}$
\tabularnewline
\hline
\end{tabular}
\caption{ NNLO Valence distribution for the up quark with the Les Houches benchmark model}
\label{tab2}
\end{footnotesize}
\end{table}

\begin{table}
\begin{footnotesize}
\hspace{-1cm}
\begin{tabular}{|c||c|c|c|c|c|c|c|}
\hline
\multicolumn{8}{|c|}{$xg(x)$ \textsc{Candia} evolution at NLO, Les Houches input, $N_f=4$, $Q=\mu_{F}=\mu_{R}=100$ GeV}
\tabularnewline
\hline
$ x $                       &
$xg(x)^{\textsc{Candia}}_{asym}$     &
$\kappa=1$                 &
$\kappa=2$                 &
$\kappa=3$                 &
$\kappa=4$                 &
$\kappa=5$                 &
$\kappa=6$  \tabularnewline
\hline
\hline
$1e-05$&
$2.3578\cdot10^{+2}$&
$2.2829\cdot10^{+2}$&
$2.3693\cdot10^{+2}$&
$2.3560\cdot10^{+2}$&
$2.3580\cdot10^{+2}$&
$2.3577\cdot10^{+2}$&
$2.3578\cdot10^{+2}$
\tabularnewline
\hline
$0.0001$&
$9.3027\cdot10^{+1}$&
$9.1414\cdot10^{+1}$&
$9.3237\cdot10^{+1}$&
$9.2998\cdot10^{+1}$&
$9.3031\cdot10^{+1}$&
$9.3026\cdot10^{+1}$&
$9.3027\cdot10^{+1}$
\tabularnewline
\hline
$0.001$&
$3.1540\cdot10^{+1}$&
$3.1320\cdot10^{+1}$&
$3.1562\cdot10^{+1}$&
$3.1537\cdot10^{+1}$&
$3.1540\cdot10^{+1}$&
$3.1540\cdot10^{+1}$&
$3.1540\cdot10^{+1}$
\tabularnewline
\hline
$0.01$&
$8.1120\cdot10^{+0}$&
$8.1098\cdot10^{+0}$&
$8.1116\cdot10^{+0}$&
$8.1122\cdot10^{+0}$&
$8.1120\cdot10^{+0}$&
$8.1120\cdot10^{+0}$&
$8.1120\cdot10^{+0}$
\tabularnewline
\hline
$0.1$&
$8.9872\cdot10^{-1}$&
$9.0284\cdot10^{-1}$&
$8.9826\cdot10^{-1}$&
$8.9877\cdot10^{-1}$&
$8.9871\cdot10^{-1}$&
$8.9872\cdot10^{-1}$&
$8.9872\cdot10^{-1}$
\tabularnewline
\hline
$0.2$&
$2.5540\cdot10^{-1}$&
$2.5695\cdot10^{-1}$&
$2.5523\cdot10^{-1}$&
$2.5542\cdot10^{-1}$&
$2.5539\cdot10^{-1}$&
$2.5540\cdot10^{-1}$&
$2.5540\cdot10^{-1}$
\tabularnewline
\hline
$0.3$&
$8.3414\cdot10^{-2}$&
$8.4005\cdot10^{-2}$&
$8.3353\cdot10^{-2}$&
$8.3422\cdot10^{-2}$&
$8.3414\cdot10^{-2}$&
$8.3415\cdot10^{-2}$&
$8.3414\cdot10^{-2}$
\tabularnewline
\hline
$0.4$&
$2.7017\cdot10^{-2}$&
$2.7232\cdot10^{-2}$&
$2.6995\cdot10^{-2}$&
$2.7020\cdot10^{-2}$&
$2.7017\cdot10^{-2}$&
$2.7017\cdot10^{-2}$&
$2.7017\cdot10^{-2}$
\tabularnewline
\hline
$0.5$&
$8.0411\cdot10^{-3}$&
$8.1131\cdot10^{-3}$&
$8.0338\cdot10^{-3}$&
$8.0420\cdot10^{-3}$&
$8.0410\cdot10^{-3}$&
$8.0411\cdot10^{-3}$&
$8.0411\cdot10^{-3}$
\tabularnewline
\hline
$0.6$&
$2.0343\cdot10^{-3}$&
$2.0552\cdot10^{-3}$&
$2.0321\cdot10^{-3}$&
$2.0345\cdot10^{-3}$&
$2.0342\cdot10^{-3}$&
$2.0343\cdot10^{-3}$&
$2.0343\cdot10^{-3}$
\tabularnewline
\hline
$0.7$&
$3.8654\cdot10^{-4}$&
$3.9132\cdot10^{-4}$&
$3.8603\cdot10^{-4}$&
$3.8660\cdot10^{-4}$&
$3.8653\cdot10^{-4}$&
$3.8654\cdot10^{-4}$&
$3.8654\cdot10^{-4}$
\tabularnewline
\hline
$0.8$&
$4.1712\cdot10^{-5}$&
$4.2399\cdot10^{-5}$&
$4.1634\cdot10^{-5}$&
$4.1722\cdot10^{-5}$&
$4.1710\cdot10^{-5}$&
$4.1712\cdot10^{-5}$&
$4.1712\cdot10^{-5}$
\tabularnewline
\hline
$0.9$&
$1.8310\cdot10^{-6}$&
$1.8598\cdot10^{-6}$&
$1.8272\cdot10^{-6}$&
$1.8315\cdot10^{-6}$&
$1.8309\cdot10^{-6}$&
$1.8310\cdot10^{-6}$&
$1.8310\cdot10^{-6}$
\tabularnewline
\hline
\end{tabular}
\caption{NLO gluon density in the Les Houches model}
\label{tab3}
\end{footnotesize}
\end{table}

\begin{table}
\begin{footnotesize}
\hspace{-1cm}
\begin{tabular}{|c||c|c|c|c|c|c|c|}
\hline
\multicolumn{8}{|c|}{$xg(x)$ \textsc{Candia} evolution at NNLO, Les Houches input, $N_f=4$, $Q=\mu_{F}=\mu_{R}=100$ GeV}
\tabularnewline
\hline
$ x $                       &
$xg(x)^{\textsc{Candia}}_{asymp}$     &
$\kappa=2$                 &
$\kappa=3$                 &
$\kappa=4$                 &
$\kappa=5$                 &
$\kappa=6$                 &
$\kappa=7$  \tabularnewline
\hline
\hline
$1e-05$&
$2.2328\cdot10^{+2}$&
$2.2090\cdot10^{+2}$&
$2.2351\cdot10^{+2}$&
$2.2330\cdot10^{+2}$&
$2.2328\cdot10^{+2}$&
$2.2328\cdot10^{+2}$&
$2.2328\cdot10^{+2}$
\tabularnewline
\hline
$0.0001$&
$9.0763\cdot10^{+1}$&
$9.0383\cdot10^{+1}$&
$9.0795\cdot10^{+1}$&
$9.0765\cdot10^{+1}$&
$9.0762\cdot10^{+1}$&
$9.0763\cdot10^{+1}$&
$9.0763\cdot10^{+1}$
\tabularnewline
\hline
$0.001$&
$3.1371\cdot10^{+1}$&
$3.1336\cdot10^{+1}$&
$3.1372\cdot10^{+1}$&
$3.1371\cdot10^{+1}$&
$3.1371\cdot10^{+1}$&
$3.1371\cdot10^{+1}$&
$3.1371\cdot10^{+1}$
\tabularnewline
\hline
$0.01$&
$8.1407\cdot10^{+0}$&
$8.1433\cdot10^{+0}$&
$8.1403\cdot10^{+0}$&
$8.1407\cdot10^{+0}$&
$8.1407\cdot10^{+0}$&
$8.1407\cdot10^{+0}$&
$8.1407\cdot10^{+0}$
\tabularnewline
\hline
$0.1$&
$9.0545\cdot10^{-1}$&
$9.0662\cdot10^{-1}$&
$9.0538\cdot10^{-1}$&
$9.0545\cdot10^{-1}$&
$9.0545\cdot10^{-1}$&
$9.0545\cdot10^{-1}$&
$9.0545\cdot10^{-1}$
\tabularnewline
\hline
$0.2$&
$2.5753\cdot10^{-1}$&
$2.5793\cdot10^{-1}$&
$2.5750\cdot10^{-1}$&
$2.5752\cdot10^{-1}$&
$2.5753\cdot10^{-1}$&
$2.5753\cdot10^{-1}$&
$2.5753\cdot10^{-1}$
\tabularnewline
\hline
$0.3$&
$8.4120\cdot10^{-2}$&
$8.4266\cdot10^{-2}$&
$8.4112\cdot10^{-2}$&
$8.4119\cdot10^{-2}$&
$8.4120\cdot10^{-2}$&
$8.4120\cdot10^{-2}$&
$8.4120\cdot10^{-2}$
\tabularnewline
\hline
$0.4$&
$2.7238\cdot10^{-2}$&
$2.7288\cdot10^{-2}$&
$2.7235\cdot10^{-2}$&
$2.7238\cdot10^{-2}$&
$2.7238\cdot10^{-2}$&
$2.7238\cdot10^{-2}$&
$2.7238\cdot10^{-2}$
\tabularnewline
\hline
$0.5$&
$8.1019\cdot10^{-3}$&
$8.1176\cdot10^{-3}$&
$8.1009\cdot10^{-3}$&
$8.1018\cdot10^{-3}$&
$8.1019\cdot10^{-3}$&
$8.1019\cdot10^{-3}$&
$8.1019\cdot10^{-3}$
\tabularnewline
\hline
$0.6$&
$2.0476\cdot10^{-3}$&
$2.0518\cdot10^{-3}$&
$2.0473\cdot10^{-3}$&
$2.0476\cdot10^{-3}$&
$2.0477\cdot10^{-3}$&
$2.0476\cdot10^{-3}$&
$2.0476\cdot10^{-3}$
\tabularnewline
\hline
$0.7$&
$3.8845\cdot10^{-4}$&
$3.8929\cdot10^{-4}$&
$3.8837\cdot10^{-4}$&
$3.8845\cdot10^{-4}$&
$3.8845\cdot10^{-4}$&
$3.8845\cdot10^{-4}$&
$3.8845\cdot10^{-4}$
\tabularnewline
\hline
$0.8$&
$4.1738\cdot10^{-5}$&
$4.1842\cdot10^{-5}$&
$4.1724\cdot10^{-5}$&
$4.1738\cdot10^{-5}$&

$4.1738\cdot10^{-5}$&
$4.1738\cdot10^{-5}$&
$4.1738\cdot10^{-5}$
\tabularnewline
\hline
$0.9$&
$1.8861\cdot10^{-6}$&
$1.8899\cdot10^{-6}$&
$1.8853\cdot10^{-6}$&
$1.8862\cdot10^{-6}$&
$1.8861\cdot10^{-6}$&
$1.8861\cdot10^{-6}$&
$1.8861\cdot10^{-6}$
\tabularnewline
\hline
\end{tabular}
\caption{NNLO gluon density in the Les Houches model}
\label{tab4}
\end{footnotesize}
\end{table}


\begin{table}
\begin{footnotesize}
\hspace{-1cm}
\begin{tabular}{|c||c|c|c|c|c|c|c|}
\hline
\multicolumn{8}{|c|}{$xu_v(x)$ \textsc{Candia} evolution at NLO, MRST input, $\mu_0=1$ GeV, $Q=\mu_{F}=\mu_{R}=100$ GeV}
\tabularnewline
\hline
$ x $                       &
$x{u_v(x)}^{\textsc{Candia}}_{asymp}$     &
$\kappa=1$                 &
$\kappa=2$                 &
$\kappa=3$                 &
$\kappa=4$                 &
$\kappa=5$                 &
$\kappa=6$  \tabularnewline
\hline
\hline
$1e-05$&
$1.3874\cdot10^{-2}$&
$1.4223\cdot10^{-2}$&
$1.4015\cdot10^{-2}$&
$1.4049\cdot10^{-2}$&
$1.4043\cdot10^{-2}$&
$1.4044\cdot10^{-2}$&
$1.4044\cdot10^{-2}$
\tabularnewline
\hline
$0.0001$&
$2.9241\cdot10^{-2}$&
$3.0230\cdot10^{-2}$&
$2.9654\cdot10^{-2}$&
$2.9753\cdot10^{-2}$&
$2.9734\cdot10^{-2}$&
$2.9738\cdot10^{-2}$&
$2.9737\cdot10^{-2}$
\tabularnewline
\hline
$0.001$&
$7.2512\cdot10^{-2}$&
$7.5810\cdot10^{-2}$&
$7.4246\cdot10^{-2}$&
$7.4530\cdot10^{-2}$&
$7.4472\cdot10^{-2}$&
$7.4485\cdot10^{-2}$&
$7.4482\cdot10^{-2}$
\tabularnewline
\hline
$0.01$&
$2.1394\cdot10^{-1}$&
$2.2294\cdot10^{-1}$&
$2.2014\cdot10^{-1}$&
$2.2068\cdot10^{-1}$&
$2.2057\cdot10^{-1}$&
$2.2059\cdot10^{-1}$&
$2.2059\cdot10^{-1}$
\tabularnewline
\hline
$0.1$&
$5.1946\cdot10^{-1}$&
$5.1264\cdot10^{-1}$&
$5.1673\cdot10^{-1}$&
$5.1598\cdot10^{-1}$&
$5.1613\cdot10^{-1}$&
$5.1610\cdot10^{-1}$&
$5.1611\cdot10^{-1}$
\tabularnewline
\hline
$0.2$&
$4.7098\cdot10^{-1}$&
$4.4773\cdot10^{-1}$&
$4.5514\cdot10^{-1}$&
$4.5372\cdot10^{-1}$&
$4.5402\cdot10^{-1}$&
$4.5395\cdot10^{-1}$&
$4.5397\cdot10^{-1}$
\tabularnewline
\hline
$0.3$&
$3.3696\cdot10^{-1}$&
$3.1056\cdot10^{-1}$&
$3.1763\cdot10^{-1}$&
$3.1627\cdot10^{-1}$&
$3.1655\cdot10^{-1}$&
$3.1649\cdot10^{-1}$&
$3.1650\cdot10^{-1}$
\tabularnewline
\hline
$0.4$&
$2.0581\cdot10^{-1}$&
$1.8429\cdot10^{-1}$&
$1.8952\cdot10^{-1}$&
$1.8851\cdot10^{-1}$&
$1.8872\cdot10^{-1}$&
$1.8868\cdot10^{-1}$&
$1.8869\cdot10^{-1}$
\tabularnewline
\hline
$0.5$&
$1.0726\cdot10^{-1}$&
$9.3306\cdot10^{-2}$&
$9.6490\cdot10^{-2}$&
$9.5870\cdot10^{-2}$&
$9.6001\cdot10^{-2}$&
$9.5972\cdot10^{-2}$&
$9.5979\cdot10^{-2}$
\tabularnewline
\hline
$0.6$&
$4.5848\cdot10^{-2}$&
$3.8656\cdot10^{-2}$&
$4.0225\cdot10^{-2}$&
$3.9918\cdot10^{-2}$&
$3.9983\cdot10^{-2}$&
$3.9968\cdot10^{-2}$&
$3.9972\cdot10^{-2}$
\tabularnewline
\hline
$0.7$&
$1.4636\cdot10^{-2}$&
$1.1901\cdot10^{-2}$&
$1.2479\cdot10^{-2}$&
$1.2365\cdot10^{-2}$&
$1.2390\cdot10^{-2}$&
$1.2384\cdot10^{-2}$&
$1.2385\cdot10^{-2}$
\tabularnewline
\hline
$0.8$&
$2.7935\cdot10^{-3}$&
$2.1669\cdot10^{-3}$&
$2.2968\cdot10^{-3}$&
$2.2710\cdot10^{-3}$&
$2.2765\cdot10^{-3}$&
$2.2753\cdot10^{-3}$&
$2.2756\cdot10^{-3}$
\tabularnewline
\hline
$0.9$&
$1.5488\cdot10^{-4}$&
$1.1138\cdot10^{-4}$&
$1.2032\cdot10^{-4}$&
$1.1852\cdot10^{-4}$&
$1.1891\cdot10^{-4}$&
$1.1882\cdot10^{-4}$&
$1.1884\cdot10^{-4}$
\tabularnewline
\hline
\end{tabular}
\caption{ NLO distribution of the $u_v$ quark with MRST input evolved with \textsc{Candia}}
\label{tab5}
\end{footnotesize}
\end{table}

\begin{table}
\begin{footnotesize}
\hspace{-1cm}
\begin{tabular}{|c||c|c|c|c|c|c|c|}
\hline
\multicolumn{8}{|c|}{$xu_v(x)$ \textsc{Candia} evolution at NNLO, MRST input, $\mu_0=1$ GeV, $Q=\mu_{F}=\mu_{R}=100$ GeV}
\tabularnewline
\hline
$ x $                       &
$x{u_{v}(x)}^{\textsc{Candia}}_{asymp}$     &
$\kappa=2$                 &
$\kappa=3$                 &
$\kappa=4$                 &
$\kappa=5$                 &
$\kappa=6$                 &
$\kappa=7$  \tabularnewline
\hline
\hline
$1e-05$&
$1.2678\cdot10^{-2}$&
$1.3052\cdot10^{-2}$&
$1.2939\cdot10^{-2}$&
$1.2940\cdot10^{-2}$&
$1.2945\cdot10^{-2}$&
$1.2944\cdot10^{-2}$&
$1.2944\cdot10^{-2}$
\tabularnewline
\hline
$0.0001$&
$2.9047\cdot10^{-2}$&
$2.9858\cdot10^{-2}$&
$2.9616\cdot10^{-2}$&
$2.9627\cdot10^{-2}$&
$2.9636\cdot10^{-2}$&
$2.9634\cdot10^{-2}$&
$2.9634\cdot10^{-2}$
\tabularnewline
\hline
$0.001$&
$7.3813\cdot10^{-2}$&
$7.5701\cdot10^{-2}$&
$7.5337\cdot10^{-2}$&
$7.5391\cdot10^{-2}$&
$7.5399\cdot10^{-2}$&
$7.5395\cdot10^{-2}$&
$7.5395\cdot10^{-2}$
\tabularnewline
\hline
$0.01$&
$2.1492\cdot10^{-1}$&
$2.2015\cdot10^{-1}$&
$2.1997\cdot10^{-1}$&
$2.2009\cdot10^{-1}$&
$2.2008\cdot10^{-1}$&
$2.2007\cdot10^{-1}$&
$2.2007\cdot10^{-1}$
\tabularnewline
\hline
$0.1$&
$5.2810\cdot10^{-1}$&
$5.2525\cdot10^{-1}$&
$5.2612\cdot10^{-1}$&
$5.2599\cdot10^{-1}$&
$5.2597\cdot10^{-1}$&
$5.2598\cdot10^{-1}$&
$5.2598\cdot10^{-1}$
\tabularnewline
\hline
$0.2$&
$4.8293\cdot10^{-1}$&
$4.6760\cdot10^{-1}$&
$4.6869\cdot10^{-1}$&
$4.6841\cdot10^{-1}$&
$4.6841\cdot10^{-1}$&
$4.6842\cdot10^{-1}$&
$4.6842\cdot10^{-1}$
\tabularnewline
\hline
$0.3$&
$3.4556\cdot10^{-1}$&
$3.2689\cdot10^{-1}$&
$3.2781\cdot10^{-1}$&
$3.2753\cdot10^{-1}$&
$3.2753\cdot10^{-1}$&
$3.2755\cdot10^{-1}$&
$3.2755\cdot10^{-1}$
\tabularnewline
\hline
$0.4$&
$2.0973\cdot10^{-1}$&
$1.9409\cdot10^{-1}$&
$1.9473\cdot10^{-1}$&
$1.9451\cdot10^{-1}$&
$1.9452\cdot10^{-1}$&
$1.9453\cdot10^{-1}$&
$1.9453\cdot10^{-1}$
\tabularnewline
\hline
$0.5$&
$1.0794\cdot10^{-1}$&
$9.7714\cdot10^{-2}$&
$9.8084\cdot10^{-2}$&
$9.7952\cdot10^{-2}$&
$9.7958\cdot10^{-2}$&
$9.7964\cdot10^{-2}$&
$9.7963\cdot10^{-2}$
\tabularnewline
\hline
$0.6$&
$4.5227\cdot10^{-2}$&
$3.9989\cdot10^{-2}$&
$4.0165\cdot10^{-2}$&
$4.0101\cdot10^{-2}$&
$4.0104\cdot10^{-2}$&
$4.0107\cdot10^{-2}$&
$4.0106\cdot10^{-2}$
\tabularnewline
\hline
$0.7$&
$1.4011\cdot10^{-2}$&
$1.2057\cdot10^{-2}$&
$1.2121\cdot10^{-2}$&
$1.2097\cdot10^{-2}$&
$1.2098\cdot10^{-2}$&
$1.2099\cdot10^{-2}$&
$1.2099\cdot10^{-2}$
\tabularnewline
\hline
$0.8$&
$2.5508\cdot10^{-3}$&
$2.1205\cdot10^{-3}$&
$2.1343\cdot10^{-3}$&
$2.1291\cdot10^{-3}$&
$2.1294\cdot10^{-3}$&
$2.1297\cdot10^{-3}$&
$2.1296\cdot10^{-3}$
\tabularnewline
\hline
$0.9$&
$1.2957\cdot10^{-4}$&
$1.0200\cdot10^{-4}$&
$1.0290\cdot10^{-4}$&
$1.0257\cdot10^{-4}$&
$1.0258\cdot10^{-4}$&
$1.0260\cdot10^{-4}$&
$1.0260\cdot10^{-4}$
\tabularnewline
\hline
\end{tabular}
\caption{NNLO $u_v$ quark distribution evolved with \textsc{Candia} using the MRST input.}
\label{tab6}
\end{footnotesize}
\end{table}

\begin{table}
\begin{footnotesize}
\hspace{-1cm}
\begin{tabular}{|c||c|c|c|c|c|c|c|}
\hline
\multicolumn{8}{|c|}{$xg(x)$ \textsc{Candia} evolution at NLO, MRST input, $\mu_0=1$ GeV, $Q=\mu_{F}=\mu_{R}=100$ GeV}
\tabularnewline
\hline
$ x $                       &
$xg(x)^{\textsc{Candia}}_{asymp}$     &
$\kappa=1$                 &
$\kappa=2$                 &
$\kappa=3$                 &
$\kappa=4$                 &
$\kappa=5$                 &
$\kappa=6$  \tabularnewline
\hline
\hline
$1e-05$&
$1.9288\cdot10^{+2}$&
$1.8441\cdot10^{+2}$&
$1.9429\cdot10^{+2}$&
$1.9262\cdot10^{+2}$&
$1.9293\cdot10^{+2}$&
$1.9287\cdot10^{+2}$&
$1.9288\cdot10^{+2}$
\tabularnewline
\hline
$0.0001$&
$8.1300\cdot10^{+1}$&
$7.9222\cdot10^{+1}$&
$8.1613\cdot10^{+1}$&
$8.1245\cdot10^{+1}$&
$8.1310\cdot10^{+1}$&
$8.1297\cdot10^{+1}$&
$8.1300\cdot10^{+1}$
\tabularnewline
\hline
$0.001$&
$2.9001\cdot10^{+1}$&
$2.8675\cdot10^{+1}$&
$2.9043\cdot10^{+1}$&
$2.8995\cdot10^{+1}$&
$2.9002\cdot10^{+1}$&
$2.9001\cdot10^{+1}$&
$2.9001\cdot10^{+1}$
\tabularnewline
\hline
$0.01$&
$7.8335\cdot10^{+0}$&
$7.8282\cdot10^{+0}$&
$7.8328\cdot10^{+0}$&
$7.8339\cdot10^{+0}$&
$7.8334\cdot10^{+0}$&
$7.8336\cdot10^{+0}$&
$7.8335\cdot10^{+0}$
\tabularnewline
\hline
$0.1$&
$9.3962\cdot10^{-1}$&
$9.4520\cdot10^{-1}$&
$9.3881\cdot10^{-1}$&
$9.3976\cdot10^{-1}$&
$9.3959\cdot10^{-1}$&
$9.3963\cdot10^{-1}$&
$9.3962\cdot10^{-1}$
\tabularnewline
\hline
$0.2$&
$2.7632\cdot10^{-1}$&
$2.7832\cdot10^{-1}$&
$2.7605\cdot10^{-1}$&
$2.7636\cdot10^{-1}$&
$2.7631\cdot10^{-1}$&
$2.7632\cdot10^{-1}$&
$2.7632\cdot10^{-1}$
\tabularnewline
\hline
$0.3$&
$9.1622\cdot10^{-2}$&
$9.2347\cdot10^{-2}$&
$9.1533\cdot10^{-2}$&
$9.1635\cdot10^{-2}$&
$9.1619\cdot10^{-2}$&
$9.1622\cdot10^{-2}$&
$9.1622\cdot10^{-2}$
\tabularnewline
\hline
$0.4$&
$2.9629\cdot10^{-2}$&
$2.9876\cdot10^{-2}$&
$2.9600\cdot10^{-2}$&
$2.9633\cdot10^{-2}$&
$2.9628\cdot10^{-2}$&
$2.9629\cdot10^{-2}$&
$2.9629\cdot10^{-2}$
\tabularnewline
\hline
$0.5$&
$8.6249\cdot10^{-3}$&
$8.6999\cdot10^{-3}$&
$8.6170\cdot10^{-3}$&
$8.6260\cdot10^{-3}$&
$8.6248\cdot10^{-3}$&
$8.6249\cdot10^{-3}$&
$8.6249\cdot10^{-3}$
\tabularnewline
\hline
$0.6$&
$2.0714\cdot10^{-3}$&
$2.0901\cdot10^{-3}$&
$2.0697\cdot10^{-3}$&
$2.0716\cdot10^{-3}$&
$2.0714\cdot10^{-3}$&
$2.0714\cdot10^{-3}$&
$2.0714\cdot10^{-3}$
\tabularnewline
\hline
$0.7$&
$3.5704\cdot10^{-4}$&
$3.6038\cdot10^{-4}$&
$3.5677\cdot10^{-4}$&
$3.5707\cdot10^{-4}$&
$3.5704\cdot10^{-4}$&
$3.5704\cdot10^{-4}$&
$3.5704\cdot10^{-4}$
\tabularnewline
\hline
$0.8$&
$3.2860\cdot10^{-5}$&
$3.3189\cdot10^{-5}$&
$3.2839\cdot10^{-5}$&
$3.2861\cdot10^{-5}$&
$3.2861\cdot10^{-5}$&
$3.2860\cdot10^{-5}$&
$3.2860\cdot10^{-5}$
\tabularnewline
\hline
$0.9$&
$6.4223\cdot10^{-7}$&
$6.4990\cdot10^{-7}$&
$6.4192\cdot10^{-7}$&
$6.4218\cdot10^{-7}$&
$6.4225\cdot10^{-7}$&
$6.4222\cdot10^{-7}$&
$6.4223\cdot10^{-7}$
\tabularnewline
\hline
\end{tabular}
\caption{NLO gluon density with MRST input, evolved with \textsc{Candia}.}
\label{tab7}
\end{footnotesize}
\end{table}

\begin{table}
\begin{footnotesize}
\hspace{-1cm}
\begin{tabular}{|c||c|c|c|c|c|c|c|}
\hline
\multicolumn{8}{|c|}{$xg(x)$ \textsc{Candia} evolution at NNLO, MRST input, $\mu_0=1$ GeV, $Q=\mu_{F}=\mu_{R}=100$ GeV}
\tabularnewline
\hline
$ x $                       &
$xg(x)^{\textsc{Candia}}_{asymp}$     &
$\kappa=2$                 &
$\kappa=3$                 &
$\kappa=4$                 &
$\kappa=5$                 &
$\kappa=6$                 &
$\kappa=7$  \tabularnewline
\hline
\hline
$1e-05$&
$1.6068\cdot10^{+2}$&
$1.5727\cdot10^{+2}$&
$1.6089\cdot10^{+2}$&
$1.6075\cdot10^{+2}$&
$1.6063\cdot10^{+2}$&
$1.6066\cdot10^{+2}$&
$1.6066\cdot10^{+2}$
\tabularnewline
\hline
$0.0001$&
$7.1188\cdot10^{+1}$&
$7.0542\cdot10^{+1}$&
$7.1229\cdot10^{+1}$&
$7.1197\cdot10^{+1}$&
$7.1176\cdot10^{+1}$&
$7.1181\cdot10^{+1}$&
$7.1181\cdot10^{+1}$
\tabularnewline
\hline
$0.001$&
$2.6591\cdot10^{+1}$&
$2.6526\cdot10^{+1}$&
$2.6596\cdot10^{+1}$&
$2.6591\cdot10^{+1}$&
$2.6589\cdot10^{+1}$&
$2.6590\cdot10^{+1}$&
$2.6590\cdot10^{+1}$
\tabularnewline
\hline
$0.01$&
$7.5377\cdot10^{+0}$&
$7.5426\cdot10^{+0}$&
$7.5373\cdot10^{+0}$&
$7.5374\cdot10^{+0}$&
$7.5377\cdot10^{+0}$&
$7.5376\cdot10^{+0}$&
$7.5376\cdot10^{+0}$
\tabularnewline
\hline
$0.1$&
$9.8748\cdot10^{-1}$&
$9.8964\cdot10^{-1}$&
$9.8738\cdot10^{-1}$&
$9.8748\cdot10^{-1}$&
$9.8755\cdot10^{-1}$&
$9.8753\cdot10^{-1}$&
$9.8753\cdot10^{-1}$
\tabularnewline
\hline
$0.2$&
$3.0191\cdot10^{-1}$&
$3.0265\cdot10^{-1}$&
$3.0188\cdot10^{-1}$&
$3.0192\cdot10^{-1}$&
$3.0194\cdot10^{-1}$&
$3.0193\cdot10^{-1}$&
$3.0193\cdot10^{-1}$
\tabularnewline
\hline
$0.3$&
$1.0210\cdot10^{-1}$&
$1.0235\cdot10^{-1}$&
$1.0210\cdot10^{-1}$&
$1.0211\cdot10^{-1}$&
$1.0211\cdot10^{-1}$&
$1.0211\cdot10^{-1}$&
$1.0211\cdot10^{-1}$
\tabularnewline
\hline
$0.4$&
$3.3271\cdot10^{-2}$&
$3.3345\cdot10^{-2}$&
$3.3270\cdot10^{-2}$&
$3.3274\cdot10^{-2}$&
$3.3275\cdot10^{-2}$&
$3.3275\cdot10^{-2}$&
$3.3275\cdot10^{-2}$
\tabularnewline
\hline
$0.5$&
$9.6593\cdot10^{-3}$&
$9.6776\cdot10^{-3}$&
$9.6592\cdot10^{-3}$&
$9.6606\cdot10^{-3}$&
$9.6607\cdot10^{-3}$&
$9.6607\cdot10^{-3}$&
$9.6607\cdot10^{-3}$
\tabularnewline
\hline
$0.6$&
$2.2882\cdot10^{-3}$&
$2.2915\cdot10^{-3}$&
$2.2882\cdot10^{-3}$&
$2.2885\cdot10^{-3}$&
$2.2885\cdot10^{-3}$&
$2.2885\cdot10^{-3}$&
$2.2885\cdot10^{-3}$
\tabularnewline
\hline
$0.7$&
$3.8337\cdot10^{-4}$&
$3.8368\cdot10^{-4}$&
$3.8339\cdot10^{-4}$&
$3.8345\cdot10^{-4}$&
$3.8343\cdot10^{-4}$&
$3.8343\cdot10^{-4}$&
$3.8343\cdot10^{-4}$
\tabularnewline
\hline
$0.8$&
$3.3492\cdot10^{-5}$&
$3.3487\cdot10^{-5}$&
$3.3495\cdot10^{-5}$&
$3.3501\cdot10^{-5}$&
$3.3497\cdot10^{-5}$&
$3.3498\cdot10^{-5}$&
$3.3498\cdot10^{-5}$
\tabularnewline
\hline
$0.9$&
$5.8939\cdot10^{-7}$&
$5.8839\cdot10^{-7}$&
$5.8951\cdot10^{-7}$&
$5.8960\cdot10^{-7}$&
$5.8947\cdot10^{-7}$&
$5.8950\cdot10^{-7}$&
$5.8950\cdot10^{-7}$
\tabularnewline
\hline
\end{tabular}
\caption{NNLO gluon density evolved with \textsc{Candia} using the MRST input.}
\label{tab8}
\end{footnotesize}
\end{table}

\begin{table}
\begin{footnotesize}
\hspace{-1 cm}
\begin{tabular}{|c||c|c|c|c|c|c|c|}
\hline
\multicolumn{8}{|c|}{$d\sigma^{NLO}/dQ$ [pb/GeV] with MRST input, $\mu_0^2=1.25$ GeV$^2$, \textsc{Candia} evolution, $\mu_R=\mu_F=Q$, $\sqrt{S}=14$ TeV}
\tabularnewline
\hline
$ Q $&
$\sigma_{NLO}$ asym.&
$\kappa=1$&
$\kappa=2$&
$\kappa=3$&
$\kappa=4$&
$\kappa=5$&
$\kappa=6$\tabularnewline
\hline
\hline
$50.0000$&
$6.8121\cdot10^{+0}$&
$7.0043\cdot10^{+0}$&
$6.7706\cdot10^{+0}$&
$6.8020\cdot10^{+0}$&
$6.7972\cdot10^{+0}$&
$6.7980\cdot10^{+0}$&
$6.7978\cdot10^{+0}$
\tabularnewline
\hline
$60.0469$&
$4.1554\cdot10^{+0}$&
$4.2645\cdot10^{+0}$&
$4.1317\cdot10^{+0}$&
$4.1495\cdot10^{+0}$&
$4.1468\cdot10^{+0}$&
$4.1473\cdot10^{+0}$&
$4.1472\cdot10^{+0}$
\tabularnewline
\hline
$70.0938$&
$3.8112\cdot10^{+0}$&
$3.9073\cdot10^{+0}$&
$3.7916\cdot10^{+0}$&
$3.8071\cdot10^{+0}$&
$3.8047\cdot10^{+0}$&
$3.8051\cdot10^{+0}$&
$3.8051\cdot10^{+0}$
\tabularnewline
\hline
$80.1407$&
$7.9374\cdot10^{+0}$&
$8.1327\cdot10^{+0}$&
$7.9013\cdot10^{+0}$&
$7.9327\cdot10^{+0}$&
$7.9278\cdot10^{+0}$&
$7.9287\cdot10^{+0}$&
$7.9285\cdot10^{+0}$
\tabularnewline
\hline
$90.1876$&
$3.0658\cdot10^{+2}$&
$3.1385\cdot10^{+2}$&
$3.0529\cdot10^{+2}$&
$3.0646\cdot10^{+2}$&
$3.0627\cdot10^{+2}$&
$3.0630\cdot10^{+2}$&
$3.0630\cdot10^{+2}$
\tabularnewline
\hline
$91.1876$&
$5.0243\cdot10^{+2}$&
$5.1429\cdot10^{+2}$&
$5.0033\cdot10^{+2}$&
$5.0223\cdot10^{+2}$&
$5.0193\cdot10^{+2}$&
$5.0198\cdot10^{+2}$&
$5.0197\cdot10^{+2}$
\tabularnewline
\hline
$120.0701$&
$9.5681\cdot10^{-1}$&
$9.7592\cdot10^{-1}$&
$9.5264\cdot10^{-1}$&
$9.5582\cdot10^{-1}$&
$9.5531\cdot10^{-1}$&
$9.5540\cdot10^{-1}$&
$9.5538\cdot10^{-1}$
\tabularnewline
\hline
$146.0938$&
$2.6563\cdot10^{-1}$&
$2.7026\cdot10^{-1}$&
$2.6442\cdot10^{-1}$&
$2.6522\cdot10^{-1}$&
$2.6509\cdot10^{-1}$&
$2.6511\cdot10^{-1}$&
$2.6511\cdot10^{-1}$
\tabularnewline
\hline
$172.1175$&
$1.1382\cdot10^{-1}$&
$1.1558\cdot10^{-1}$&
$1.1328\cdot10^{-1}$&
$1.1360\cdot10^{-1}$&
$1.1355\cdot10^{-1}$&
$1.1356\cdot10^{-1}$&
$1.1355\cdot10^{-1}$
\tabularnewline
\hline
$200.0000$&
$5.5942\cdot10^{-2}$&
$5.6694\cdot10^{-2}$&
$5.5662\cdot10^{-2}$&
$5.5805\cdot10^{-2}$&
$5.5782\cdot10^{-2}$&
$5.5786\cdot10^{-2}$&
$5.5785\cdot10^{-2}$
\tabularnewline
\hline
\end{tabular}
\caption{Drell-Yan cross section at NLO computed with the MRST parametric input ($\mu_0^2=1.25$ GeV$^2$)
and the evolution performed using \textsc{Candia}. Shown are the cross sections for the truncated
solutions and the asymptotic cross section.}
\label{sigma_1}
\end{footnotesize}
\end{table}

\begin{table}
\begin{footnotesize}
\hspace{-1 cm}
\begin{tabular}{|c||c|c|c|c|c|c|c|}
\hline
\multicolumn{8}{|c|}{$d\sigma^{NNLO}/dQ$ [pb/GeV] with MRST input, $\mu_0^2=1.25$, GeV$^2$ \textsc{Candia} evolution, $\mu_R=\mu_F=Q$, $\sqrt{S}=14$ TeV}
\tabularnewline
\hline
$ Q $&
$\sigma_{NNLO}$ asym.&
$\kappa=2$&
$\kappa=3$&
$\kappa=4$&
$\kappa=5$&
$\kappa=6$&
$\kappa=7$\tabularnewline
\hline
\hline
$50.0000$&
$6.4940\cdot10^{+0}$&
$6.5052\cdot10^{+0}$&
$6.4758\cdot10^{+0}$&
$6.4807\cdot10^{+0}$&
$6.4805\cdot10^{+0}$&
$6.4803\cdot10^{+0}$&
$6.4804\cdot10^{+0}$
\tabularnewline
\hline
$60.0469$&
$3.9989\cdot10^{+0}$&
$4.0040\cdot10^{+0}$&
$3.9886\cdot10^{+0}$&
$3.9911\cdot10^{+0}$&
$3.9911\cdot10^{+0}$&
$3.9910\cdot10^{+0}$&
$3.9910\cdot10^{+0}$
\tabularnewline
\hline
$70.0938$&
$3.6948\cdot10^{+0}$&
$3.6995\cdot10^{+0}$&
$3.6868\cdot10^{+0}$&
$3.6888\cdot10^{+0}$&
$3.6888\cdot10^{+0}$&
$3.6887\cdot10^{+0}$&
$3.6887\cdot10^{+0}$
\tabularnewline
\hline
$80.1407$&
$7.6740\cdot10^{+0}$&
$7.6871\cdot10^{+0}$&
$7.6600\cdot10^{+0}$&
$7.6641\cdot10^{+0}$&
$7.6642\cdot10^{+0}$&
$7.6640\cdot10^{+0}$&
$7.6640\cdot10^{+0}$
\tabularnewline
\hline
$90.1876$&
$2.9335\cdot10^{+2}$&
$2.9395\cdot10^{+2}$&
$2.9283\cdot10^{+2}$&
$2.9299\cdot10^{+2}$&
$2.9299\cdot10^{+2}$&
$2.9299\cdot10^{+2}$&
$2.9299\cdot10^{+2}$
\tabularnewline
\hline
$91.1876$&
$4.8027\cdot10^{+2}$&
$4.8124\cdot10^{+2}$&
$4.7940\cdot10^{+2}$&
$4.7966\cdot10^{+2}$&
$4.7967\cdot10^{+2}$&
$4.7966\cdot10^{+2}$&
$4.7966\cdot10^{+2}$
\tabularnewline
\hline
$120.0701$&
$9.0552\cdot10^{-1}$&
$9.0678\cdot10^{-1}$&
$9.0326\cdot10^{-1}$&
$9.0373\cdot10^{-1}$&
$9.0376\cdot10^{-1}$&
$9.0373\cdot10^{-1}$&
$9.0373\cdot10^{-1}$
\tabularnewline
\hline
$146.0938$&
$2.5318\cdot10^{-1}$&
$2.5334\cdot10^{-1}$&
$2.5243\cdot10^{-1}$&
$2.5255\cdot10^{-1}$&
$2.5256\cdot10^{-1}$&
$2.5255\cdot10^{-1}$&
$2.5255\cdot10^{-1}$
\tabularnewline
\hline
$172.1175$&
$1.0963\cdot10^{-1}$&
$1.0963\cdot10^{-1}$&
$1.0927\cdot10^{-1}$&
$1.0931\cdot10^{-1}$&
$1.0932\cdot10^{-1}$&
$1.0931\cdot10^{-1}$&
$1.0931\cdot10^{-1}$
\tabularnewline
\hline
$200.0000$&
$5.4572\cdot10^{-2}$&
$5.4533\cdot10^{-2}$&
$5.4367\cdot10^{-2}$&
$5.4388\cdot10^{-2}$&
$5.4390\cdot10^{-2}$&
$5.4389\cdot10^{-2}$&
$5.4389\cdot10^{-2}$
\tabularnewline
\hline
\end{tabular}
\caption{NNLO Drell-Yan cross section with the MRST initial conditions and the evolution performed with \textsc{Candia}.
We present the results for the various truncated solutions and for the asymptotic one.}
\label{sigma_2}
\end{footnotesize}
\end{table}

\end{document}